\pdfoutput=1
\documentclass[12pt]{iopart}

\usepackage{iopams}  
\expandafter\let\csname equation*\endcsname=\relax
\expandafter\let\csname endequation*\endcsname=\relax

\usepackage{physics}                            
\usepackage{bm}									
\usepackage{enumerate}							
\usepackage{amssymb}							
\usepackage[dvipsnames]{xcolor}					
\usepackage[colorlinks=true,allcolors=RoyalBlue]{hyperref}
\hypersetup{colorlinks=true}
\usepackage[capitalize]{cleveref}               
\usepackage{graphicx,color}
\usepackage{xcolor}
\usepackage{bbold}                              
\usepackage{cite}
\usepackage{tikz}
\usepackage[utf8]{inputenc}
\usepackage{braket}
\usepackage[normalem]{ulem} 
\usepackage{comment}

\newcommand{\rhoemp}{\rho^{(\mathrm{e})}}

\makeatletter
\renewcommand{\thefootnote}{\arabic{footnote}}
\def\@fnsymbol#1{\arabic{#1}}
\renewcommand\@makefnmark{\hbox{\textsuperscript{\thefootnote}}}
\long\def\@makefntext#1{\noindent\hbox{\textsuperscript{\thefootnote}}#1} 
\makeatother
\usepackage{etoolbox}
\makeatletter
\newrobustcmd{\fixappendix}{%
  \patchcmd{\l@section}{1.5em}{7em}{}{}%
  \patchcmd{\l@subsection}{2.3em}{7em}{}{}%
}
\makeatother

\newcommand{\thetitle}{The Wishart--Rosenzweig--Porter random matrix ensemble}

\begin{document}
\title[\thetitle]{\thetitle}
\author{Victor Delapalme$^{1,2}$, Leticia F.~Cugliandolo$^{3,4}$, Grégory Schehr$^{3}$, Marco Tarzia$^{1,4}$, and Davide Venturelli$^{1}$}
\address{$^1$ Sorbonne Université, Laboratoire de Physique Théorique de la Matière Condensée, CNRS-UMR 7600, 4 Place Jussieu, 75252 Paris Cedex 05, France}
\address{$^2$ Department of Mathematics, King’s College London, Strand, London, WC2R 2LS, UK}
\address{$^3$ Sorbonne Université, Laboratoire de Physique Théorique et Hautes Energies, CNRS-UMR 7589, 4 Place Jussieu, 75252 Paris Cedex 05, France}
\address{$^4$ Institut Universitaire de France, 1 rue Descartes, 75005 Paris, France}

\begin{abstract}
In recent years the Rosenzweig--Porter (RP) ensemble, obtained by adding a diagonal matrix with independent and identically distributed elements to a Gaussian random matrix, has been widely used as a minimal model for the emergence of fractal eigenstates in complex many-body systems. 
A key open question concerns the robustness of its phase diagram when the assumption of independent  and uncorrelated entries is relaxed --- an assumption that simplifies its analysis, but is generally violated in realistic quantum systems. 
In this work, we take a first step in this direction by considering a deformed Wishart (rather than Gaussian) random matrix, which we dub the ``Wishart--RP'' ensemble. 
Using perturbation theory, as well as the cavity and replica methods and the Dyson Brownian motion approach, we characterize its phase diagram and localization properties. 
Remarkably, we show that the level compressibility, which quantifies spectral correlations in the fractal phase, coincides with that of the Gaussian RP model, thereby extending the universality conjectured in [SciPost Phys.~{\bf 14}, 110~(2023)] beyond the fully uncorrelated setting. 
We confirm our results with numerical tests.
\end{abstract} 
\vspace{10pt}
\begin{indented}
\item[]\today
\end{indented}

\tableofcontents

\markboth{\thetitle}{\thetitle}

\setcounter{footnote}{0} 

\section{Introduction}
\label{sec:introduction}

Random matrix theory (RMT)~\cite{Livan_2018,potters2020first,Mingo2017} has long been an invaluable framework for describing and understanding complex physical systems. Its power lies in universality: many RMT results are largely independent of the specific distribution of matrix entries, making them applicable across a broad variety of physical contexts. One prominent example is quantum chaos and its breakdown. In particular, RMT underpins our current understanding of quantum
ergodicity, formalized through the eigenstate thermalization hypothesis (ETH)~\cite{srednicki1994chaos,rigol2008thermalization}.

In recent years, a large body of work has suggested that disordered interacting quantum systems may violate ETH under certain conditions. A paradigmatic example is many-body localization (MBL~\cite{Gornyi_2005,basko2006metal}, see Refs.~\cite{reviewMBL,reviewMBL2,reviewMBL3,reviewMBL4,reviewMBL5,sierant2024manybody} for recent reviews), which occurs at strong disorder. 
More generally, the emergence of multifractal eigenstates --- i.e.~states that do not uniformly explore the accessible Hilbert space, thereby violating ETH and quantum ergodicity, hence often called \textit{non-ergodic}\footnote{Similarly, fully-delocalized eigenstates that satisfy ETH are often called \textit{ergodic}~\cite{srednicki1994chaos,rigol2008thermalization}. Strictly speaking, however, in the context of single-particle non-interacting problems such as the one considered here, the concept of ergodicity is not sharply defined (especially because the eigenvalues lack the extensive scaling typical of interacting systems). Nevertheless, following the common usage in the literature, throughout this paper we will use the term non-ergodic as a synonym of (multi)fractal.} --- has been identified as a robust feature of the phase diagram of such systems~\cite{mace2019multifractal,de2021rare,gornyi2017spectral,tarzia2020many,Luitz_2015,Serbyn_2017,Tikhonov_2018,Luitz_2020,Faoro_2019,baldwin2018quantum,biroli2021out,parolini2020multifractal,kechedzhi2018efficient,pino2017multifractal,pino2016nonergodic,cugliandolo2024multifractal,biroli2024large}, 
at the origin of several of their unconventional properties. 
To capture these phenomena, simple random-matrix models have been proposed and extensively studied~\cite{Kravtsov_2015,vonSoosten_2019,Facoetti_2016,Truong_2016,Bogomolny_2018,DeTomasi_2019,amini2017spread,pino2019ergodic,berkovits2020super,kravtsov2020localization,khaymovich2020fragile,monthus2017multifractality,biroli2021levy,buijsman2022circular,khaymovich2021dynamical,roy2018multifractality,wang2016phase,nosov2019correlation,duthie2022anomalous,kutlin2021emergent,motamarri2021localization,tang2022non,tarzia2022fully,Skvortsov_2022,Cai_2013,DeGottardi_2013,Liu_2015,Das_2022,Ahmed_2022,Lee_2022,Venturelli_2023,Das2023,Buijsman2024,cadez2024,LeDoussal2025,Baron2025,jahnke2025,Lunkin2025,safonova2025}. The philosophy behind these approaches is that RMT offers analytically tractable models capable of explaining general and universal features of systems that otherwise resist analytic treatment.

The most prominent such model is the (generalized) Gaussian Rosenzweig--Porter (GRP)
ensemble~\cite{RP_1960,Kravtsov_2015}. It is defined as the sum of two independent $N \times N$ matrices, namely a diagonal random matrix  $\textbf{A}$ with 
independent and identically distributed (i.i.d.) entries, and a  matrix $\textbf{B}$ from the 
Gaussian Orthogonal Ensemble (GOE) 
with random elements with zero mean and variance of $\mathcal{O}(1)$,
\begin{equation}
    \textbf{H}_{\rm GRP} = \textbf{A} + \nu N^{-\gamma/2}\, \textbf{B} \;.
\end{equation}
In the prefactor $\nu N^{-\gamma/2}$, the parameter $\nu$ is of $\mathcal{O}(1)$.
The physical interpretation is intuitive: each site of the reference space (matrix index) corresponds to a configuration of the system with a random on-site energy drawn from $\textbf{A}$, while transitions between configurations are mediated by Gaussian-distributed amplitudes from $\textbf{B}$. The spectral properties of the model are controlled by the parameter $\gamma$. This model provides a prototypical example of a system that exhibits an intermediate non-ergodic extended phase ($1 < \gamma<2$), characterized by fractal eigenstates and unconventional spectral properties, lying between a fully delocalized phase ($\gamma<1$) and a fully Anderson-localized phase ($\gamma>2$)~\cite{Kravtsov_2015}.

For this reason, the RP model and its generalizations have recently received renewed attention as a playground to explore the nature and properties of non-ergodic extended states~\cite{Kravtsov_2015,vonSoosten_2019,Facoetti_2016,Truong_2016,Bogomolny_2018,DeTomasi_2019,amini2017spread,pino2019ergodic,berkovits2020super,kravtsov2020localization,khaymovich2020fragile,monthus2017multifractality,biroli2021levy,buijsman2022circular,khaymovich2021dynamical,sarkar2021mobility,roy2018multifractality,wang2016phase,nosov2019correlation,duthie2022anomalous,kutlin2021emergent,motamarri2021localization,tang2022non,tarzia2022fully,Skvortsov_2022,Cai_2013,DeGottardi_2013,Liu_2015,Das_2022,Ahmed_2022,Lee_2022,Venturelli_2023,Das2023,Sarkar_2023,Buijsman2024,cadez2024,LeDoussal2025,Baron2025,jahnke2025,Ghosh_2025}. A key open question is the extent to which the spectral properties of the RP ensemble are robust under modifications of the distribution of the Hamiltonian matrix elements. One of the main motivations of this work is to address precisely this issue. In particular, we will focus on the spectral compressibility, closely related to the two-point spectral correlation function. This quantity,
the definition of which is recalled in Sec.~\ref{sec:summary},
displays distinct behaviors in the delocalized regime (with level repulsion described by RMT) and in the localized regime (with uncorrelated Poisson statistics).

In a recent work~\cite{Venturelli_2023}, some of us derived the exact scaling function describing the crossover between these two regimes in the intermediate phase of the GRP model. There, we showed that this scaling function is universal with respect to the distribution of the entries of $\textbf{A}$. More recently, a generalization of the RP model where the matrix $\textbf{B}$ is 
drawn from the Lévy ensemble~\cite{monthus2017multifractality,biroli2021levy,Lunkin2025} (with i.i.d.~entries and power-law tails) was also shown\footnote{Strictly speaking, the authors of Ref.~\cite{safonova2025} actually computed the two-point density-density correlation function $\langle \rho (\omega_1) \rho (\omega_2) \rangle_c$, which is related to the level compressibility via the relation in Eq.~(\ref{eq:k2-rhorho}) below.\label{foot:levy}} to yield the same universal scaling function~\cite{safonova2025}.

In this paper, we introduce another generalization of the RP model: the Wishart--Rosenzweig--Porter (WRP) ensemble, in which $\textbf{B}$ is 
drawn from the Wishart ensemble~\cite{Wishart1928}, a fundamental class of random matrices~\cite{Livan_2018,potters2020first}.
An ensuing difference with respect to the standard GRP case is that in the Wishart case the matrix $\textbf{B}$ is positive definite, with all positive eigenvalues, thereby producing an asymmetric rightward shift of the energy levels of $\textbf{A}$. More importantly, the entries of $\textbf{B}$ are statistically dependent, featuring non-zero higher order correlations, even though the pairwise correlations vanish (see below). Using a combination of complementary approaches, including perturbation theory, the cavity method, the replica method, and the Dyson Brownian motion, we obtain
the full phase diagram of the WRP ensemble, and study its spectral properties in its different phases. We then compute the spectral compressibility and the two-point correlation function in the intermediate regime, showing that the same universal crossover scaling function
as in the GRP model
emerges, although the elements of $\textbf{B}$ are not independent. 
These predictions are fully supported by exact diagonalization numerics.

Our results support the idea that, at least for models where the intermediate phase is characterized by a simple fractal (rather than multifractal) spectrum with compact mini-bands (see Sec.~\ref{sec:summary} for a precise definition), the crossover from RMT universality to Poisson statistics is genuinely universal. 
This calls for a systematic numerical analysis of the crossover function in more realistic many-body systems.

Finally, we stress that the implications of our results extend beyond toy models of ergodicity breaking in disordered quantum systems. Indeed, variants of the WRP model are relevant in other scientific contexts as well. For example, in denoising problems, whose aim is to recover
a signal hidden in a noisy covariance matrix of many correlated time series, this ensemble provides a mathematically controlled framework to characterize the statistical structure of noise at different scales, and to separate it from the meaningful signal~\cite{bun2017cleaning,Bouchaud_Les_Houches,Semerjian2024,ledoit2011eigenvectors,bun2016rotational}. Another example are generative machine learning models: from the perspective of generative diffusion models, which gradually transform an initial random state through a learned reverse diffusion process, the WRP model naturally emerges in the late stages of the backward diffusion process (see e.g.~Ref.~\cite{Marc_2023}).

The 
rest of the presentation
is organized as follows. In Sec.~\ref{sec:WRP-model} we define the
WRP
ensemble, and in Sec.~\ref{sec:summary} we present a 
summary of our results.  
Next, we 
employ multiple methods, namely perturbation theory 
in Sec.~\ref{sec:criteria}, the cavity method in Sec.~\ref{sec:cavity}, 
the replica method in Sec.~\ref{sec:replica}, and the Dyson Brownian motion
in Sec.~\ref{sec:Dyson}, to derive the 
phase diagram, the average density of eigenvalues, 
and the spectral compressibility. 
We support our results with exact diagonalization data, which we discuss throughout the paper.
Finally, in Sec.~\ref{sec:conclusions} 
we conclude and mention some directions for future research.

\section{The Wishart--Rosenzweig--Porter ensemble}
\label{sec:WRP-model}

The WRP ensemble is defined as a sum of two $N \times N$ random matrices $\textbf{A}$ and $\textbf{B}$:
\begin{equation} 
    \label{eq:AB}
    \textbf{H} = \textbf{A} + \nu \textbf{B} \, ,
\end{equation}
where $\nu $ is a constant of 
$\mathcal{O}(1)$.
Here, $\textbf{A}$ is a random diagonal matrix $A_{ij} = a_i \delta_{ij}$, where the $a_i$'s are $N$ independent and identically distributed random variables with probability density $p_a$ (we assume $\langle a \rangle = 0$, 
and $p_a(0) > 0$). 
In Eq.~(\ref{eq:AB}), $\textbf{B}$ is a Wishart matrix:
\begin{equation} \label{eq:WW}
    \textbf{B} = M^{- \gamma} \,  \textbf{WW}^T \, ,
\end{equation}
where $\textbf{W}$ is an $N \times M$ rectangular matrix with i.i.d.~Gaussian distributed entries, i.e.
\begin{equation}
    \rho_W[\textbf{W}] = \frac{1}{(2 \pi)^{NM/2}} \prod_{i=1}^{N} \prod_{\ell=1}^{M}  e^{-W_{i \ell}^2/2} 
    \, , 
    \label{eq:pdf-W}
\end{equation}
such that $\langle W_{i\ell} \rangle = 0$ and $\langle W_{i\ell}^2 \rangle = 1$. 
We define 
$c = N/M \leq 1$, 
and 
we restrict ourselves to the case in which $N$ and $M$ are of the same order, i.e.~$c$ is of 
$\mathcal{O}(1)$.
The elements of the matrix $\textbf{B}$ are given by
\begin{equation} \label{eq:Bij}
\begin{aligned}
    B_{ij} & = M^{- \gamma} \sum_{\ell = 1}^M W_{i \ell} W_{j \ell} \, , 
\end{aligned}
\end{equation}
hence one has 
\begin{equation}
\langle B_{ii} \rangle = M^{1 - \gamma}, \qquad 
\langle B_{ij} \rangle = 0, \qquad 
\langle B_{ij}^2 \rangle = M^{1 - 2 \gamma} \quad \text{for } i \neq j \, .
\end{equation}
Note that the matrix $\textbf{B}$ in Eq.~(\ref{eq:Bij}) is positive definite, i.e.~with all positive eigenvalues.
In the large-$N$ limit, the eigenvalues of $\textbf{B}$ are distributed according to the celebrated Mar\v{c}enko--Pastur law~\cite{Livan_2018,potters2020first}. 
Since $\langle \mathrm{Tr}\, \textbf{B} \rangle \propto N M^{1-\gamma}  \propto N^{2 - \gamma}$,
it follows that the eigenvalues of $\textbf{B}$ (and, consequently, the support of the Mar\v{c}enko--Pastur distribution) scale as $N^{1 - \gamma}$. In particular, they grow with $N$ for $\gamma < 1$, decrease with $N$ for $\gamma > 1$, and remain of 
$\mathcal{O}(1)$ for $\gamma = 1$.

At first sight, the ensemble~\eqref{eq:AB} appears similar to the model introduced in Ref.~\cite{kutlin2021emergent}, where the matrix $\mathbf{B}$ is likewise constructed as a sum of independent projectors. However, there is an important distinction between the two cases: in Ref.~\cite{kutlin2021emergent}, the matrix $\mathbf{B}$ is defined so that its energy levels are independent random variables, and therefore follow the Poisson statistics. In contrast, in our setting, $\mathbf{B}$ 
has a rotationally invariant statistical weight,
and its energy levels exhibit the universal correlations of random matrix theory at all scales. Although the phase diagrams of the two models share qualitative similarities (see Sec.~\ref{sec:spectrum}below), this distinction results in fundamentally different spectral correlations. In this sense, the WRP model studied here is more closely related to the standard GRP ensemble than to the model of Ref.~\cite{kutlin2021emergent}.

Nonetheless, the WRP ensemble also differs from the standard GRP ensemble in a key aspect: the entries of $\mathbf{B}$ are statistically dependent. 
In particular, from Eq.~\eqref{eq:Bij} it is straightforward to show that,  albeit the connected two-point correlations vanish, the correlations between $n$-tuples of matrix elements with repeated (but different) indices are given by  
\begin{equation} \label{eq:correlations}
\langle B_{i_1 i_2} B_{i_2 i_3} \cdots B_{i_n i_1} \rangle = M^{\,1 - n \gamma} \, .
\end{equation}
Note that, since $M = N/c$, the parameter $c$ has the effect of tuning these correlations (at fixed $N$), the maximally correlated case corresponding to the $c \to 0$ limit.

\subsection{Summary of the main results}
\label{sec:summary}

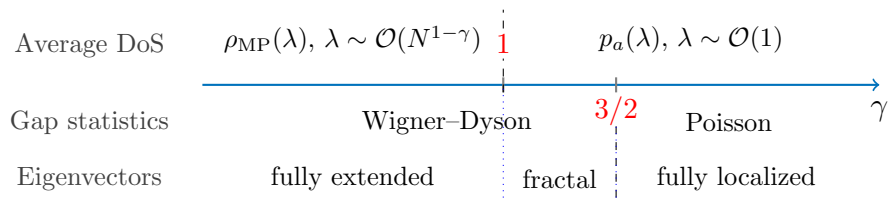
\begin{figure}[t]
\begin{center}
\begin{tikzpicture}
\node[] at (-6.5,2.52) {\textcolor{darkgray}{\footnotesize Average DoS}};
\node[] at (-6.5,1.5) {\textcolor{darkgray}{\footnotesize Gap statistics}};
\node[] at (-6.5,0.75) {\textcolor{darkgray}{\footnotesize Eigenvectors}};
\draw[RoyalBlue, thick,->] (-5,2) -- (4,2);
\draw[gray, thick] (-1,1.9) -- (-1,2.1);
\draw[gray, thick] (0.5,1.9) -- (0.5,2.1);
\draw[black, dashed] (0.5,0.5) -- (0.5,1.5);
\draw[black, dashed] (-1,1.9) -- (-1,2.35);
\draw[black, dashed] (-1,2.75) -- (-1,3);
\draw[blue, dotted] (0.5,0.5) -- (0.5,1.5);
\draw[blue, dotted] (-1,0.5) -- (-1,2);
\node[] at (-3,2.6) {\textcolor{black}{ \footnotesize $\rho_{\rm MP}(\lambda)$, $\lambda \sim {\cal O}(N^{1-\gamma})$}};
\node[] at (1.5,2.6) {\textcolor{black}{ \footnotesize $p_a(\lambda)$, $\lambda \sim {\cal O}(1)$}};
\node[] at (-3,0.75) {\textcolor{black}{\footnotesize fully extended}};
\node[] at (-0.25,0.75) {\textcolor{black}{\footnotesize fractal}};
\node[] at (2.1,0.75) {\textcolor{black}{\footnotesize fully localized}};
\node[] at (-1.75,1.5) {\textcolor{black}{\footnotesize Wigner--Dyson}};
\node[] at (2,1.5) {\textcolor{black}{\footnotesize Poisson}};
\node[] at (4,1.65) {\textcolor{black}{$\gamma$}};
\node[] at (-1,2.55) {\textcolor{red}{\small $1$}};
\node[] at (0.5,1.65) {\textcolor{red}{\small $3/2$}};
\end{tikzpicture}
\end{center}
\caption{The phase diagram of the Wishart--Rosenzweig--Porter (WRP) model (see Sec.~\ref{sec:summary} for its description). }
\label{fig:phase-diagram}
\end{figure}

\begin{figure}[t]
    \centering
    \includegraphics[width=0.9\linewidth]{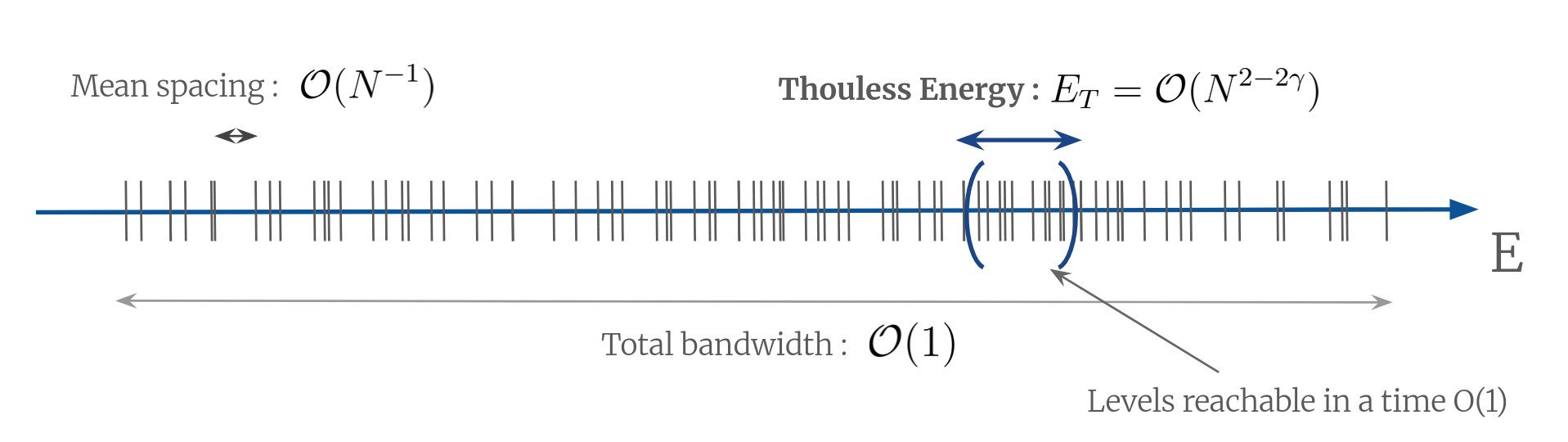}
    \caption{Illustration of the different scales in the spectrum in the intermediate phase of the WRP model ($1<\gamma < 3/2$). The blue window illustrates the concept of mini-bands in the spectrum, extending over the scale of the Thouless energy $E_T$. 
    } 
    \label{fig:thouless}
\end{figure}

Here we present a brief summary of the main 
findings of our work. The first concerns the phase diagram of the model, shown in Fig.~\ref{fig:phase-diagram},
which features three distinct phases separated by two transition points, as detailed below:

\begin{itemize}
\item $\gamma < 1$: in this regime and for large $N$, 
the matrix $\textbf{A}$ is subleading with respect to $\textbf{B}$, and thus the latter completely dominates the spectral properties of $\textbf{H}$. The average density of states (DoS) is given by the Mar\v{c}enko--Pastur distribution~\cite{Marchenko1967}, with eigenvalues and support growing as $N^{1 - \gamma}$, up to subleading finite-$N$ corrections which depend on $\textbf{A}$.  This corresponds to a fully delocalized phase with Wigner--Dyson statistics. 

\item $\gamma > 3/2$: here, the matrix $\textbf{A}$ dominates over $\textbf{B}$ when $N$ is large,
and fully controls the spectral properties of $\textbf{H}$. The eigenvalues of $\textbf{H}$ are close to the diagonal entries of $\textbf{A}$, up to small perturbative corrections induced by $\textbf{B}$, which are not strong enough to hybridize more than $\mathcal{O}(1)$ energy levels. This corresponds to a fully localized phase, in which the eigenstates are localized around those of $\textbf{A}$, and the energy levels obey Poisson statistics.

\item $1 < \gamma < 3/2$: this intermediate regime is the most interesting one. The matrix $\textbf{B}$ can still be treated as a perturbation, but in this case it hybridizes energy levels on a scale that is parametrically much larger than the spectral gap (i.e.~the average level spacing), yet still much smaller than the total bandwidth. As a result, the average DoS is still given by the distribution of the diagonal entries of $\textbf{A}$, but correlations between energy levels emerge on an energy scale known as the \textit{Thouless energy} $E_T$, involving a number of hybridized states that grows with $N$ as $N^{2 - 2 \gamma}$ (see Fig.~\ref{fig:thouless}). This corresponds to a partially ergodic {\it fractal} phase, analogous to that of the standard GRP model. The eigenvectors are partially delocalized on a fractal support set that grows with $N$ as $N^D$, with $D=3 - 2 \gamma$, but is much smaller than $N$.
\end{itemize}

The subleading corrections to the average DoS (which converges to the Mar\v{c}enko--Pastur distribution for $\gamma<1$ and to $p_a$ for $\gamma>1$) are given by the Zee formula~\cite{voiculescu1992free,Zee_1996} for the average density of eigenvalues of the sum of two random matrices, for which we provide a derivation within the replica approach (see \cref{eq:DoS_cav,eq:DoSreplicas,eq:resolvantWRP} below).

The second and main result of this work concerns the (super)universal behavior of the \textit{level compressibility} (or, equivalently, of the density-density correlation function) in the intermediate phase on the scale of the Thouless energy. The 
level compressibility $\chi(E)$ is a simple indicator of the degree of level repulsion and is defined as follows~\cite{Mirlin_2000}. Let 
\begin{equation}
    \rhoemp(\lambda) = \frac{1}{N} \sum_{i=1}^N \delta(\lambda-\lambda_i)
    \label{eq:emp_density}
\end{equation}
denote the ``empirical'' eigenvalues density,
and
\begin{equation}
    I_N[\omega_1,\omega_2] \equiv N \int_{\omega_1}^{\omega_2} \dd{\lambda} \rhoemp(\lambda)
    \label{eq:levels_number}
\end{equation}
denote the number of eigenvalues $\lambda_i$ lying in the interval $ [\omega_1,\omega_2]\subseteq \mathbb{R}$, which is a random variable. 
Denoting $E = (\omega_2 - \omega_1)/2$ the width of the energy window, and $E_0 = (\omega_2 + \omega_1)/2$ 
its middle point,
the level compressibility is defined as 
\begin{equation} 
    \chi(E) \equiv \frac{\kappa_2(E)}{\kappa_1(E)} = \frac{\expval{I_N^2}-\expval{I_N}^2}{\expval{I_N}} = \frac{\expval{I_N^2}_c}{\expval{I_N}} \, ,
    \label{eq:level_compress}
\end{equation}
where $\kappa_1$ and $\kappa_2$ are the first two cumulants of $I_N$. For Poisson statistics\footnote{\samepage However, for energy separations of the order of the total bandwidth, $\chi(E)$ actually decreases from $1$ to $0$ at very large $E$ (see e.g.~Appendix~A of~\cite{Venturelli_2023}). In fact, one 
has $\kappa_1 =  \expval{I_N[\omega_1,\omega_2]}$ and $\kappa_2 = \expval{I_N[\omega_1,\omega_2]} \left(1 - \expval{I_N[\omega_1,\omega_2]}/N\right)$,
and thus the level compressibility reads
\begin{equation}
    \label{eq:chiPoisson}
    \chi(E) = 1 - \frac{\expval{I_N[\omega_1,\omega_2]}}{N} = 1 - \int_{\omega_1}^{\omega_2} \dd{\lambda} \rho(\lambda)  .
\end{equation}
We thus generically expect $\chi(E)\sim 1$ for small $E$, and $\chi(E)\to 0$ for large $E$.}, one has $\kappa_2(E) \simeq \kappa_1(E)$, and then $\chi(E) \simeq 1$.
On the contrary, for a rigid spectrum like that of the Wishart matrix $\textbf{B}$, the mean number of eigenvalues behaves for small $E$ as $\expval{I_N} \propto N \rho(E_0)  E$, where 
\begin{equation}
    \rho(\lambda) \equiv \expval{\rhoemp(\lambda)},
    \label{eq:avg_rho}
\end{equation}
while $\expval{I_N^2}_c \propto \ln ( N \rho(E_0) E)$~\cite{Mirlin_2000}. Hence, in this case one finds $\chi(E) \to 0$  for $E \gg [N \rho(E_0)]^{-1}$ (but still much smaller than $E\sim\order{1}$).

\begin{figure}[t]
    \centering
    \includegraphics[width=0.7\linewidth]{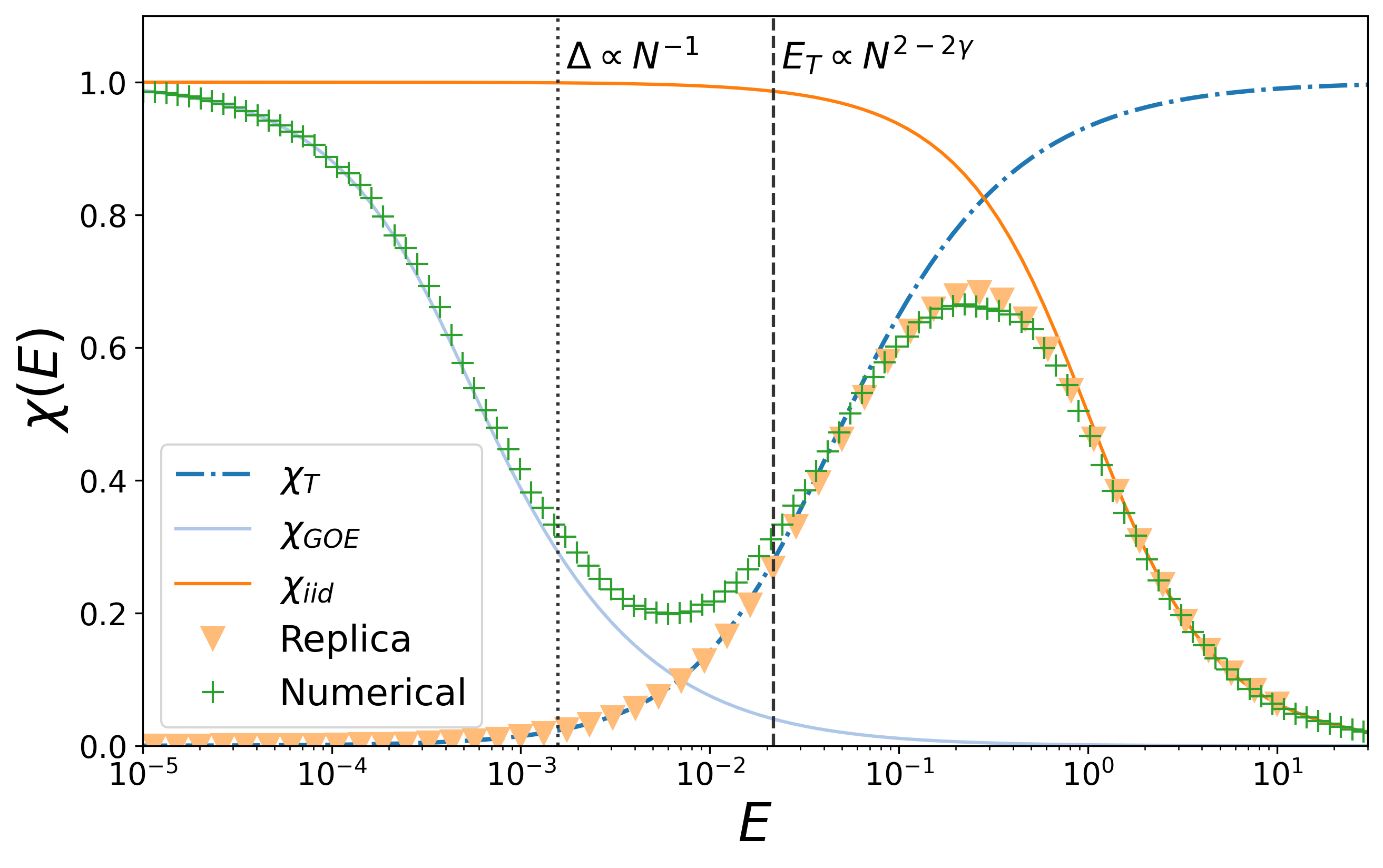}
    \caption{Level compressibility of the 
    WRP ensemble. The numerical data obtained via exact diagonalization (green symbols) are compared to analytical predictions in the distinct regimes of the model, as detailed below.
    First, at small energy scales (comparable to the mean level spacing $\Delta = 1 / (N \rho)$), the level compressibility decreases in accordance with the universal behavior $\chi_{\mathrm{GOE}}$ of Wigner--Dyson statistics (see \cref{chi_GOE_final}). 
    In this regime, the spectrum is said to be rigid because the variance of the number of eigenvalues in an interval 
    is small
    due to
    level repulsion (and thus, so is the level compressibility). Conversely, for intervals of $ \mathcal{O}(1)$, the level compressibility matches 
    that of i.i.d.~random variables~\eqref{eq:chiPoisson}, meaning that there are no correlations between energy levels at this scale. Finally, for intervals close to the Thouless energy $E_T$, 
    the numerical curve approaches
    the (super)universal crossover function $\chi_T$ which connects the two regimes, see \cref{eq:comp_universal}. 
    Note that the replica method provides an analytical prediction (yellow symbols, see Sec.~\ref{sec:replica}) that is valid for all regimes, except the one at $E\lesssim \Delta$, where it breaks down.
    For this plot, we used $\gamma=1.25$, $c\approx 0.998$, $\nu= 1$, $N =2000$, and $p_a$ is the standard Cauchy distribution. 
    }
    \label{fig:LC_all_E}
\end{figure}

As we show below, in the intermediate fractal phase of the WRP ensemble, the level compressibility assumes a scaling form.
This function, 
which we plot in \cref{fig:LC_all_E}, 
describes the crossover from the universal RMT behavior at small energy separations to Poisson statistics at large separations, and it reads 
\begin{equation}
    \chi \left( y= \frac{E}{E_T} \right) = \frac{1}{\pi y}\left[2y \atan(y) -\ln (1+y^2)   \right] \;.
    \label{eq:comp_universal}
\end{equation}
The small- and large-$y$ asymptotics of this scaling function are given in \cref{asympt_chi} below.
Remarkably, this expression is identical to that of the GRP ensemble, both when $\textbf{B}$ is real and symmetric (GOE), and when it is Hermitian (GUE). 

In Ref.~\cite{Venturelli_2023} some of us showed that, on the scale of the Thouless energy, the level compressibility of the GRP model is insensitive to the specific distribution of the i.i.d.\ diagonal entries of $\textbf{A}$ (up to a rescaling by $E_T$ and provided $p_a(0)>0$). More recently, the same crossover function was found in the L\'evy-RP ensemble, where $\textbf{B}$ is a L\'evy matrix with entries drawn from a power-law distribution~\cite{safonova2025} (see however footnote~\ref{foot:levy}). 
Here, we demonstrate (using both the cavity method and the replica approach) that the very same scaling function~\eqref{eq:comp_universal} also governs the crossover in the WRP ensemble, which  
differs in two key aspects from the Gaussian and L\'evy RP ensembles: (i)~$\textbf{B}$ has positive-definite eigenvalues, inducing an asymmetric rightward shift of the spectrum of $\textbf{A}$, 
and (ii)~its matrix elements are not independent (see Eq.~\eqref{eq:correlations}).

\begin{figure}[t]
    \centering 
    \includegraphics[width = 7.5cm]{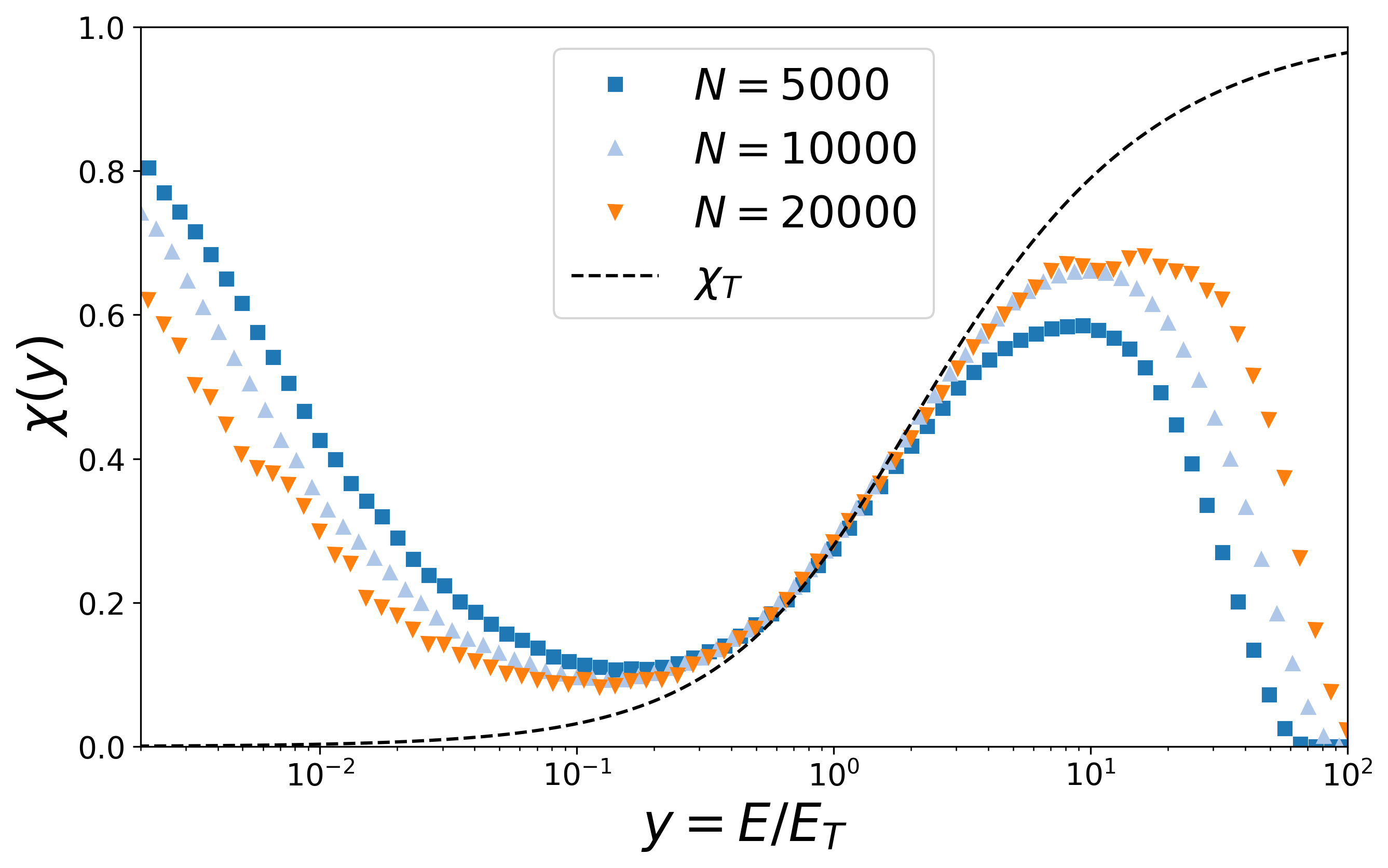}
    \put(-188,115){(a)}
    \includegraphics[width = 7.5cm]{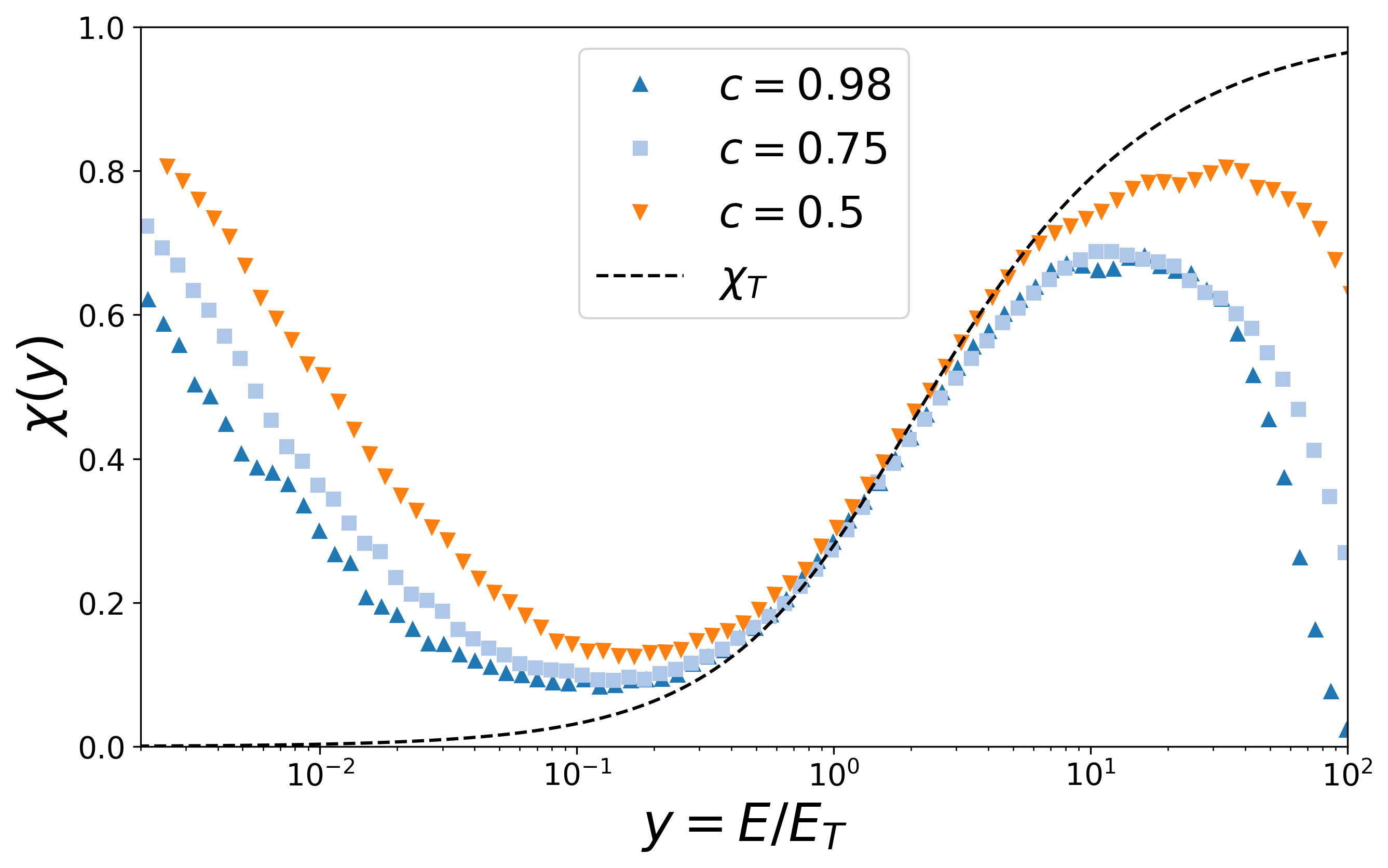}
    \put(-188,115){(b)}
    \caption{Numerical data for the level compressibility 
    of the WRP model
    at the Thouless energy scale, obtained by exact diagonalization of large random matrices, for (a)~different values of $N$ with $c = 0.98$ fixed, and (b)~different values of $c$ with $N= 20000$ fixed. We chose $\gamma=1.25$ and $\nu= 1$; moreover, we took $p_a$ to be uniform between $-1$ and $1$, 
    which allowed us to average the level compressibility over multiple
    non-overlapping
    intervals
    (which are statistically equivalent since the density of states is constant). 
    The 
    dashed line indicates the universal scaling form $\chi_T$ given in \cref{eq:comp_universal}.
    }
    \label{fig:LC_ET_numerics}
\end{figure}

We have tested this theoretical prediction by performing 
exact numerical diagonalization of large random matrices from the WRP ensemble around the Thouless energy $E_T$, varying the matrix size $N$ while keeping the ratio $c = N/M$ fixed, in the intermediate phase ($\gamma = 1.25$). 
In Fig.~\ref{fig:LC_all_E}, we compare the numerical results with the analytical scaling prediction~\eqref{eq:comp_universal}, finding an excellent agreement. In Fig.~\ref{fig:LC_ET_numerics}(a) we show that the range of validity of~\eqref{eq:comp_universal} increases with the system size when the energy separation is measured in units of the Thouless energy. The crossover function~\eqref{eq:comp_universal} is also found to be independent of the parameter $c$ that controls the correlation strength, see \cref{fig:LC_ET_numerics}(b).

These observations are quite remarkable, since the same crossover scaling function is found in many distinct random matrix ensembles, for which indeed the level compressibilities do not necessarily coincide for $E \gg E_T$ or $E\ll E_T$. This finding strongly suggests that such crossover function is in fact fully and genuinely universal, and that such universality originates from the structural properties of the model, rather than from the specific choice of the matrices $\textbf{A}$ and $\textbf{B}$, at least across all models in which the intermediate phase is fractal rather than multifractal. It also motivates numerical studies of this crossover function in more realistic many-body disordered quantum systems that exhibit (multi)fractal phases, to test whether the same universal behavior emerges in that context as well.

\section{Perturbation theory: Mott's criteria for localization and ergodicity}
\label{sec:criteria}

The simplest and most intuitive way to analyze the phase diagram of the model is through first and second-order perturbation theory for the eigenvalues and the eigenvectors of $\textbf{H}$ in \cref{eq:AB}, which yield the so-called \emph{Mott's criteria} for localization and ergodicity~\cite{mott1961theory}. 
These have been successfully applied to the GRP model and its generalizations (see e.g.~Ref.~\cite{khaymovich2021dynamical} for a detailed explanation). 

In the limit where the off-diagonal matrix $\textbf{B}$ is absent, all eigenvectors $|\psi_i\rangle$ are trivially localized on a single site, i.e.~$|\psi_i\rangle = |i\rangle$ (where $|i\rangle$ is the position basis), with corresponding eigenvalues $\lambda_i = a_i$.  The first criterion, known as \emph{Mott's criterion for localization}, states that Anderson localization around a single matrix index occurs when the average level spacing $\Delta = 1/ (N \rho)$ is much larger than the tunneling amplitude between different indices. The second criterion, known as \emph{Mott's criterion for ergodicity}, provides a sufficient condition for ergodicity. The idea is to estimate the average escape rate $\Gamma$ of a particle localized on a given site using Fermi's Golden Rule, and to compare it to the spread of energy levels. When the average spreading width $\Gamma$ is much larger than the 
energy bandwidth,
the different indices are fully hybridized: starting from a given site, the wave packet spreads to any other site with the same energy on a timescale of $\mathcal{O}(1)$.

In equations, the Mott's criterion for localization reads: $\langle \vert H_{ij} \vert \rangle \ll \Delta$, where $\Delta = 1/(N \rho)$ is the average gap, $\rho$ being the average spectral density at the considered value of the energy.
Since for $\gamma > 1$ the eigenvalues of the matrix $\textbf{B}$ are of order $N^{1-\gamma}$ and they all vanish in the large-$N$ limit, the average density of states 
is asymptotically given by the probability distribution of the entries of the matrix $\textbf{A}$, namely $\rho(\lambda) = p_a(\lambda)$. 
For $\gamma < 1$, instead, the entries of $\textbf{A}$ are much smaller than the eigenvalues of $\textbf{B}$ in the large-$N$ limit, and the average DoS is given by a Mar\v{c}enko--Pastur distribution with support growing as $N^{1-\gamma}$. We thus obtain that the average level spacing is 
\begin{eqnarray}
    && 
    \left\{ 
    \begin{array}{ll}
    \Delta= 1/(N p_a(\lambda)) = 1/(M c \, p_a(\lambda)), \qquad & {\rm for} \;\;\; \gamma>1
    \, , 
    \\
    [5pt]
    \Delta\propto N^{-\gamma}, \qquad & {\rm for} \;\;\; \gamma<1
    \, . 
    \end{array}
    \right.
\end{eqnarray}

Since the spectral properties of the model do not depend on the specific energy at which they are probed, in the following, without loss of generality, we focus on the center of the energy band, i.e.
\begin{eqnarray}
    &&
    \left\{ 
    \begin{array}{ll}
    E_0 = 0 + \mathcal{O}(N^{1-\gamma}), & \qquad {\rm for} \;\;\; \gamma > 1
    \, , 
    \\
    [5pt]
    E_0 \propto N^{1-\gamma},  & \qquad {\rm for} \;\;\; \gamma < 1
    \, . 
    \end{array}
    \right.
\end{eqnarray}
From the definition of the model one immediately obtains
\begin{equation} \label{eq:mott}
    \langle \vert H_{ij} \vert \rangle = \nu M^{- \gamma} \left \langle \left \vert \sum_{\ell = 1}^M W_{i \ell} W_{j \ell} \right \vert \right \rangle = \nu \, \sqrt{\frac{2}{\pi}} M^{1/2 - \gamma} 
    \, .
\end{equation}
Mott's criterion then states that Anderson localization 
close to the eigenvectors of $\textbf{A}$ occurs provided that 
\begin{equation}
\langle \vert H_{ij} \vert \rangle = \nu \, \sqrt{\frac{2}{\pi}} M^{1/2 - \gamma} \ll 
\dfrac{1}{c p_a(0) M}
\ \ \ \ \Rightarrow \ \ \ \ \gamma > 3/2
\, . 
\end{equation}

Applying second-order perturbation theory to the eigenvalues and the Fermi golden rule, we now compute the average bandwidth $\Gamma$ that corresponds to the energy window within which hybridization occurs:
\begin{equation} \label{eq:FGR}
\Gamma = 2 \pi p_a(0) \left \langle \sum_j H_{ij}^2 \right \rangle = 2 \pi p_a(0) N\nu ^2 M^{1 - 2 \gamma} = 2 \pi p_a(0)\nu ^2 c M^{2 - 2 \gamma} \, .
\end{equation}
The quantity $\hbar \Gamma$ can be interpreted as the bandwidth that can be reached in a time of $\mathcal{O}(1)$
from a given site $i$, and is often called the Thouless energy $E_T$. This implies that the eigenvectors within this energy window are hybridized by the 
Wishart perturbation.  

For $1 < \gamma < 3/2$, such energy band decreases with the system size as $E_T \propto N^{2-2\gamma}$, but is still much larger than the mean level spacing $\Delta \propto N^{-1}$. This entails that the system is not Anderson localized; nevertheless, $E_T$ remains much smaller than the total bandwidth, which is of $\mathcal{O}(1)$. This signifies that the particle can only explore a subextensive portion of the total Hilbert space.  

The Anderson localization transition occurs when $E_T$ becomes smaller than the mean level spacing, i.e.~for $\gamma > 3/2$. This implies that the average escape time from site $i$, defined as $\Delta t \equiv \hbar/E_T$, grows at least linearly with $N$, and thus the eigenfunctions remain localized on $\mathcal{O}(1)$ sites. In contrast, the transition to the fully delocalized phase takes place when $E_T$ becomes of the order of the total bandwidth, i.e.~for $\gamma \leq 1$. For $\gamma<1$ one has that $E_T$ is much larger than the total bandwidth $N^{1-\gamma}$. Hence, starting at site $i$, a wave packet can reach any other site in a time of $\mathcal{O}(1)$, corresponding to full delocalization.

In the intermediate phase, $1 < \gamma < 3/2$, the support set of the eigenvectors (i.e.~the number of sites hybridized by the perturbation) is given by the ratio between the width of the hybridized energy window and the average gap between adjacent energy levels. It therefore scales as
\begin{equation}
\frac{E_T}{\Delta} \sim N^{D}, \qquad D = 3 - 2 \gamma.
\end{equation}
The partially extended but fractal eigenstates are thus linear combinations of $N^{D}$ localized states associated with nearby energy levels. The wave-function of one of these eigenstates can thus be represented as~\cite{Venturelli_2023,Bogomolny_2018,monthus2017multifractality}
\begin{equation}
    \vert \psi \rangle \approx \sum_{i : |a_i| < E_T} c_i \, N^{-D/2} \vert i \rangle \, ,
    \label{eq:eigen-D}
\end{equation}
with $c_i$ being a Gaussian random variable with zero mean and variance $1$. These eigenstates give rise to the so-called \emph{mini-bands} in the local spectrum (see Fig.~\ref{fig:thouless}). The Thouless energy thus corresponds to the energy window within which  Wigner--Dyson-like spectral correlations (and in particular level repulsion) establish.  

All the moments $I_q$ of the wave-function coefficients (the so-called generalized inverse participation ratios, IPRs) behave as
\begin{equation}
I_{ q} = \sum_i |\langle i|\psi\rangle|^{2{q}} \propto N^{D_{ q} (1-{q})} 
\, ,
\end{equation}
which defines
the fractal dimensions $D_q$.
%
%
The schematic picture presented in Eq.~\eqref{eq:eigen-D} corresponds to a situation in which all fractal dimensions $D_q$ are degenerate and equal to $D$. In Sec.~\ref{sec:spectrum} below, we provide a more detailed analysis of the statistics of the eigenstate coefficients, showing that this is only true for positive integer values of $q \ge 1$. A more precise expression for the full $q$-dependence of the fractal dimensions is given in Eq.~\eqref{eq:Dfirst} below.
Hence, similarly
to the GRP model, the intermediate phase of the WRP ensemble is fractal but not multifractal~\cite{Kravtsov_2015,khaymovich2021dynamical,kutlin2021emergent} (in which case $D_q$ would actually vary with $q$).
As discussed above in Sec.~\ref{sec:introduction}, the emergence of such fractal phase is particularly relevant in many contexts.

In summary, the phase diagram of the WRP ensemble is schematically shown in Fig.~\ref{fig:phase-diagram} and contains three phases: fully delocalized for $\gamma < 1$, Anderson localized for $\gamma > 3/2$, and fractal for $1 < \gamma < 3/2$. The fractal dimensions (for ${ q} \ge 1$) are given by
\begin{equation}
    D_{q} = D = \left \{ 
    \begin{array}{ll}
    1 & \quad  \textrm{for~} \gamma < 1 \, , \\
    3 - 2 \gamma  & \quad \textrm{for~} 1 < \gamma < 3/2 \, , \\
    0 & \quad \textrm{for~} \gamma > 3/2 \, .
    \end{array}
    \right .
\end{equation}

\subsection{Spectrum of fractal dimensions}
\label{sec:spectrum}

In this Section we derive the full multifractal spectrum of eigenstate amplitudes using standard perturbation theory.  
To this end, we follow closely the
perturbative calculation done in Ref.~\cite{Kravtsov_2015} for the GRP model, by adapting it to the WRP ensemble.
We denote by $w_{ij} = |\psi_i(j)|^2$ the squared amplitude of the $i$-th eigenvector on site $j$, with $\psi_i (j)=\langle j|\psi_i\rangle$.  
The first-order correction to the eigenvectors of $A$ reads
\begin{equation}
|\psi_i\rangle = |i\rangle + \sum_{j(\neq i)} \frac{H_{ij}}{a_i - a_j}\,|j\rangle.
\end{equation}
Consequently, for $j \neq i$, the amplitude 
is
\begin{equation}
w_{ij} = \frac{H_{ij}^2}{(a_i - a_j)^2}.
\label{eq:first}
\end{equation}
The convergence of the perturbative series is ensured for $\gamma > 3/2$, where eigenstates are fully localized.  
For $1 < \gamma < 3/2$, convergence still holds due to the random signs of both $H_{ij}$ and $\Delta_{ij} = a_i - a_j$, as in the RP ensemble.

Equation~\eqref{eq:first} shows that $w_{ij}$ results from the product of two independent random factors:  
$x_{ij} = H_{ij}^2$, and $y_{ij} = \Delta_{ij}^{-2}$.  
The first is the square of a Gaussian variable with variance $\langle H_{ij}^2 \rangle = \nu^2 N^{1 - 2\gamma}$, while the second has a heavy-tailed distribution $P(y_{ij}) \sim y_{ij}^{-3/2}$ and a typical value of $\order{1}$.  
Hence, $w_{ij}$ inherits a power-law tail with exponent $3/2$ and a typical scale $w_{\mathrm{typ}} \sim N^{1-2\gamma}$.  
We can thus represent its probability density as
\begin{equation}
P(w_{ij}) = \frac{1}{w_{\mathrm{typ}}}\, P_{\mathrm{reg}}\!\left(\frac{w_{ij}}{w_{\mathrm{typ}}}\right)
 + C\, \frac{\Theta(w_{ij} > N^{1-2\gamma})}{N^{\gamma - 1/2}\,w_{ij}^{3/2}},
\end{equation}
where $P_{\mathrm{reg}}$ denotes the regular part, and $C$ is fixed by normalization.  
Imposing the normalization condition $\sum_j |\psi_i (j)|^2 = 1 \Leftrightarrow \langle w_{ij}\rangle = N^{-1}$ introduces an upper cutoff $w_{\max}$ on the singular tail, obtained from
\begin{equation}
N \langle w_{ij} \rangle \!\sim\! N^{2-2\gamma}
+ \frac{C}{N^{\gamma - 3/2}}\!\int_{N^{1-2\gamma}}^{w_{\max}}\!dw_{ij}\,w_{ij}^{-1/2} = 1,
\end{equation}
which yields $w_{\max} \sim N^{2(\gamma - 3/2)}$.
Because amplitudes cannot exceed unity, this expression holds only for $\gamma < 3/2$; when $\gamma > 3/2$, in the localized phase, one must set $w_{\max} = 1$.  
To restore proper normalization in the localized regime, an additional delta peak at $w_{ij} = 1$ must be included:
\begin{equation}
\hat P(w_{ij}) = P(w_{ij}) + A\,\delta(w_{ij}-1),
\qquad \text{for } \gamma > 3/2,
\label{eq:hatPW}
\end{equation}
with $A = N^{-1}$, yielding the dominant contribution from the fully localized site.

\begin{figure}[t]
    \centering 
    \includegraphics[width = 7.5cm]{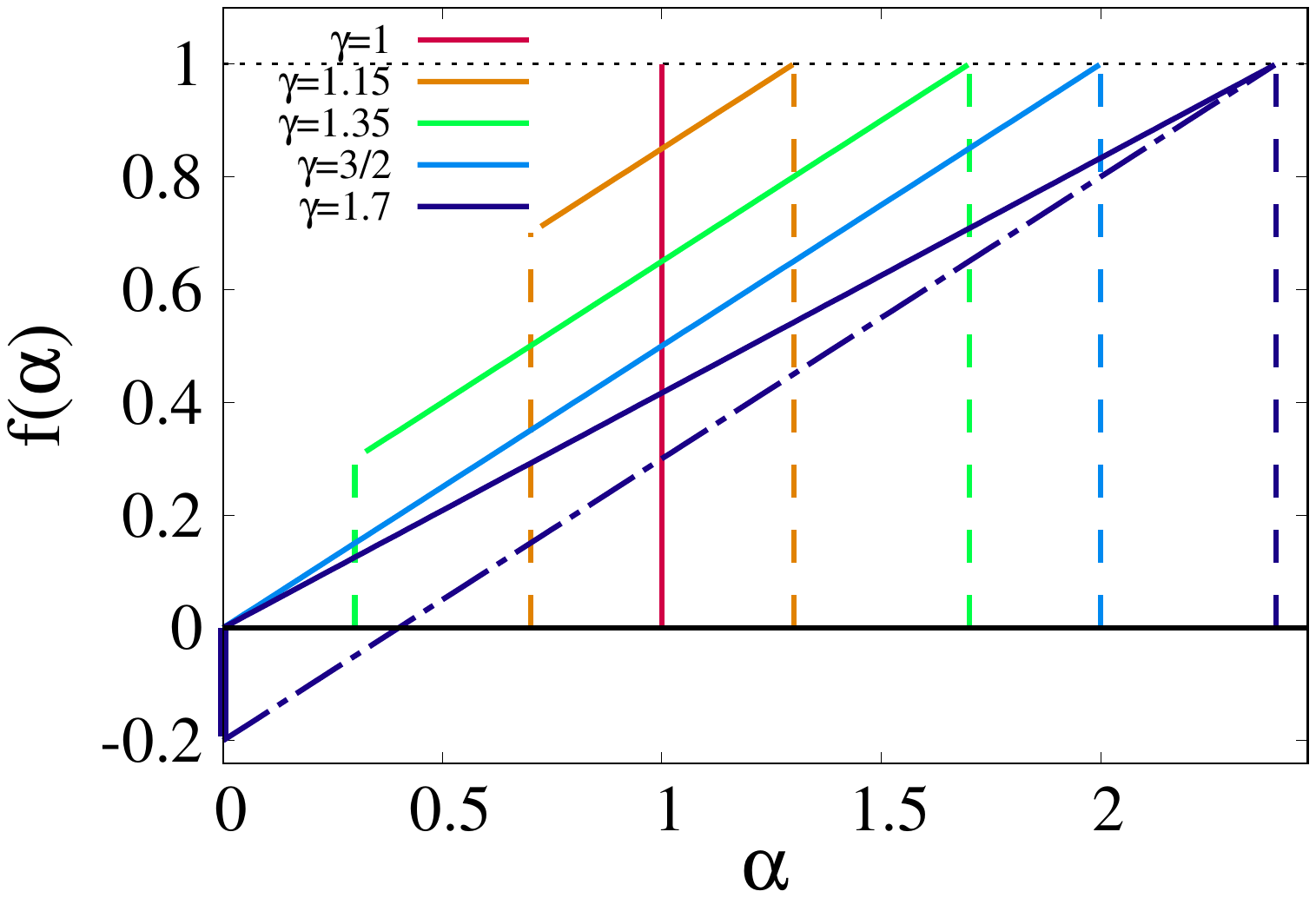}
    \put(-45,33){(a)}
    \includegraphics[width = 7.5cm]{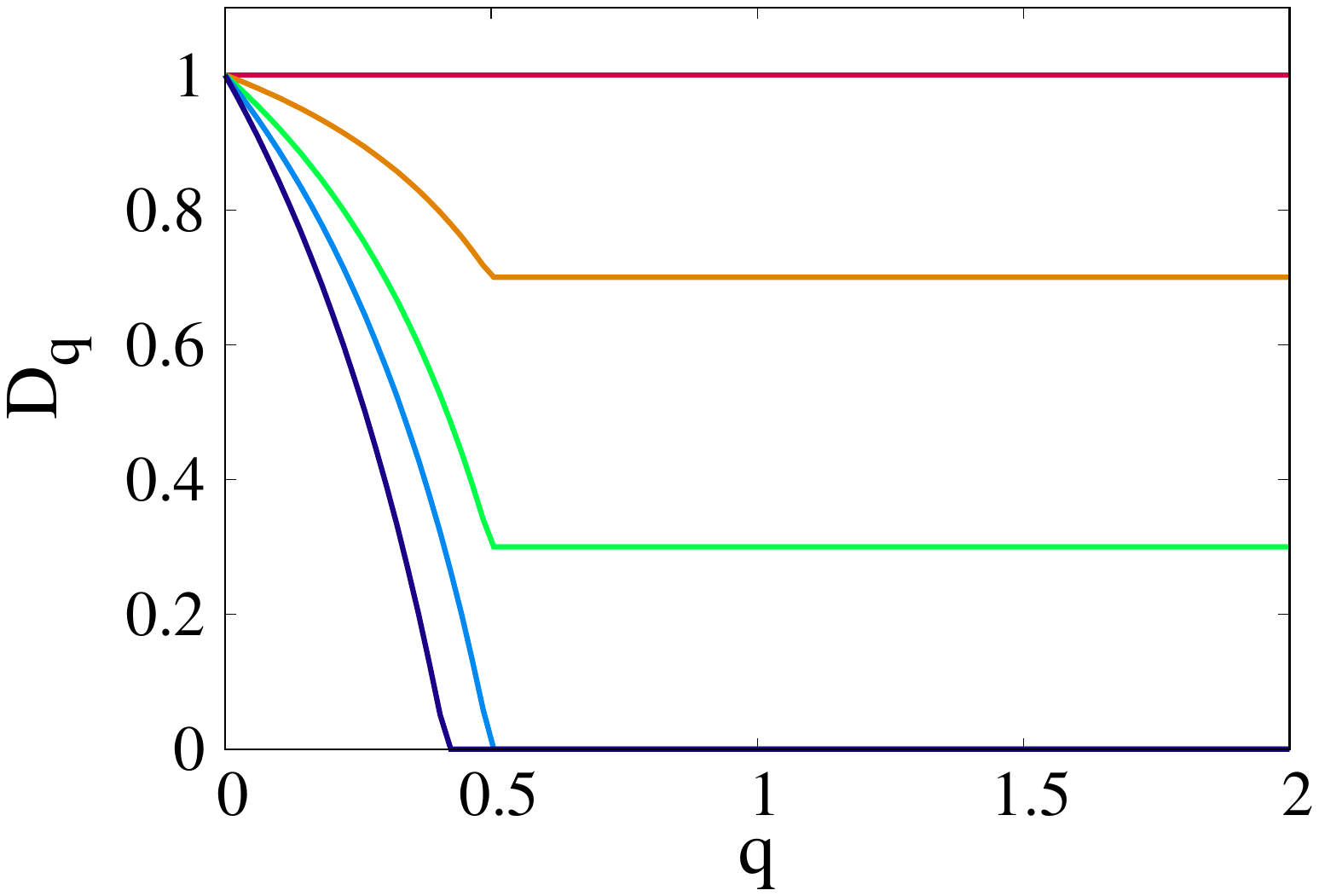}
    \put(-45,33){(b)}
    \caption{(a) Multifractal spectrum $f(\alpha)$, Eq.~\eqref{eq:falpha}, and (b) fractal dimensions $D_q$, Eqs.~\eqref{eq:Dfirst} and~\eqref{eq:Dloc}, 
    for several values of $\gamma$ spanning the three regimes.}
    \label{fig:falpha}
\end{figure}

Next, we define the function $f(\alpha)$ (known as the spectrum of fractal dimensions)
in such a way that $\sim N^{f(\alpha)}$ nodes have wavefunction amplitudes 
$|\psi_i(j)|^2 \sim N^{-\alpha}$.
For $1 < \gamma < 3/2$, the integration of the tail of $P(w_{ij})$ gives
\begin{equation}
N^{f(\alpha)} = \frac{C}{N^{\gamma - 3/2}}
\!\int_{N^{-\alpha}}^{N^{2(\gamma - 3/2)}}\!dw_{ij}\,w_{ij}^{-3/2}
\sim N^{\alpha/2 + 3/2 - \gamma},
\end{equation}
valid for
$\alpha$ in between
$\alpha_{\min} = 2(3/2 - \gamma)$ and $\alpha_{\max} = 2\gamma - 1$.  
Thus,
\begin{equation}
f(\alpha) = \frac{\alpha}{2} + \frac{3}{2} - \gamma,
\qquad (\alpha_{\min} < \alpha < \alpha_{\max}).
\label{eq:falpha}
\end{equation}
For $\gamma > 3/2$, the minimum exponent becomes $\alpha_{\min} = 0$, and the triangular shape of $f(\alpha)$ at the transition, $f(\alpha) = \alpha/2$, coincides with that of the Anderson model on the Bethe lattice~\cite{deluca2013support,garcia2020two}.

Alternatively, one may directly compute the moments of the wave-function intensities,
$N\langle |\psi_i(j)|^{2q} \rangle \propto N^{-\tau_q}$, using
\begin{equation}
\langle w_{ij}^q \rangle
\sim N^{q(1-2\gamma)} 
+ \frac{C}{N^{\gamma-1/2}}
\!\int_{N^{1-2\gamma}}^{N^{2(\gamma - 3/2)}}\!dw_{ij}\,w_{ij}^{q-3/2}.
\end{equation}
For $q < 1/2$, the integral is dominated by typical amplitudes, leading to $\tau_q = q(2\gamma - 1) - 1$,  
while for $q > 1/2$ large fluctuations dominate, yielding $\tau_q = (q - 1)(3 - 2\gamma)$.  
The corresponding generalized fractal dimensions, $D_q = \tau_q / (q-1)$, are then (for $1<\gamma<3/2$)
\begin{equation}
D_q =
\begin{cases}
3 - 2\gamma, & q > 1/2, \\[5pt]
\dfrac{1 - q(2\gamma - 1)}{1 - q}, & q < 1/2.
\end{cases}
\label{eq:Dfirst}
\end{equation}
In the localized regime ($\gamma > 3/2$), a similar computation gives:
\begin{equation}
D_q =
\begin{cases}
0, & q > 1/(2\gamma - 1), \\[5pt]
\dfrac{1 - q(2\gamma - 1)}{1 - q}, & q < 1/(2\gamma - 1).
\end{cases}
\label{eq:Dloc}
\end{equation}
The full set of $f(\alpha)$ and $D_q$ curves is displayed in Fig.~\ref{fig:falpha}.

For $\gamma > 3/2$, the distribution $\hat P(w_{ij})$ in Eq.~\eqref{eq:hatPW} produces a delta peak at $w_{ij}=1$, resulting in a singular contribution $N^{-1}\delta(w_{ij}-1)$.  
The corresponding multifractal spectrum exhibits a spike at $\alpha=0$, see Fig.~\ref{fig:falpha}(a).  
In this case, Eq.~\eqref{eq:falpha} must be modified: one can check that the scaling exponents $\tau_q$ extracted from Eq.~\eqref{eq:Dloc} match those of the convex envelope of $f(\alpha)$, namely $f(\alpha)=\alpha/(2\gamma-1)$, for $0\le\alpha\le 2\gamma-1$.
An analogous triangular shape with slope smaller than $1/2$ is also observed in the localized regime on random regular graphs~\cite{deluca2013support}.

\section{The cavity method}
\label{sec:cavity}

In this Section, we apply the cavity method to determine the diagonal elements of the resolvent matrix of our model~(\ref{eq:AB}). This approach not only enables us to re-derive the phase diagram 
shown in Fig.~\ref{fig:phase-diagram}
beyond perturbation theory, but also provides access to more complex spectral observables, such as the two-point correlations of the energy levels and the spectral compressibility.

The basic idea of this approach is to obtain a 
self-consistent relation for the resolvent matrix $\textbf{G} = (\lambda_\epsilon \mathbb{1} - \textbf{H})^{-1}$ of $\textbf{H}$, which becomes asymptotically exact in the large-$N$ limit. Here $\lambda_\epsilon = \lambda - {\rm i} \epsilon$, $\lambda$ being the real energy at which we probe the spectral properties of $\textbf{H}$, and $\epsilon$ being an imaginary regulator that will be sent to zero at the end of the calculation (after taking the $N \to \infty$ limit). 
To set the stage, let
us assume that we know the resolvent of an $(N-1) \times (N-1)$ matrix 
$\textbf{H}^{(1)}$, and let us add an extra row and an extra column (say row $1$ and column $1$),  thereby obtaining an $N  \times N$ matrix that we denote $\textbf{H}$.
Using the standard formula of matrix inversion one immediately deduces
\begin{equation}
\left[ G_{11} (\lambda_\epsilon) \right]^{-1} =\frac{\textrm{minor}(\lambda_\epsilon \mathbb{1} - \textbf{H})_{11}}{\det(\lambda_\epsilon \mathbb{1} - \textbf{H})} \, . 
\end{equation}
We now use the Schur complement formula (also known as the block matrix inversion formula) to expand the determinant in the denominator in terms of the minor along the first row. After simple algebra one immediately obtains 
\begin{equation}
    \left[ G_{11} (\lambda_\epsilon) \right]^{-1} = \lambda_\epsilon - H_{11} - \sum_{i,j=2}^N H_{1i} G^{(1)}_{ij} (\lambda_\epsilon) H_{j1} \, ,
\end{equation}
where the superscript of $G^{(1)}_{ij}$ indicates the element of the resolvent of the $(N-1) \times (N-1)$  matrix 
$\textbf{H}^{(1)}$ in the absence of the first row and column, with indices $i,j$ going from $2$ to $N$. 

This relation is {\it general} and {\it exact}, without any assumption on the elements $H_{ij}$. However, in order to close these equations and obtain useful relations for the diagonal terms, one must introduce certain approximations. The standard approach consists in neglecting the contributions of the off-diagonal terms $\sum_{i \neq j = 2}^N H_{1i} G^{(1)}_{ij}(\lambda_\epsilon) H_{j1}$, which leads to a closed set of equations for the diagonal elements. Yet, applying this approximation to the Wishart case (even to the pure Wishart ensemble, without the matrix $\mathbf{A}$) yields an incorrect self-consistent relation --- that even fails to reproduce the correct Mar\v{c}enko--Pastur distribution for the average DoS. As discussed in Ref.~\cite{potters2020first}, a different treatment is required for the Wishart ensemble. By substituting $H_{1i}$ and $H_{j1}$ with their explicit definitions from Eqs.~\eqref{eq:AB} and~\eqref{eq:Bij}, and separating the resulting double sum over $\ell$ and $\ell'$ into contributions with equal and distinct indices, we obtain:
\begin{equation}
   \frac{\nu }{M^\gamma} \sum_{\ell,\ell^\prime = 1}^M W_{1\ell} W_{i \ell} W_{1 \ell^\prime} W_{j \ell^\prime} = \frac{\nu }{M^\gamma} \sum_{\ell= 1}^M W_{1\ell}^2 W_{i \ell} W_{j \ell} + \frac{\nu }{M^\gamma} \sum_{\ell \neq \ell^\prime }^M W_{1\ell} W_{i \ell} W_{1 \ell^\prime} W_{j \ell^\prime} 
   \, .
\end{equation}
In this expression, $W_{1\ell}^2$ is a random variable with mean $1$ and variance $2$, while $W_{1\ell} W_{1\ell^\prime}$ are Gaussian random variables with zero mean and unit variance. As discussed in Ref.~\cite{potters2020first}, for a fixed realization of the elements of $\mathbf{H}^{(1)}$, the left-hand side of the equation above converges in the thermodynamic limit to its average, $\nu M^{-\gamma} \sum_{\ell=1}^M W_{i\ell} W_{j\ell} = \nu B_{ij}$. We thus obtain, to the leading order,
\begin{equation}
\begin{aligned}
    & \frac{\nu }{M^\gamma} \sum_{i,j = 2}^N G^{(1)}_{ij} \frac{\nu }{M^\gamma} \sum_{\ell,\ell^\prime = 1}^M W_{1\ell} W_{i \ell} W_{1 \ell^\prime} W_{j \ell^\prime} \simeq \frac{\nu }{M^\gamma} \sum_{i,j = 2}^N G^{(1)}_{ij} \left( \nu B_{ij} + O(M^{1 - \gamma}) \right ) \\
    & \qquad \qquad = \frac{\nu }{M^\gamma} \textrm{Tr} \left[ \textbf{G}^{(1)} \nu \textbf{B}^{(1)} \right] + O(M^{2(1 - \gamma)}) \, .
\end{aligned}
\end{equation}
The trace in the expression above can be rewritten as (we omit the superscript $(1)$ to simplify the notation):
\begin{equation}
\begin{aligned}
    & \textrm{Tr} \left[ (\lambda_\epsilon \mathbb{1} - \textbf{A} - \nu \textbf{B})^{-1} (\nu \textbf{B} + \textbf{A} - \lambda_\epsilon \mathbb{1} - \textbf{A} + \lambda_\epsilon \mathbb{1}) \right] 
   = \textrm{Tr} \left[ - \mathbb{1} + (\lambda_\epsilon \mathbb{1} -  \textbf{A} )  \textbf{G}\right] \\
   & \qquad\qquad\qquad = - (N-1) + \sum_{i=2}^N (\lambda_\epsilon-a_i) G_{ii} (\lambda_\epsilon) \, . 
   \end{aligned}
\end{equation}
In the large-$N$ limit, the resolvent of the $N \times N$ matrix  converges to the resolvent of the $(N-1) \times (N-1)$ matrix 
(up to corrections of $\mathcal{O}(1/N$)). Furthermore $H_{11} = a_1 + \nu M^{-\gamma} \sum_l W_{1\ell}^2 = a_1 + \nu M^{1 - \gamma} + \mathcal{O}(M^{1/2 - \gamma})$. Neglecting all terms smaller than $M^{1 - \gamma}$, one finally obtains the cavity equations for the diagonal elements of the resolvent matrix:
\begin{equation} \label{eq:cavity}
    G_{11}^{-1}(\lambda_\epsilon)  
    = \lambda_\epsilon - a_1 + \nu M^{1 - \gamma} \left[ c - 1 - \frac{c}{N} \sum_i (\lambda_\epsilon - a_i) G_{ii} (\lambda_\epsilon) \right] \, .
\end{equation}
This equation is asymptotically exact for the WRP ensemble in the large-$N$ limit.
In the following, we will use it to extract information on the density of states.

\subsection{Density of states, local density of states, and phase diagram}

First, the knowledge of the diagonal elements of the resolvent matrix immediately yields the spectral density~\eqref{eq:emp_density}, 
\begin{equation}
    \rhoemp (\lambda) = 
    \frac{1}{N \pi} \lim_{\epsilon \to 0} \Im \left[ \Tr \textbf{G}(\lambda_\epsilon) \right]
    \, . 
    \label{eq:density-resolvent-matrix}
\end{equation}
For $\gamma<1$ 
the eigenvalues of $\textbf{H}$ scale as $M^{1-\gamma}\gg 1$, and the support of the DoS  grows with $N$. Indeed, upon neglecting the subleading $a_i$ terms, defining $\lambda = M^{1 - \gamma} \tilde{\lambda}_\epsilon$ with ${\rm Re}(\tilde{\lambda}_\epsilon) \sim \mathcal{O}(1)$, and introducing $g(\lambda_\epsilon) = 1/N \sum_i G_{ii} (\lambda_\epsilon)$, in the large-$N$ limit one obtains~\cite{potters2020first}
\begin{equation}
    \frac{1}{g(\tilde{\lambda}_\epsilon)} = \tilde{\lambda}_\epsilon + \nu (c-1) - \nu c \tilde{\lambda}_\epsilon g(\tilde{\lambda}_\epsilon) \, ,
\end{equation}
from which the Mar\v{c}enko--Pastur distribution is immediately recovered from \cref{eq:density-resolvent-matrix}.

In the regime $\gamma>1$, we have that $M^{1 - \gamma} \ll 1$. In this case, it is thus convenient to introduce the small parameter 
\begin{equation}
    \eta \equiv \nu M^{1-\gamma} \ll 1,
\end{equation}
and the (rescaled) self-energies as
\begin{equation} \label{eq:Gii_cav}
    G_{ii} (\lambda_\epsilon) = \frac{1}{\lambda_\epsilon-a_i + \eta \, \Sigma_i(\lambda_\epsilon)}  \, ,
\end{equation}
where the $\Sigma_i$'s are of $\mathcal{O}(1)$ (and should not be confused with a summation symbol over $i$).
The imaginary part of $G_{ii}$ gives the local density of states (LDoS), representing the contribution of site $i$ to the total density of states:
\begin{equation} \label{eq:LDoS}
    \rho_i (\lambda) 
     = \sum_{\alpha=1}^N |\psi_\alpha(i)|^2 \delta(\lambda-\lambda_\alpha)
    = \frac{1}{\pi} \lim_{\epsilon \to 0} \Im G_{ii} (\lambda_\epsilon) \, , 
\end{equation}
so that $\rhoemp (\lambda) = \frac{1}{N} \sum_{i=1}^N \rho_i (\lambda)$ (compare with \cref{eq:emp_density}).
Physically, $\rho_i (\lambda)$ corresponds to the inverse lifetime of a particle created at site $i$ with energy $\lambda$, and provides the order parameter distribution function for Anderson localization.

As shown self-consistently below, the self-energy $\Sigma_i$ turns out to be independent of the index $i$ at leading order for $\gamma>1$. Plugging the expression~\eqref{eq:Gii_cav} into the cavity equation~\eqref{eq:cavity} and expanding the terms proportional to $\eta$, one deduces the following self-consistent equation for the self-energy: 
\begin{equation} \label{eq:Sigma_cav}
    \Sigma(\lambda_\epsilon) =   \frac{\eta \, c \, \Sigma(\lambda_\epsilon)}{N} \sum_i \frac{1}{\lambda_\epsilon - a_i} - 1  + \mathcal{O}(M^{2(1 - \gamma)} ) 
    \, ,
\end{equation}
which immediately yields
\begin{equation} \label{eq:Sigma}
    \Sigma(\lambda_\epsilon) \simeq - 
    \left(1 - \dfrac{\eta \, c }{N} \sum\limits_{i=1}^N \dfrac{1}{\lambda_\epsilon - a_i}\right)^{-1} \, .
\end{equation}
In the large-$N$ limit the average of the self-energy over the probability distribution of the diagonal elements can be conveniently rewritten in a more compact form using the definition of the Stieltjes transform of $p_a$
\begin{equation}
    \mathcal{G}_a(\lambda_\epsilon) = \int \mathrm{d}a \; \frac{p_a(a)}{\lambda_\epsilon - a} \, .
    \label{eq:stieltjes}
\end{equation}
From Eq.~\eqref{eq:Sigma} and the definition above, for large $N$ one finds
\begin{equation}
\Sigma(\lambda_\epsilon) \approx - \frac{1}{1 - \eta \, c \, {\cal G}_a (\lambda_\epsilon)} \, ,
\end{equation}
and from Eq.~\eqref{eq:Gii_cav} one obtains
\begin{equation}
g(\lambda)  = \frac{1}{N} \sum_{i=1}^N G_{ii} (\lambda)  = {\cal G}_a (\lambda + \eta  \Sigma ) = {\cal G}_a \left(\lambda - \frac{\eta} {1 - \eta \, c \, {\cal G}_a (\lambda)} \right) \, .
\end{equation}
Finally, the average DoS in the large-$N$ limit can be written as
\begin{equation} \label{eq:DoS_cav}
\rho ( \lambda) = \frac{1}{\pi } {\rm Im} \, {\cal G}_a \left(\lambda - \frac{\eta} {1 - \eta \, c \, {\cal G}_a (\lambda)} \right) \, .
\end{equation}
Hence, for $\gamma>1$, the average DoS of the WRP ensemble coincides with the probability distribution of the diagonal entries, $\rho(\lambda) = p_a(\lambda)$, up to subleading corrections of $\mathcal{O}(\eta)$. 
In fact, Eq.~(\ref{eq:DoS_cav}) follows directly from the Zee formula~\cite{Zee_1996}, 
which expresses the resolvent (i.e.~the Stieltjes transform) of the sum of two independent random matrices in terms of their individual resolvents. The same result will be re-derived in Sec.~\ref{sec:DoS_replicas} using the replica method. We will also return to the finite-$N$ corrections to the average DoS of the WRP ensemble in the different regimes.

This result shows that the cavity approach captures the transition of the average DoS taking place at $\gamma = 1$: for $\gamma < 1$, $\textbf{H}$ is dominated by $\textbf{B}$, and the average DoS follows, to leading order, the spectral density of $\textbf{B}$ (i.e.~the Mar\v{c}enko--Pastur law), up to subleading corrections. Conversely, for $\gamma > 1$, $\textbf{A}$ dominates, and the leading contribution to the average DoS is given by the distribution of the diagonal entries, again up to subleading corrections.

We now show that the cavity approach, and in particular Eqs.~\eqref{eq:Gii_cav} and~\eqref{eq:Sigma}, also allow one to obtain the second phase transition, occurring at $\gamma = 3/2$, between the intermediate fractal phase and the Anderson-localized one. To this end, we return to Eq.~\eqref{eq:Sigma} and introduce the real and imaginary parts of the self-energy as $\Sigma = \sigma - \mathrm{i}\,\hat{\sigma}$ (recall that $\lambda_\epsilon = \lambda - \mathrm{i}\epsilon$). For $\gamma > 1$, expanding to leading order in $\eta \ll 1$, one obtains
\begin{equation} \label{eq:sigma_sol}
    \begin{aligned}
    \sigma (\lambda) & \simeq -1 - \frac{\eta \, c }{N} \sum_{i=1}^N \dfrac{1}{\lambda - a_i} + o(\eta) \, , \\
    \hat{\sigma} (\lambda) & \simeq  \frac{\eta \, c }{N} \sum_{i=1}^N \dfrac{\epsilon}{(\lambda - a_i)^2 + \epsilon^2} + o(\eta) \, .
    \end{aligned}
\end{equation}
Note that we have neglected the imaginary regulator $\epsilon$ with respect to the imaginary part of the self-energy, since the natural scale of the former is of $\mathcal{O}(N^{-1})$, while the latter is of order $\eta \propto N^{1 - \gamma} \gg N^{-1}$.
As discussed in Sec.~\ref{sec:criteria}, localization occurs when the corrections to the Green's function due to the imaginary part of the self-energy 
(and hence the Thouless energy) are smaller than the mean level spacing. From Eq.~\eqref{eq:sigma_sol} one finds
\begin{equation}
    \hat{\sigma} (\lambda) \simeq  \eta \,  c \, \pi \, p (\lambda) \, .
\end{equation}
From Eq.~\eqref{eq:Gii_cav}, one immediately obtains the corrections to the imaginary part of the Green's function due to the perturbation, yielding the Thouless energy $E_T = \eta \hat{\sigma}(\lambda) =\nu ^2 c \pi p_a(\lambda) M^{2 - 2 \gamma}$. Anderson localization is then realized when this Thouless energy is smaller than the average gap $1/[N p_a(\lambda)]$, i.e.~when
\begin{equation}
    E_T =\nu ^2 c \pi M^{2 - 2 \gamma} p_a(\lambda) \ll \frac{1}{N p_a(\lambda) } \, ,
\end{equation}
which reproduces the condition $\gamma > 3/2$ obtained from the Fermi golden rule, Eq.~\eqref{eq:FGR}, and Mott's criterion, Eq.~\eqref{eq:mott}. The transition to the fully delocalized regime occurs when the Thouless energy exceeds the total bandwidth, i.e.~when $\gamma<1$.

In conclusion, the cavity method allows one to re-derive the phase diagram discussed in Sec.~\ref{sec:criteria} and schematically presented in Fig.~\ref{fig:phase-diagram}, in a more rigorous and controlled way.
 
\subsection{Density-density correlations and level compressibility in the intermediate regime}
We now apply the cavity approach to compute the two-point density-density correlation function in the intermediate fractal regime, $1 < \gamma < 3/2$, which constitutes one of the main focuses of this work. The two-point function is defined as $\langle \rho(\omega_1)\rho(\omega_2)\rangle_c$. From Eqs.~\eqref{eq:Gii_cav},~\eqref{eq:LDoS} and~\eqref{eq:sigma_sol} we first obtain~\cite{Lunkin2025}
\begin{equation} \label{eq:rhorho}
\begin{aligned}
   &\langle \rhoemp(\omega_1) \rhoemp(\omega_2) \rangle =  \frac{1}{\pi^2 N^2} \sum_{i,j=1}^N \left \langle \rho_i (\omega_1) \rho_j (\omega_2)  \right \rangle \, , \\
   &\rho_i (\omega)  = \frac{\eta \hat{\sigma}(\omega)}{(\omega - a_i + \eta \sigma (\omega))^2 + (\eta \hat{\sigma}(\omega))^2}  \, ,
   \end{aligned}
\end{equation}
where $\sigma(\omega)$ and $\hat{\sigma}(\omega)$ are given in Eq.~\eqref{eq:sigma_sol}. It is convenient to introduce the discrete version of the Stieltjes transform~\eqref{eq:stieltjes} of $p_a$ at finite $N$ as
\begin{equation}
\begin{aligned}
    S(\lambda_\epsilon) &= \frac{1}{N} \sum_i \frac{1}{\lambda_\epsilon - a_i} = s(\lambda_\epsilon) + {\rm i} \hat{s} (\lambda_\epsilon) \, , \\
    \sigma(\omega) & = -1 - \eta c s(\omega) \, , \;\quad \hat{\sigma} (\omega)  = \eta c \hat{s} (\omega) \, .
    \end{aligned}
\end{equation}
In the intermediate phase, for energy separations of the order of the mean level spacing, $\omega_2 - \omega_1 \sim \Delta \propto N^{-1}$, the level statistics is governed by the universal GOE ensemble. (This regime actually lies outside the range of validity of the approximations underlying the cavity approach.) For energy separations of order $\omega_2 - \omega_1 \sim \mathcal{O
}(1)$, i.e.~much larger than the width of the mini-bands in the local spectrum, the levels are instead expected to be uncorrelated. 

The physically relevant regime in the fractal phase is the crossover between RMT behavior at small energy separation and Poisson statistics at large separations, which is expected to occur on the scale of the Thouless energy. For this reason, we consider energies $\omega_1$ and $\omega_2$
separated by intervals of the order of the Thouless energy. For convenience (although this is not strictly necessary), we shift the interval by $\eta$, which slightly simplifies the formulas below. We thus set 
\begin{equation}
    \omega_{1} = \eta - x\,\eta^2 \,, \qquad \omega_{2} = \eta + x\,\eta^2 \,, \qquad x = \mathcal{O}(1) \, .
    \label{eq:shift}
\end{equation}
On these energy scales, one can approximate 
$s(\omega_{j})$ and $\hat{s}(\omega_{j})$ (with $j=1$ or 2) as independent of $x$ at the leading order,
namely 
\begin{equation}
    s(\omega_{j}) \simeq s(\eta + o(\eta)) \approx s(\eta) \,, 
    \qquad 
    \hat{s}(\omega_{j}) \simeq \hat{s}(\eta + o(\eta)) \approx \hat{s}(\eta) \approx \pi \big[ p_a(0) + \eta p'(0) \big] \, .
\end{equation}
Within this approximation, the $\rho_i$'s become random variables that depend only on the set of all $\{ a_i \}$'s, and are therefore uncorrelated. One thus obtains
\begin{equation} \label{eq:rhorho1}
   \langle \rhoemp(\omega_1) \rhoemp(\omega_2) \rangle_c \simeq  \frac{1}{\pi^2 N^2} \sum_{i=1}^N \left \langle \rho_i (\omega_1) \rho_i (\omega_2)  \right \rangle_c \, = \frac{1}{\pi^2 N} \left \langle \rho_i (\omega_1) \rho_i (\omega_2)  \right \rangle_c ,
\end{equation}
where the local density of states $\rho_i (\omega)$ is given by the second line of Eq.~\eqref{eq:rhorho},
and the remaining average on the r.h.s.~of \cref{eq:rhorho1} is intended over $p_a(a)$.
Note that $\langle \rhoemp(\omega_1) \rhoemp(\omega_2) \rangle_c $ actually becomes $i$-independent after averaging over $p_a(a)$.
This approximation is only valid on the scale of the Thouless energy, and breaks down at the scale of the average spectral gap $1/N$.

The expression above of the density-density correlator allows one to compute the level compressibility $\chi(E)$ introduced in \cref{eq:level_compress}, using the following relation between the two-point density-density correlation function and the second cumulant of the number of eigenvalues in a given energy window 
(see e.g.~App.~F in Ref.~\cite{Venturelli_2023}):
\begin{equation}
    \kappa_2 (\omega_2 - \omega_1) = N \int_{\omega_1}^{\omega_2} {\rm d} \tilde{\omega}_1 \int_{\omega_1}^{\omega_2} {\rm d} \tilde{\omega}_2  \; \langle \rho(\tilde{\omega}_1) \rho(\tilde{\omega}_2) \rangle_c \, .
    \label{eq:k2-rhorho}
\end{equation}
Introducing the energy shift $x$ as in \cref{eq:shift}, using Eqs.~\eqref{eq:rhorho} and~\eqref{eq:rhorho1}, and performing the integral over $\tilde{\omega}_{1}$ and $\tilde{\omega}_{2}$ (with the change of variable $\tilde{\omega}_{1} = \eta + x \eta^2$ and $\tilde{\omega}_{2} = \eta -x \eta^2$) before the one over the $a_i$'s, one finally obtains
\begin{align}
    \pi^2 \kappa_2 (x)  =& 
    \int {\rm d} a \, p_a(a) \left[ \atan \left( \frac{\eta^2 x + a + \eta^2 s(\eta)}{\eta^2 c \hat{s} (\eta)} \right) - \atan \left( \frac{a - \eta^2 x + \eta^2 s(\eta)}{\eta^2 c \hat{s} (\eta)} \right) \right]^2 \\
    &   -  
    \left\{ \int {\rm d} a \, p_a(a) \left[ \atan \left( \frac{\eta^2 x + a + \eta^2 s(\eta)}{\eta^2 c \hat{s} (\eta)} \right) - \atan \left( \frac{a - \eta^2 x + \eta^2 s(\eta)}{\eta^2 c \hat{s} (\eta)} \right) \right]
    \right\}^2 \, .
    \nonumber
\end{align} 
Since the difference
of the two atan's is non-zero only in a small interval centered around zero and of width $\eta^2$, one can perform the change of variable $a = \eta^2 \tilde{a}$ 
and replace $p_a(\eta^2 \tilde{a})$ by $p_a(0)$ in the integrals above. 
The disconnected part of $\kappa_2$, corresponding to the integral in the second line, gives $\kappa_1^2(x) = [2 \eta^2 p_a(0) (x + s(\eta))]^2$. 
The term in the second line is thus of $\mathcal{O}(\eta^4)$, and can be neglected with respect to the term in the first line, which is of $\mathcal{O}(\eta^2)$. 
For a symmetric distribution $p_a$, the real part of the Stieltjes transform~\eqref{eq:stieltjes} computed at energy $\eta$ is of $\mathcal{O}(\eta)$, and can be also neglected at the leading order. Furthermore, it is convenient to change again variable as $\tilde{a} \to c \hat{s} \tilde{a}$. Since $\hat{s} \approx \pi p_a(0)$, we have:
\begin{equation}
    \kappa_2 (x)  \approx \frac{\eta^2 p_a^2(0) c }{\pi}  \int_{-\infty}^{+\infty} {\rm d} \tilde{a} \,  \left[ \atan \left( \frac{x}{c \pi p_a(0)} + \tilde{a}  \right) - \atan \left( \tilde{a} - \frac{x}{c \pi p_a(0)}  \right) \right]^2 \\
     \, .
\end{equation}
Plugging this result into Eq.~\eqref{eq:level_compress} and dividing by the first cumulant $\kappa_1(x) \simeq 2 \eta^2 p_a(0) x$, one finally obtains the spectral compressibility on the scale of the Thouless energy:
\begin{equation}
    \chi (x) \approx \frac{c \, p_a(0)}{2 \pi x}  \int_{-\infty}^{+\infty} {\rm d} \tilde{a} \, \left[ \atan \left( \frac{x}{c\pi  p_a(0)} + \tilde{a}  \right) - \atan \left( \tilde{a} - \frac{x}{c \pi p_a(0)}  \right) \right]^2 \\
     \, .
     \label{eq:cavity-scaling-comp}
\end{equation}
Using the property (see e.g.~1.625 in~\cite{table})
\begin{equation}
    \atan(a) - \atan(b) = \atan(\frac{a-b}{1+ab}) \, ,
    \label{eq:property}
\end{equation}
 upon calling $y\equiv \frac{x}{c \pi  p_a(0)}$ in Eq.~\eqref{eq:cavity-scaling-comp} and changing variables in the integral as $u= \tilde a \sqrt{1+y^2}$, we obtain
\begin{equation}
    \chi \left( y= \frac{\omega_2 - \omega_1}{2\pi c p_a(0) \eta} \right) =\frac{\sqrt{1+y^2}}{\pi^2 y} \int_0^\infty \dd{u} \left\lbrace \atan \left[\frac{2y}{u^2(1+y^2)+1-y^2 }\right] \right\rbrace^2  \, ,
    \label{eq:comp_scaling_calculation}
\end{equation}
where we stress that we have chosen the branch $\atan(z) \in [0,\pi]$. Upon integrating by parts and performing some algebra~\cite{stack,table}, the integral over $u$ can be computed explicitly to give Eq.~\eqref{eq:comp_universal}, which is exactly the same universal scaling function that some of us found for the GRP ensemble (see Eqs.~(126) and~(127) of Ref.~\cite{Venturelli_2023}). 
Alternatively (and equivalently), in \ref{app:scaling-function} we provide a derivation of the scaling function~\eqref{eq:comp_universal} starting from \cref{eq:cavity-scaling-comp}.
In particular, the asymptotics of the function $\chi(y)$ can be checked to give
\begin{equation} \label{asympt_chi}
    \chi(y) \simeq \left \{
    \begin{array}{ll}
        \dfrac{y}{\pi} \, , & \qquad y \ll 1 \, , 
        \\
        [5pt]
        1- \dfrac{2(1+\ln y)}{\pi y} \, , & \qquad y \gg 1 \, ,
    \end{array} \right .
\end{equation}
showing that $\chi(y)$ interpolates between Wigner--Dyson statistics at low energy, and Poisson statistics at higher energy.
This function is plotted as a dot-dashed line in \cref{fig:LC_all_E,fig:LC_ET_numerics}.

\section{The replica method}
\label{sec:replica}

In this Section we derive the average density of states reported in Eq.~(\ref{eq:DoS_cav}) using the replica method~\cite{Mezard_1987}.
We also apply the replica strategy to calculate the level compressibility 
given in Eq.~(\ref{eq:comp_universal}). In passing, we derive in Sec.~\ref{sec:number-eigen-replica} the full counting statistics of the WRP model.

\subsection{Density of states} 
\label{sec:DoS_replicas}

We report here the main steps of the derivation, while we defer its details to~\ref{app:replicas}.
Using the Edwards--Jones formula~\cite{Edwards_1976}, we 
first
express the average density of states as
\begin{equation}
    \label{eq:edwardsJones}
    \rho(\lambda) = -\frac{2}{N\pi} \lim_{\epsilon \to 0^{+}} 
    \operatorname{Im} \frac{\partial}{\partial \lambda} \langle \ln \; \mathcal{Z}(\lambda - {\rm i}\epsilon)\rangle,
\end{equation} 
where we 
introduced
the partition function 
\begin{equation}
    \label{eq:edwardsjonesZ}
    \mathcal{Z}(\lambda) = \int_{\mathbb{R}^N} \dd[N]{r} \; 
    \mathrm e^{-\frac{{\rm i}}{2} {\bm r}^{\; T}(\lambda \mathbb{1} - \textbf{H}) \bm r }.
\end{equation}
The average of the logarithm 
in Eq.~\eqref{eq:edwardsJones}
can be expressed using the replica trick as 
\begin{equation}
    \label{eq:replicatrick}
    \langle \ln \mathcal{Z}(\lambda)\rangle = \lim_{n \to 0} \frac{1}{n} \ln \langle \mathcal{Z}^n(\lambda) \rangle.
\end{equation}
Using standard techniques, one can then recast the average as
\begin{align}
    \langle \mathcal{Z}^n(\lambda) \rangle  = 
    & \int \mathcal{D}\phi \mathcal{D}\hat{\phi} \mathcal{D}\psi \mathcal{D}\hat{\psi} \text{ } \exp \left\{ - \sqrt{NM} \mathcal{S}_n [ \phi, \hat{\phi}, \psi, \hat{\psi} ; \lambda]  \right\} 
    \, , 
    \label{eq:Zn-lambda-RPW}
\end{align}
with the action $\mathcal{S}_n[ \phi, \hat{\phi}, \psi, \hat{\psi} ; \lambda]$  given by
\begin{align}
    \label{eq:action_RP}
    & \mathcal{S}_n[ \phi, \hat{\phi}, \psi, \hat{\psi} ; \lambda] =  
      \frac{{\rm i}}{\sqrt{c}} \int
    \mathrm{d}\vec{u} \; \hat{\phi}(\vec{u}) \phi(\vec{u}) 
    + {\rm i} \sqrt{c} \, \int
     \mathrm{d}\vec{r} \; \hat{\psi}(\vec{r}) \psi(\vec{r}) \nonumber \\
    & \quad - \sqrt{c} \; \ln 
    \int
     \mathrm{d}\vec{r} \; \varphi_a(-\vec{r}^{\, 2}/2) \; 
     \exp \left[-\frac{{\rm i}}{2} \lambda  \vec{r}^{\, 2} + i \hat{\psi}(\vec{r}) \right] 
     \nonumber \\
    & \quad - \frac{1}{\sqrt{c}} \; \ln 
    \int
     \mathrm{d}\vec{u} \; \exp \left[- \frac{1}{2} \vec{u}^{\, 2} + {\rm i} \hat{\phi}(\vec{u}) \right]  + \frac{{\rm i}}{2} \sqrt{c} \eta \int \mathrm{d}\vec{r} \mathrm{d}\vec{u} \text{ }\phi(\vec{u}) \psi(\vec{r}) \left( \vec{u} \cdot \vec{r} \right)^2,
\end{align} 
where $\eta = \nu M^{(1-\gamma)} $ and  $\varphi_a$ is the characteristic function of $p_a$, i.e.
\begin{equation}
    \varphi_a(y) = \int \; \mathrm{d}a \; p_a(a)  \; \mathrm e^{-{\rm i} ay}
    \, . 
    \label{eq:varphia}
\end{equation}
(Here and in \ref{app:replicas}, we denote by $\bm r$ a vector in $\mathbb{R}^N$, and by $\vec r$ a vector in the replica space $\mathbb{R}^n$.)
Next, by using saddle-point evaluation, we get that $\langle \mathcal{Z}^n(\lambda) \rangle \approx \exp \left( - \sqrt{NM} \mathcal{S}_n ^{\rm sp}  \right) $, where $\mathcal{S}_n ^{\rm sp}  $ denotes the action evaluated at the saddle point. Combining this with Eqs.~(\ref{eq:edwardsJones}) and~(\ref{eq:replicatrick}), we finally obtain the density of states as
\begin{equation}
    \label{eq:DoSreplicas}
    \rho(\lambda) = \frac{1}{\pi} \lim_{\epsilon \rightarrow 0^+} \Im  G(\lambda_\epsilon)
     + \mathcal{O}(1/N)\, ,
\end{equation} 
where $G(\lambda) \propto  \lim_{n \to 0} \partial _{\lambda} \mathcal{S}_n ^{\rm sp} $ 
(which actually coincides with the resolvent associated to the WRP model, compare with \cref{eq:density-resolvent-matrix})
satisfies the 
self-consistent equation 
\begin{align}
    \label{eq:resolvantWRP}
    G(\lambda) = {\rm i} \int_0^{\infty} \mathrm{d}z \; \; \varphi_a(-z) \exp \left[ -{\rm i} z \left(  \lambda - \frac{\eta}{1 -  c \, \eta \, G(\lambda)} \right) \right]
    \, . 
\end{align} 
Equivalently, upon rewriting the characteristic function as 
in \cref{eq:varphia},
switching the order of the two integrals, computing the integral over $z$ first and then the one over $a$, we obtain
\begin{equation}
    \label{eq:WRP_resolvent}
    G(\lambda) = \mathcal{G}_a \left( \lambda - \frac{\eta}{1-c\,\eta \, G(\lambda)}\right),
\end{equation} 
where $\mathcal{G}_a$ is the Stieltjes transform of $p_a$, defined as
in \cref{eq:stieltjes}.
This is consistent with the result found with the cavity method in Eq.~(\ref{eq:DoS_cav}), since $ G(\lambda) = \mathcal{G}_a ( \lambda) + \mathcal{O}(\eta) $. We 
also note that this result is equivalent to the Zee formula~\cite{Zee_1996}, which allows one to compute the resolvent $G_{1+2}$ of the sum of two mutually free random matrices, given their respective resolvents $G_1$ and $G_2$. Indeed, the Zee formula can be written as a self-consistent equation (see e.g.~Appendix~C in~\cite{Venturelli_2023})
\begin{equation}
    \label{eq:Zee_sc}
    G_{1+2}(z) = G_1 \bigl( z-R_2(G_{1+2}(z))\bigr ),
\end{equation} 
in terms of the R-transform of the second matrix, i.e.~$R_2$. In our case, the R-transform of Wishart matrices is
$R_W(z) = 1 /(1-cz)$~\cite{potters2020first}. Now, using the fact that rescaling a random matrix by a factor $\eta$ scales its R-transform as $R_{\eta W}(z) = \eta R_W (\eta z)$, and plugging this into Eq.~\eqref{eq:Zee_sc}, we obtain again Eq.~\eqref{eq:WRP_resolvent}. 

\begin{figure}[t]
    \centering 
    \includegraphics[width = 7.5cm]{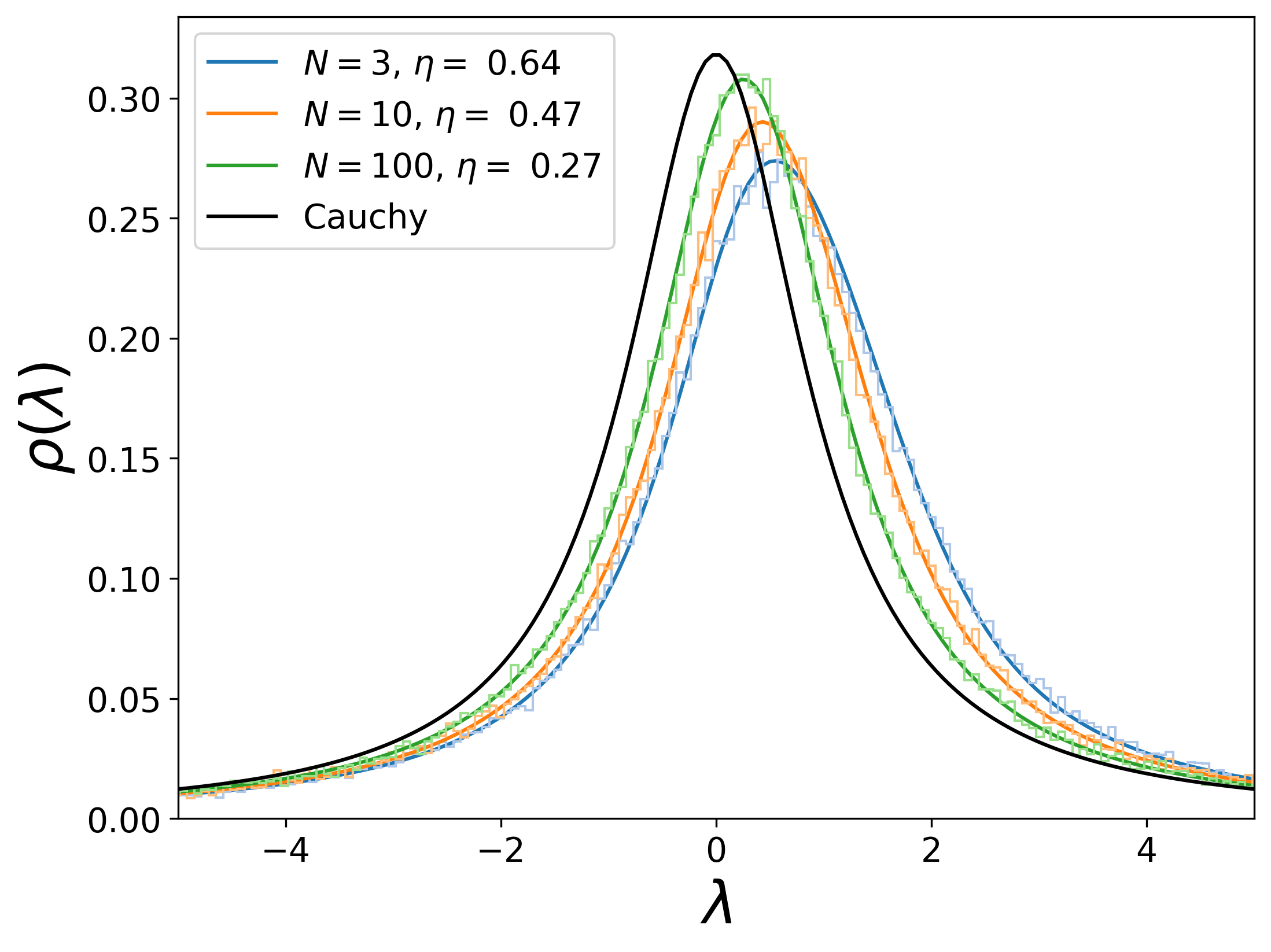}
    \put(-25,142){(a)}
    \includegraphics[width = 7.5cm]{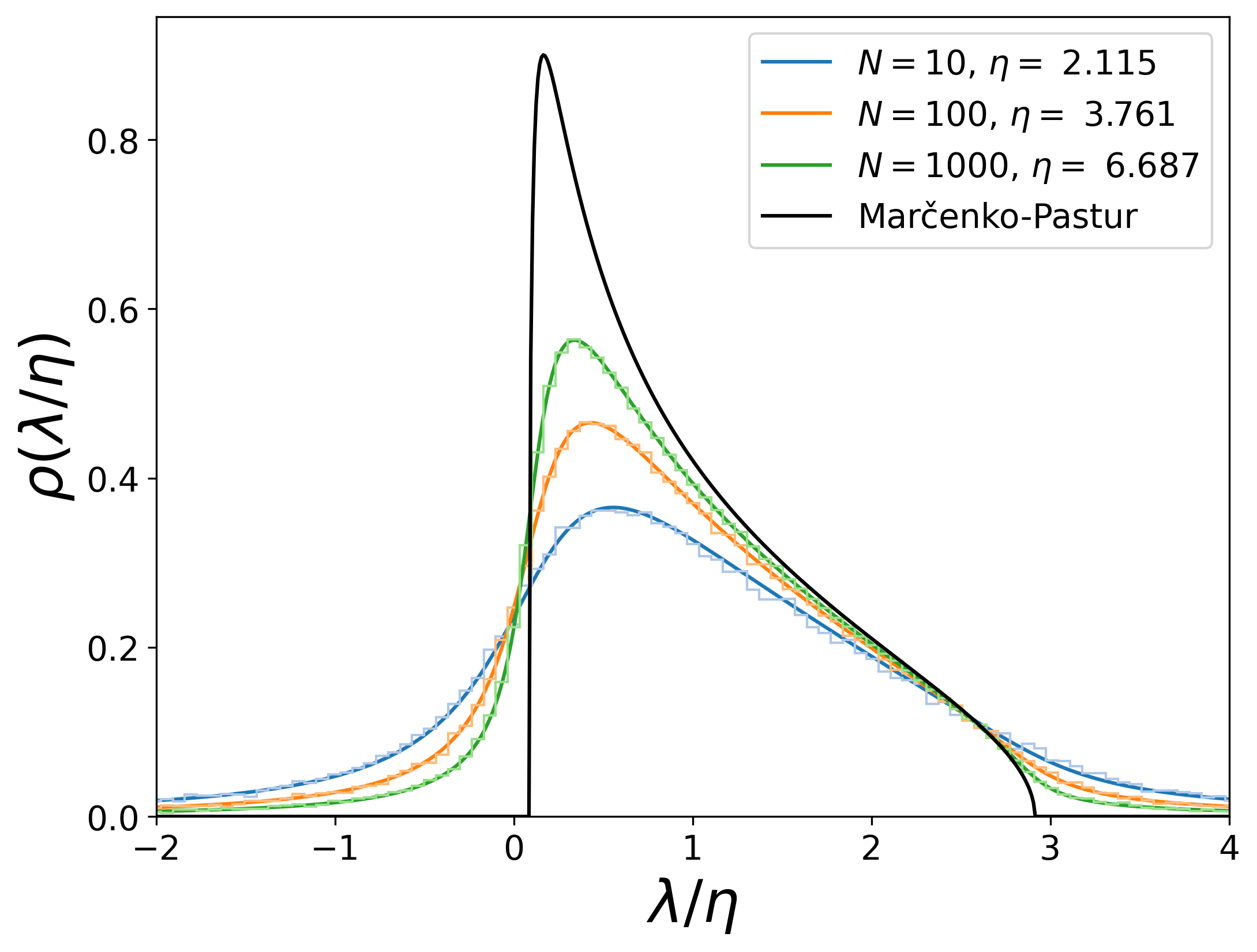}
    \put(-180,142){(b)}
    \caption{Density of states of the 
    WRP
    ensemble for a Cauchy distributed $p_a$, in (a) the intermediate phase ($1<\gamma = 5/4<3/2$), and in (b) the delocalized phase ($\gamma=3/4<1$). 
    The solid lines 
    correspond to
    our prediction~(\ref{eq:WRP_resolvent}), while the histograms are obtained from exact diagonalization of $10^5$ samples of the WRP ensemble, 
    with 
    $c=1/2$, $\nu =1$,
    and varying the matrix size $N$. 
    Upon increasing $N$,
    the densities converge  
    either to $p_a$ (which is the Cauchy distribution here) in (a), or to the Mar\v{c}enko--Pastur distribution in (b) --- this corresponds to the transition sketched in Fig.~\ref{fig:phase-diagram}.    
    Note that, in the delocalized phase, we have rescaled the eigenvalues $\lambda$ by $\eta$ in order to compensate for the growing support.}
    \label{fig:DoS_cauchy}
\end{figure}

In the case of Cauchy distributed $p_a$ (centered at $\mu$ and of width $\omega$), the Stieltjes transform is known in closed form: $\mathcal{G}_{Cauchy}(\lambda) = 1/(\lambda-\mu \pm {\rm i}\omega) $, where the $\pm $ branches correspond to $\Im \lambda> 0$ or $\Im \lambda < 0$, respectively. This 
renders Eq.~\eqref{eq:WRP_resolvent} a quadratic equation that can be easily solved (however, note that any other choice of $p_a$ could be evaluated numerically using Eq.~\eqref{eq:resolvantWRP}). The average spectral density in the particular case of Cauchy distributed $p_a$
is presented in Fig.~\ref{fig:DoS_cauchy}, where 
it is compared
to the numerical diagonalization of samples of the WRP ensemble.

\subsection{Full counting statistics}
\label{sec:number-eigen-replica}

Here, we apply the replica method to derive the level compressibility given in Eq.~\eqref{eq:comp_universal}.
To this end, we follow the procedure introduced in Refs.~\cite{Metz_2016,Metz_2017,PerezCastillo2018}, and more recently applied in Ref.~\cite{Venturelli_2023} to the case of the GRP model. Accordingly, we first obtain the cumulant generating function of the number of eigenvalues in a given interval, see Eq.~\eqref{eq:levels_number}, in the large-$N$ limit --- this quantity is also known as full counting statistics. From this expression, the level compressibility can then be retrieved using Eq.~\eqref{eq:level_compress}. Here, we will mainly focus on the intermediate phase ($ 1<\gamma<3/2$), and on intervals of the scale of the Thouless energy.

The number of eigenvalues in an interval delimited by $\alpha$ and $\beta$ is
given by 
\begin{equation}
    I_N[\alpha, \beta] = \sum_{i=1}^N \left[ \Theta(\beta-x_i) - \Theta(\alpha-x_i) \right] \, ,
\end{equation} 
where $\Theta(x)$ is the Heaviside distribution. 
Using the standard identity
\begin{equation}
    \Theta(-x) = \frac{1}{2\pi {\rm i}}\lim_{\varepsilon \to 0^+} \left[ \ln(x+{\rm i}\varepsilon) - \ln(x-{\rm i}\varepsilon) \right] 
\end{equation}
from complex analysis,
one can show that
\begin{equation}
    I_N[\alpha,\beta] = -\frac{1}{\pi {\rm i}}\lim_{\varepsilon \to 0^+} \ln \left[\frac{\mathcal{Z}(\beta-{\rm i}\varepsilon)\mathcal{Z}(\alpha+{\rm i}\varepsilon) }{\mathcal{Z}(\beta+{\rm i}\varepsilon)\mathcal{Z}(\alpha-{\rm i}\varepsilon)} \right],
    \label{eq:I_N_complex_rep}
\end{equation} 
where $\mathcal{Z} $ is the partition function appearing in the calculation
of the spectral density in Eq.~(\ref{eq:edwardsjonesZ}) 
(note that,
for $\mathcal{Z}(\alpha+{\rm i}\varepsilon)$ and $\mathcal{Z}(\beta+{\rm i} \varepsilon)$, one actually has to 
compute $ \int_{\mathbb{R}^N} \mathrm{d}^Nr \, \mathrm e^{\frac{{\rm i}}{2} \bm r^{T}(\lambda \mathbb{1} - \textbf{H}) \bm r } $ instead of $ \int_{\mathbb{R}^N} \mathrm{d}^Nr \, \mathrm e^{-\frac{{\rm i}}{2} \bm r^{T}(\lambda \mathbb{1} - \textbf{H}) \bm r } $ to ensure convergence 
of the integral). Now, assuming that one 
can exchange the limit $\epsilon \to 0^+$ and the logarithm, one writes the cumulant generating 
function of the random variable $I_N[\alpha, \beta] $ as 
\begin{equation}
    \mathcal{F}_{[\alpha,\beta]}(s) \equiv \ln \langle \mathrm e^{-sI_N[\alpha, \beta]}\rangle=   \lim_{\epsilon \to 0^+} \; \ln \; \langle \left[ \mathcal{Z}(\beta_\varepsilon^*)\mathcal{Z}(\alpha_\varepsilon)\right]^{ {\rm i} s/\pi}  \left[\mathcal{Z}(\beta_\varepsilon) \mathcal{Z}(\alpha_\varepsilon^*)\right]^{- {\rm i} s/\pi}   \rangle 
    \, . 
    \label{eq:cgf-def}
\end{equation}
Once again, 
in the spirit of the replica method, we first calculate
\begin{equation} 
\label{eq:cgf_replicas}
Q_{[\alpha, \beta]}(n_{\pm}) = \langle \left[ \mathcal{Z}(\beta_\varepsilon^*)\mathcal{Z}(\alpha_\varepsilon)\right]^{n_+}  \left[\mathcal{Z}(\beta_\varepsilon) \mathcal{Z}(\alpha_\varepsilon^*)\right]^{n_-}   \rangle 
\, , 
\end{equation} 
where $n_{\pm} $ are two independent integers. Next, we perform 
the analytic continuation of $n_{\pm} $ to the imaginary axis to 
express $\mathcal{F}_{[\alpha,\beta]}(s)$ in the form
\begin{equation}
    \label{eq:CGF_replica}
    \mathcal{F}_{[\alpha,\beta]}(s) =   \lim_{\epsilon \to 0^+} \; 
    \ln \; \lim_{n_{\pm} \to \pm  {\rm i} s/\pi} \; Q_{[\alpha, \beta]}(n_{\pm})
    \, .
\end{equation}
Following steps similar to those used in the calculation of the density of states, we obtain (see \ref{app:replicas})
\begin{align}
    \label{eq:Qalphabeta}
    Q_{[\alpha, \beta]}(n_{\pm}) \propto \int \mathcal{D}\phi \mathcal{D}\hat{\phi} \mathcal{D}\psi \mathcal{D}\hat{\psi} \text{ } \exp \left\{ - \sqrt{NM} \mathcal{S}_{n_{\pm}} [ \phi, \hat{\phi}, \psi, \hat{\psi} ; \hat{\Lambda}]  \right\},
\end{align}
with the action given by
\begin{align}
    &\mathcal{S}_{n_{\pm}} [ \phi, \hat{\phi}, \psi, \hat{\psi} ; \hat{\Lambda}] =  
      \frac{{\rm i}}{\sqrt{c}} \int
    \mathrm{d}\vec{u} \; \hat{\phi}(\vec{u}) \phi(\vec{u}) 
    + {\rm i} \sqrt{c} \; \int
     \mathrm{d}\vec{r} \; \hat{\psi}(\vec{r}) \psi(\vec{r}) \nonumber  \\
    & \qquad - \sqrt{c} \; \ln \int
     \mathrm{d}\vec{r} \; \varphi_a \left( -\frac{1}{2} \vec{r}\hat{L} \vec{r} \right) \; 
     \exp \left[-\frac{{\rm i}}{2}  \vec{r} \hat{\Lambda} \vec{r}+ {\rm i} \hat{\psi}(\vec{r}) \right] \nonumber \\
    & \qquad - \frac{1}{\sqrt{c}} \; \ln \int
     \mathrm{d}\vec{u} \; \exp \left[- \frac{1}{2} \vec{u}^{\, 2} + {\rm i} \hat{\phi}(\vec{u}) \right]  +\frac{{\rm i}}{2} \sqrt{c} \eta \int \mathrm{d}\vec{r} \mathrm{d}\vec{u} \text{ }\phi(\vec{u}) \psi(\vec{r}) \left( \vec{u} \hat{L} \vec{r} \right)^2, 
     \label{eq:action_RP_levelcompress}
\end{align} 
and where we introduced the block matrices
\begin{equation}
    \hat{\Lambda} = \mqty(\dmat{
\alpha_\epsilon \mathbb{1}_{n_+} \!\!\!\!,
  -\beta_\epsilon^*  \mathbb{1}_{n_+} \!\!\!\!,
 \beta_\epsilon \mathbb{1}_{n_-} \!\!\!\!, -\alpha_\epsilon^*  \mathbb{1}_{n_-} }),
\qquad   
\hat{L} = \mqty(\dmat{
 \mathbb{1}_{n_+}\!\!\!\!, - \mathbb{1}_{n_+} \!\!\!\!, \mathbb{1}_{n_-} \!\!\!\!, - \mathbb{1}_{n_-} }).
\end{equation}
Next, by using the saddle-point approximation, one can obtain a re-parametrisation of the action in terms of two matrices $\hat{K}$ and $\hat{C}^{-1}$ (see \ref{app:replicas}):
\begin{align}
    \label{eq:actionKC}
    \mathcal{S}_{n_{\pm}} [ \hat{K}, \hat{C}^{-1}; \hat{\Lambda}] 
     = 
    & -\sqrt{c} \ln \left[ \int_{-\infty}^{\infty} 
     \mathrm{d}a \; p_a(a) \; \exp \left\{ - \frac{1}{2} \ln \det \left( \hat{C}^{-1} - {\rm i} a \hat{L} \right)   \right\} \right] \nonumber \\ 
     & + \frac{1}{2\sqrt{c}} \Tr \log \left( \hat{\mathbb{1}} + {\rm i} q \eta   \hat{K}  \right) - \frac{{\rm i}}{2} \sqrt{c} \eta \; \Tr \left[ \hat{K} \left(  \hat{\mathbb{1}} +{\rm i} q \eta   \hat{K}\right)^{-1} \right],
\end{align} 
where 
$\hat{C}^{-1}$ and $\hat{K}$ 
have to satisfy the self-consistent relations
\begin{equation}
    \label{eq:matrixSPeq}
    \hat{C}^{-1} = {\rm i} \hat{\Lambda}+ {\rm i} \eta \hat{L} \left( \hat{\mathbb{1}} 
    + {\rm i} q \eta   \hat{K}  \right) ^{-1} \hat{L}, \qquad \text{and} \qquad \hat{K} 
    = -{\rm i}  \hat{L} \mathcal{G}_a \left( \left({\rm i} \hat{L}\hat{C}\right) ^{-1}  \right) ,
\end{equation} 
where 
$\mathcal{G}_a$ was given in \cref{eq:stieltjes}.
We now look for a block diagonal solution of Eq.~(\ref{eq:matrixSPeq}), i.e.~of the form
\begin{equation}
    \label{eq:diagK}
    \hat{K} \equiv 
    \mqty(\dmat{k_{\alpha} \mathbb{1}_{n_+} \!\!\!\!,
    \bar{k}_{\beta} \mathbb{1}_{n_+} \!\!\!\!,
    k_{\beta} \mathbb{1}_{n_-} \!\!\!\!, \bar{k}_{\alpha} \mathbb{1}_{n_-}})  
    , \;
    \hat{C}^{-1} \equiv 
    \mqty(\dmat{\Delta^{-1}_{\alpha}\mathbb{1}_{n_+}\!\!\!\!, \bar{\Delta}^{-1}_{\beta} \mathbb{1}_{n_+} \!\!\!\!, \Delta^{-1}_{\beta} \mathbb{1}_{n_-} \!\!\!\!, \bar{\Delta}^{-1}_{\alpha} \mathbb{1}_{n_-} }).
\end{equation}
This Ansatz can then be inserted in Eq.~(\ref{eq:actionKC}) to derive 
\begin{align}
    \label{eq:actiondiagK}
    \mathcal{S}_{n_{\pm}} = & -\sqrt{c} \ln \left\lbrace \int_{-\infty}^{\infty} 
     \mathrm{d}a \; p_a(a) \; \mathrm \mathrm e^{ - \frac{n_+}{2} \ln \left[  \left( \Delta^{-1}_{\alpha} - {\rm i} a \right) \left( \bar{\Delta}^{-1}_{\beta} + {\rm i} a \right) \right]  -  \frac{n_-}{2} \ln \left[  \left( \Delta^{-1}_{\beta} - {\rm i} a \right) \left( \bar{\Delta}^{-1}_{\alpha} + {\rm i} a \right) \right] } \right\rbrace \nonumber \\ 
     & + \frac{1}{2\sqrt{c}} \left\lbrace n_+ \ln\left[ (1+{\rm i}\eta c k_{\alpha}) (1+{\rm i}\eta c \bar{k}_{\beta})  \right] + n_- \ln\left[ (1+{\rm i}\eta c k_{\beta}) (1+{\rm i}\eta c \bar{k}_{\alpha})  \right]\right\rbrace \nonumber \\ 
     & - \frac{{\rm i}}{2} \sqrt{c} \eta  \left[  n_+ \left( \frac{k_{\alpha}}{1+{\rm i}\eta c k_{\alpha}} + \frac{\bar{k}_{\beta}}{1+{\rm i}\eta c \bar{k}_{\beta}}  \right) + n_- \left( \frac{k_{\beta}}{1+{\rm i}\eta c k_{\beta}} + \frac{\bar{k}_{\alpha}}{1+i\eta c \bar{k}_{\alpha}} \right) \right] .
\end{align}
Now, taking the limit $n_{\pm} \to \pm {\rm i}s/\pi$ we obtain
\begin{align}
    \mathcal{S}_{\pm \frac{{\rm i} s}{\pi}} =& -\sqrt{c} \, 
    \ln \left\{ \int_{-\infty}^{\infty} \mathrm{d}a \, p_a(a) \, \exp \left( -\frac{{\rm i} s}{2\pi} \ln \left[ \frac{ \Bigr ( \Delta^{-1}_{\alpha} - {\rm i} a \Bigl ) \left( \bar{\Delta}^{-1}_{\beta} + {\rm i} a \right) }{  \left( \Delta^{-1}_{\beta} - {\rm i} a \right) \left( \bar{\Delta}^{-1}_{\alpha} + {\rm i} a \right)  } \right] \right) \right\}
    \nonumber \\
     &+ \frac{{\rm i} s}{2\pi \sqrt{c}} \ln \left[ \frac{(1 + {\rm i} \eta c k_{\alpha})(1 + 
     {\rm i} \eta c \bar{k}_{\beta})}{(1 + {\rm i} \eta c k_{\beta})(1 + {\rm i} \eta c \bar{k}_{\alpha})} \right] 
     \nonumber \\
    &- \frac{\eta s}{2\pi \sqrt{c}} \left[ \left( \frac{k_{\alpha}}{1 + {\rm i} \eta c k_{\alpha}} + 
    \frac{\bar{k}_{\beta}}{1 + {\rm i} \eta c \bar{k}_{\beta}} \right) - \left( \frac{k_{\beta}}{1 + {\rm i} \eta c k_{\beta}} + \frac{\bar{k}_{\alpha}}{1 + {\rm i} \eta c \bar{k}_{\alpha}} \right) \right].
    \label{eq:final-action}
\end{align}
This finally allows us to obtain the cumulant generating function 
$\mathcal F_{[\alpha,\beta]}(s)$, which follows from Eqs.~\eqref{eq:CGF_replica} and~\eqref{eq:Qalphabeta} as
\begin{equation}
    \label{eq:CGF_replica-2}
    \mathcal{F}_{[\alpha,\beta]}(s) =   \sqrt{NM}\lim_{\epsilon \to 0^+} \; \mathcal{S}_{\pm {\rm i} s/\pi}
    \; + \mathcal{O}(N^{-\gamma}).
\end{equation}

This concludes the replica calculation. To get the moments of the number of eigenvalues in an interval $I_N[\alpha,\beta]$, one then proceeds as follows:
\begin{enumerate}[(i)]
    \item The self-consistent equations~\eqref{eq:matrixSPeq} can be solved numerically to find the 8 elements of the block matrices $\hat K$ and $\hat C^{-1}$ in Eq.~\eqref{eq:diagK}. Note that these are actually 8 nonlinear equations, which can be solved with minimal numerical effort for any reasonable choice of $p_a$.
    \item Inserting these matrix elements into \cref{eq:final-action} makes the cumulant generating function fully explicit, in spite of its seemingly complicated integral form.
    \item Expanding \cref{eq:final-action} in powers of $s$ and using \cref{eq:CGF_replica-2}, one can identify the cumulants $\kappa_j$ as 
    \begin{equation}
    \kappa_j[\alpha,\beta] = (-1)^{j} \eval{\partial_s^j \mathcal{F}_{[\alpha,\beta]}(s)}_{s=0} .
    \label{eq:cumulants}
\end{equation}
\end{enumerate}

We conclude by pointing out that \cref{eq:I_N_complex_rep}, i.e.~the starting point of our calculation, was actually obtained by adopting the identity $\ln (ab) = \ln a + \ln b$, which is however not satisfied in general by the complex logarithm (whose principal branch is bounded within $(-\pi,\pi]$~\cite{Vivo_2020}). This issue is usually (and quite remarkably) solved via the introduction of replicas, thanks to the so-called \textit{folding-unfolding mechanism}~\cite{PerezCastillo2018}.
Yet, the cumulant generating function obtained in \cref{eq:final-action,eq:CGF_replica-2} cannot be immediately recognized as a real quantity, as one would have hoped for in general. In other random matrix ensembles, for which the spectral density and/or the chosen interval are symmetric around the origin, the vanishing of the imaginary part of $\mathcal{F}_{[\alpha,\beta]}(s)$ can actually be proven analytically~\cite{Metz_2017,Venturelli_2023}. 
In the present case, the asymmetry of the spectral density prevented us from carrying out such a proof; however, we have checked that indeed $\mathcal{F}_{[\alpha,\beta]}(s)$ becomes real when $N\to \infty$ for $\gamma>1$, in which case the density of states also becomes symmetric, and we have considered the real part of $\mathcal{F}_{[\alpha,\beta]}(s)$ otherwise.\footnote{The method adopted here, although with a different Ansatz, was applied in Ref.~\cite{PerezCastillo2018} to characterize the \textit{index} (i.e.~the number of eigenvalues in the interval $(-\infty,\beta]$) within the diluted Wishart ensemble, whose spectrum is in fact not symmetric. Inspecting this quantity within the WRP ensemble would be insightful in the future, in view of better assessing the limitations of the replica method.}

\subsection{Level compressibility}
\label{sec:level-comp-replica}

\begin{figure}[t]
\vspace{0.5cm}
    \centering
    \includegraphics[width=0.6\linewidth]{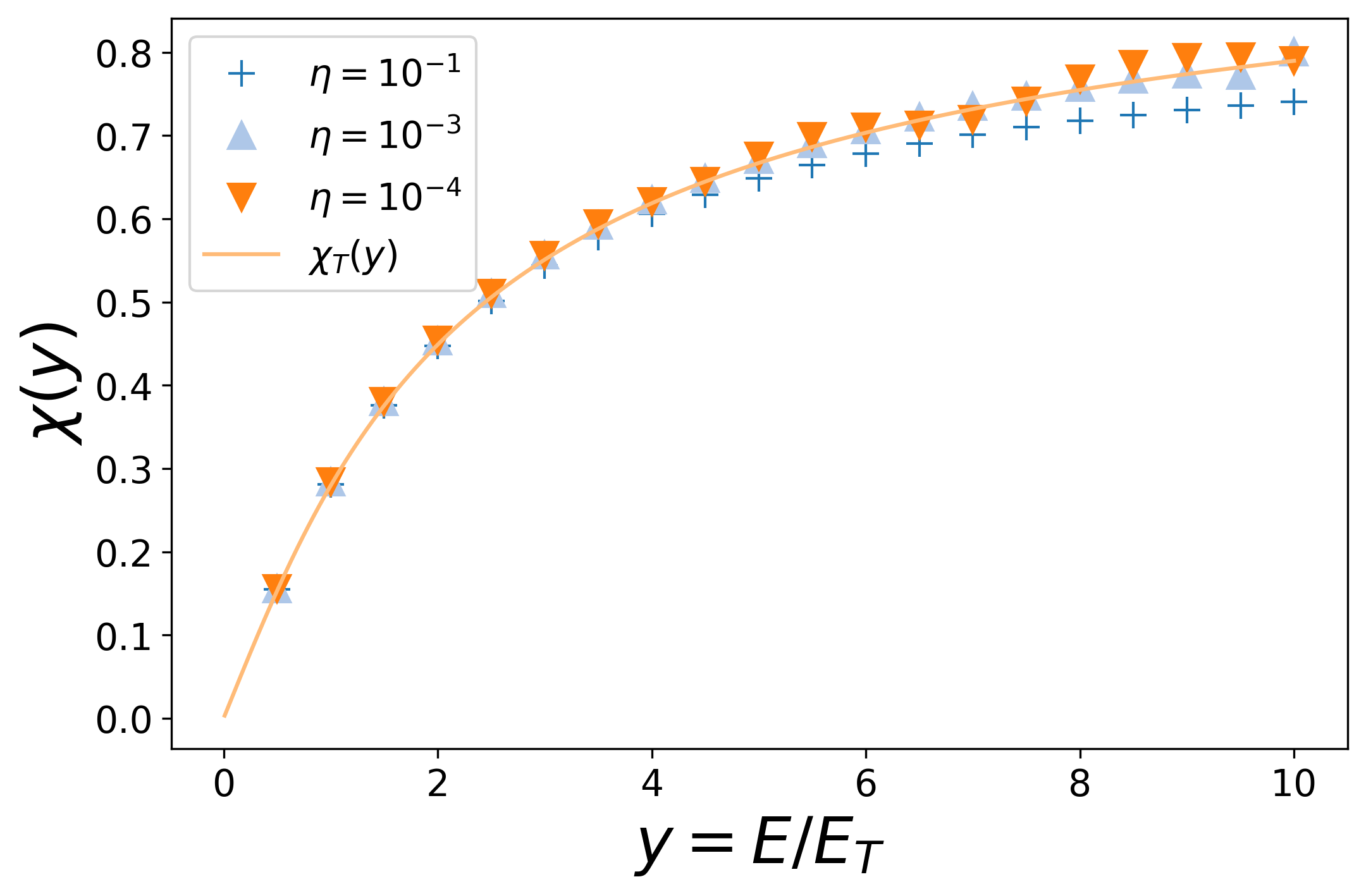}
    \caption{Comparison between the universal scaling form~\eqref{eq:comp_universal} of the level compressibility $\chi_T$ (solid line), and the analytical replica solution for $\chi(E)$ (symbols, see Sec.~\ref{sec:number-eigen-replica}), computed for 
    intervals of the order of the Thouless energy. Upon increasing the matrix size $N$ --- or equivalently, upon decreasing $\eta\propto N^{1-\gamma}$ for $\gamma>1$ --- the predicted level compressibility approaches the universal form $\chi_T$, as conjectured in Ref.~\cite{Venturelli_2023}.
    }
    \label{fig:LC_ET_replica}
\end{figure}

By specializing the interval to $[\alpha,\beta]=[-E+\eta,E+ \eta]$, we can now focus on the ratio between the first two cumulants given by \cref{eq:cumulants}, which corresponds to the level compressibility $\chi(E)$ introduced in \cref{eq:level_compress}.
The result is plotted in \cref{fig:LC_all_E}, where it is tested against numerical diagonalization of large sample random matrices. We can generically distinguish three regimes:
\begin{enumerate}[(i)]
    \item For energies $E< \Delta$, where $\Delta \simeq 1/N$ is the mean level spacing, the saddle-point calculation breaks down --- as expected, since in our derivation we have treated the eigenvalue density as a continuous distribution. At these energies, the eigenvalue statistics is dominated by level repulsion, as is typically the case in the RMT regime. Here, $\chi(E)$ turns out to be well approximated by the level compressibility of a GOE matrix (see e.g.~Appendix~E.3 in~Ref.\cite{Venturelli_2023}):
    \begin{align} 
    \label{chi_GOE_final} 
        \chi_{\rm GOE}(y) = & \, \frac{1}{2 \pi^2 y} \big\lbrace [\text{Si}(2 \pi y)]^2 -2\, \text{Ci}(4 \pi y) -\pi \, \text{Si}(2 \pi y)\nonumber\\
        &+2 \left[-4 \pi y \, \text{Si}(4 \pi y)+2 \pi ^2 y+\log (4 \pi y)-\cos (4 \pi y)+\gamma_E +1\right]\big\rbrace \, , 
    \end{align} 
    where ${\rm Ci}(z) = - \int_z^\infty \cos(t)/t\, \dd t$ and ${\rm Si}(z) = \int_0^z \sin(t)/t\, \dd t$ are the cosine-integral and sine-integral functions, respectively, while $\gamma_E$ is the Euler–Mascheroni constant. The fact that $\chi_{\rm GOE}(y)$ well describes also the level compressibility of a Wishart matrix is best rationalized within the Coulomb gas interpretation~\cite{Livan_2018}: indeed,
    at these scales, correlations between eigenvalues originate from the Coulomb gas interaction, which is the same for both GOE and real Wishart matrices.
    \item Around the Thouless energy, i.e.~for $E\sim E_T \propto N^{2(1-\gamma)}$, the level compressibility is well described by the universal scaling form~\eqref{eq:comp_universal} found in Ref.~\cite{Venturelli_2023} within the GRP model. The agreement with the numerical results improves upon increasing $N$, as we exemplified in \cref{fig:LC_ET_numerics}(a). Moreover, the independence of $\chi(E\sim E_T)$ from the ratio $c=N/M$, which is an expected but nontrivial feature, is demonstrated in \cref{fig:LC_ET_numerics}(b). Finally, in \cref{fig:LC_ET_replica} we show that the replica prediction also approaches $\chi_T$ upon increasing $N$ (i.e.~upon decreasing $\eta\propto N^{1-\gamma}$). 
    \item For energies of $\order{1}$, the eigenvalues behave as uncorrelated random variables, Eq.~\eqref{eq:chiPoisson}, whence (see e.g.~Appendix~A in~\cite{Venturelli_2023})
    \begin{equation}
    \chi(E) \sim \chi_\mathrm{iid} (E) = 1 - \frac{\langle I_N[-E, E]\rangle}{N} = 1 - \int_{-E}^{E} \dd{\lambda} p_a(\lambda) \, .
\end{equation}
\end{enumerate}

\section{Dyson Brownian motion}
\label{sec:Dyson}

In this Section we 
derive the phase diagram of the 
WRP
model 
using 
yet another
approach, similar to 
the one 
first introduced in Ref.~\cite{Dyson_1962_I,Pandey_1995}, and later
applied to
the GRP model in Ref.~\cite{Facoetti_2016}.
The idea is to 
interpret \cref{eq:AB} as a matrix-valued stochastic process, for which $\textbf{A}$ represents the initial condition, and $\textbf{B}$ a perturbation
which is turned on at 
the fictitious ``time'' $t=0$;
the matrix $\textbf{H}$ is eventually recovered at the final time $t=T$.
Under the effect of this perturbation, the eigenvalues and eigenvectors themselves become stochastic processes, and 
inspecting their behavior
can give us information on the 
limiting eigenvalue density, and the statistics of the phase (localized, delocalized or fractal).

We start by
setting up
the problem. We study the matrix $\textbf{H}(t) = \textbf{A} + \textbf{X}(t)\textbf{X}(t)^T$, where $\textbf{X}(t)$ is a $N\times M$ matrix-valued Brownian motion, and \textbf{A} is the same diagonal random matrix as in Eq.~(\ref{eq:AB}).
At each time step, we can write the evolution of $\textbf{X}(t)$ as $\textbf{X}(t+\mathrm{d}t) = \textbf{X}(t) + \textbf{g}(t) $, with $\textbf{g}(t)$ being an i.i.d.~uncorrelated $N\times M $ Gaussian noise with zero mean and variance proportional to $\mathrm{d}t$, i.e.~$ \langle g_{ik}(t) \rangle = 0$ and $ \langle g_{ik}(t) g_{jl}(t') \rangle = \sigma^2 \mathrm{d}t \; \delta_{ij} \delta_{kl} \; \delta(t-t')$. 
The stochastic evolution equation of $\textbf{H}(t)$ then follows as
\begin{align}
\label{eq:H_dbm}
       \textbf{H}(t+\mathrm{d}t) & = \textbf{A}+ \textbf{X}(t+\mathrm{d}t)\textbf{X}(t+\mathrm{d}t)^T 
       \\ 
        &= \textbf{A}+ \textbf{X}(t)\textbf{X}(t)^T + \textbf{g}(t)\textbf{X}(t)^T + \textbf{X}(t)\textbf{g}(t)^T + \textbf{g}(t) \textbf{g}(t)^T 
        = \textbf{H}(t) + \delta \textbf{H}(t),
        \nonumber
\end{align} 
where in the last step we defined $ \delta \textbf{H}(t) = \textbf{g}(t)\textbf{X}(t)^T + \textbf{X}(t)\textbf{g}(t)^T + \textbf{g}(t) \textbf{g}(t)^T $. 
Upon requiring that, at time $T$, $\textbf{H}(t=T) = \textbf{A}+ \nu M^{-\gamma}\textbf{W}\textbf{W}^T $, and by using that $\sum_{l=1}^M X_{il} X_{jl} \sim M\, t$, we get that the stopping time of this process 
must be given by $T = \nu M^{-\gamma}$. Now, applying perturbation theory up to second order, one can derive the following stochastic differential equations for the eigenvalues $ \lambda_i (t) $ and the components of the associated eigenvectors $\braket{n| \psi_i(t)} \equiv \psi_i (n ; t) $ of $\textbf{H}(t)$, starting from the initial condition $\lambda_i (t=0) = a_i$ and $\psi_i (n; \, t = 0) = \delta_{in} $ (we omitted the time dependencies for clarity, and the details of the derivation are given in \ref{app:DBM}):
\begin{equation}
    \label{eq:DBM_lambda}
    \frac{ \mathrm{d}\lambda_i}{\mathrm{d}t} = \sigma^2 \left[ M +  \sum_{j \neq i}   \frac{\lambda_i + \lambda_j  - \langle a \rangle_i - \langle a \rangle_j}{\lambda_i - \lambda_j} \right] + 2 \sigma \sqrt{\lambda_i - \langle a \rangle_i} \; \zeta_i, 
\end{equation} 
and 
\begin{align}
    \label{eq:DBM_psi}
    \frac{\mathrm{d} \psi_i (n)}{\mathrm{d} t} = &\;  \sigma^2 \sum_{j \neq i} \psi_j (n) \frac{\langle a \rangle_{ij}}{\lambda_i - \lambda_j}\Bigg( \frac{1}{\lambda_i - \lambda_j} - \sum_{l \neq i} \frac{1}{\lambda_i - \lambda_l}\Bigg) \nonumber \\
    & -\frac{1}{2} \sigma^2 \psi_i (n)  \sum_{j \neq i} \frac{\lambda_i+\lambda_j-\langle a \rangle_i - \langle a \rangle_j}{\lambda_i - \lambda_j} + \sigma \sum_{j \neq i} \psi_j (n) \frac{\zeta_{ij}}{\lambda_i - \lambda_j}.
\end{align} 
Here, $\zeta_i$ is 
a delta-correlated Gaussian white noise,
while $\zeta_{ij}$ 
is a correlated Gaussian noise with zero mean and variance
\begin{eqnarray}
    \label{eq:DBM_noise_psi}
    && 
    \langle \zeta_{ij}(t) \zeta_{kl} (t')\rangle = \; \delta (t-t') 
    \Bigl [ (\lambda_i + \lambda_j)\bigl( \delta_{ik}\delta_{jl} + \delta_{il}\delta_{jk} \bigr)
    \nonumber\\
    && 
    \qquad\qquad\qquad\qquad\quad\quad
- \bigl( 
\delta_{ik}\langle a\rangle_{jl} + 
\delta_{il}\langle a\rangle_{jk} + 
\delta_{jk}\langle a\rangle_{il} + 
\delta_{jl}\langle a\rangle_{ik} 
\bigr) \Bigr].
\end{eqnarray}
We also defined 
\begin{align}
    \langle a \rangle_i(t) &\equiv \bra{\psi_i(t)} \textbf{A}\ket{\psi_i (t)} = \sum_n a_n |\psi_i (n; t) |^2\, ,
    \label{eq:ai-def}\\
    \langle a \rangle_{ik}(t) &\equiv \bra{\psi_i (t)} \textbf{A}\ket{\psi_k (t)} = \sum_n a_n \psi_i (n; t) \psi_k (n; t)\, ,
    \label{eq:aik-def}
\end{align}
which implies that, in this model, eigenvalues are coupled to eigenvectors (this was not the case in the GRP model --- see Eqs.~(12) and~(13) in Ref.~\cite{Facoetti_2016}). 
Moreover, one can note that 
setting $\textbf{A}$ to $0$
correctly renders
the Dyson Brownian motion equation for the eigenvalues and eigenvectors of a pure Wishart process~\cite{bru_wishart_1991}. 
Although in general
these equations are complicated to solve, 
here we will merely be concerned with
their short-time 
behavior,
because the stopping time $T \sim N^{-\gamma}$ is vanishingly small for all values of $\gamma$ 
of interest.

We begin by considering the evolution of the eigenvalues in \cref{eq:DBM_lambda}.
At short times, only the first term in Eq.~\eqref{eq:DBM_lambda} 
gives a relevant contribution,
i.e.~$\frac{ \mathrm{d}\lambda_i}{\mathrm{d}t} \simeq \sigma^2 M $,
because $\lambda_i(t=0) = a_i =\expval{a}_i(t=0)$. 
Thus, after a time $T \sim N^{-\gamma}\ll 1$, the eigenvalues have moved by 
\begin{equation}
    \delta \lambda_i (T) = \lambda_i(t) -  a_i\sim \sigma^2 M T \sim N^{1-\gamma}.
\end{equation}
This means that for $\gamma > 1$ the shift of the eigenvalues is vanishingly small, and thus they 
remain
close to 
their initial configuration $\lambda_i (t=0) = a_i$:
the spectral density is then given by $p_a$, as expected.
Conversely,
for $\gamma < 1$, the $a_i = \mathcal{O}(1)$ are negligible 
with respect to
the shift of $\mathcal{O}(N^{1-\gamma})$, and thus the eigenvalues essentially follow the evolution of a pure Wishart process, whose stationary distribution is known to be the Mar\v{c}enko--Pastur distribution~\cite{bru_wishart_1991}.

For the eigenvectors, one can 
perform
a similar analysis. At short times, for $n\neq i$, the only non-zero term in Eq.~\eqref{eq:DBM_psi} is the last one:  
\begin{equation}
     \frac{\mathrm{d} \psi_i (n)}{\mathrm{d} t} \underset{t \ll 1}{\approx}  \sigma \sum_{j \neq i} \delta_{jn} \frac{\zeta_{ij}}{a_i - a_j} =   \sigma \frac{\zeta_{in}}{a_i - a_n},
\end{equation} 
and Eq.~\eqref{eq:DBM_noise_psi} simplifies to $ \langle \zeta_{ij}(t) \zeta_{kl} (t')\rangle \underset{t \ll 1}{\approx} \; \delta (t-t') \; \bigl [ \lambda_i(t) + \lambda_j(t) - a_i - a_j \bigr ] 
\bigl  ( \delta_{ik}\delta_{jl}  + \delta_{il} \delta_{jk} \bigr )$. 
The variance of the components of the eigenvectors at time $T$ is then given by
\begin{eqnarray}
\Big\langle 
\big| \psi_i (n ; T) \big|^2 
\Big\rangle
&=& 
\left( \frac{\sigma}{a_i - a_n} \right)^2
\int_0^T \mathrm{d}t \int_0^T \mathrm{d}t' 
\; 
\big\langle \zeta_{in}(t)\, \zeta_{in}(t') \big\rangle \,   
\nonumber \\
&\approx & 
\frac{\sigma^2}{(a_i - a_n)^2}
\int_0^T \mathrm{d}t \; \big( \lambda_i(t) +\lambda_n(t) - a_i - a_n \big)
\, .
\end{eqnarray} 
Now, using $\lambda_i(t) -  a_i \approx \sigma^2 M t$, and the fact that 
the mean level spacing scales as $a_i - a_n \sim 1/N$, we obtain that after a time $T \sim N^{-\gamma} $ 
\begin{equation}
    \Bigl \langle \Bigl \vert \psi_i (n ; T) \Bigr \vert ^2\Bigr \rangle \sim N^2 \times MT^2 \sim N^3T^2 \sim N^{3-2\gamma} .
\end{equation}
Therefore, the eigenvectors are delocalized whenever $\gamma < 3/2$, and are localized otherwise.

\section{Conclusions}
\label{sec:conclusions}

In this paper we introduced a new variant of the Rosenzweig--Porter model, which we called the Wishart--Rosenzweig--Porter ensemble. This model is defined as the sum of a diagonal matrix and a Wishart matrix, as in \cref{eq:AB,eq:WW}. 
We provided a comprehensive analysis of its 
spectral and localization
properties in the limit of large matrix size, combining several complementary analytical approaches. In particular, we 
characterized
the phase diagram of the model using perturbation theory (Sec.~\ref{sec:criteria}), the cavity method (Sec.~\ref{sec:cavity}), the replica formalism (Sec.~\ref{sec:replica}), and a Dyson Brownian motion approach (Sec.~\ref{sec:Dyson}). 

The WRP ensemble displays two distinct transitions (see \cref{fig:phase-diagram}): a transition in the spectral density at $\gamma = 1$, and a localization transition in the eigenvector statistics at $\gamma = 3/2$. In the intermediate regime ($1 < \gamma < 3/2$), the eigenvectors are neither fully localized nor extended, but instead occupy a fractal support of size $N^D$, with fractal dimension $D = 3 - 2\gamma < 1$, which is thus much smaller than $N$.

We also analyzed the spectral correlations through the full counting statistics and the level compressibility defined in Eq.~\eqref{eq:level_compress}. The key result of our work is that, in the intermediate phase ($1 < \gamma < 3/2$), for energy scales between the mean level spacing and the Thouless energy ($1/N \ll E_T \sim N^{2-2\gamma} \ll 1$), the level compressibility follows the same scaling function as in other RP-type models (see Eq.~\eqref{eq:comp_universal}). This agreement, observed both analytically and numerically, strongly supports the hypothesis that spectral correlations on this scale are (super)universal. While their behavior at smaller and larger energy scales depends on the specific form of the underlying matrix distributions, the intermediate-scale correlations appear to be independent of any microscopic detail.

This finding motivates further numerical investigations of the crossover function~\eqref{eq:level_compress} in realistic many-body disordered quantum systems exhibiting (multi)fractal phases, to test whether the same universal behavior persists in those contexts as well.
The most natural framework to begin this investigation is provided, in our view, by the quantum random energy model~\cite{Faoro_2019,parolini2020multifractal,baldwin2018quantum,biroli2021out,tarzia2022fully}.

Finally, several open questions emerge from our study. 
First,
Wishart matrices often describe covariance data where the number of samples greatly exceeds the number of variables~\cite{Bouchaud_Les_Houches,bun2017cleaning}; this motivates extending the WRP ensemble to the regime $ M \ll N $, e.g.~$ M \sim N^{\gamma'} $ with $ \gamma' < 1 $.
Second,
both GOE and Wishart ensembles share the key feature of having a spectral density of the form $\exp[-\mathrm{Tr}\, V(M)]$, leading to Haar-distributed eigenvectors. This common structure may underlie the observed universality and suggests possible extensions to generic matrix models with potentials $V(M)$.
Third, having found that
correlations between the entries of $\mathbf{B}$ preserve the conjectured universality of the level compressibility, it would be interesting to explore what happens when correlations are also introduced among the entries of $\mathbf{A}$~\cite{Sarkar_2023,Ghosh_2025}. Another direction is to study spectral correlations in models with explicitly multifractal eigenstates, for example by considering sums of several random matrices with distinct fractal dimension spectra. 
Finally, it would be highly desirable to develop a more systematic framework to study spectral correlations, possibly using tools from free probability theory, which would constitute a
promising avenue for future research.

 \section*{Acknowledgments}
 We thank 
Ivan Khaymovich, Guilhem Semerjian and
 Pierpaolo Vivo for useful 
 discussions, and
 Victor Sitzmann for his help with the numerical 
 calculations.
We acknowledge financial support from the ANR research grant ManyBodyNet
ANR-24-CE30-5851. GS acknowledges support from ANR Grant No.~ANR-23-CE30-0020-01 EDIPS.
 
\appendix
\addtocontents{toc}{\fixappendix}

\section{Scaling function of the level compressibility}
\label{app:scaling-function}

In this Appendix, we provide the details of the computation 
that leads to the closed-form scaling function~\eqref{eq:comp_universal}, assumed by the level compressibility $\chi(E)$ in the vicinity of the Thouless energy $E_T$.
To this end, we analyze
the integral that enters the definition of $\chi(x)$ given in Eq.~(\ref{eq:cavity-scaling-comp}). We introduce the function 
\begin{equation} \label{F_chi}
F_{\chi}(b) = \int_{-\infty}^\infty {\rm d} \tilde a \left[ \atan \left( b + \tilde{a}  \right) - \atan \left( \tilde{a} - b \right) \right]^2,
\end{equation}
such that $\chi(x) \approx  \frac{c \, p_a(0)}{2 \pi x} F_{\chi}(x/(c \pi p_a(0)))$. We first notice that $F_\chi(0) =0$ and compute its derivative $F_\chi'(b)$, which reads
\begin{equation} \label{Fprime1}
F_{\chi}'(b) = 2 \int_{-\infty}^\infty {\rm d} \tilde a \left( \frac{1}{1+(\tilde a -b)^2} + \frac{1}{1+(\tilde a +b)^2} \right)\left( {\rm atan}(\tilde a +b)-{\rm atan}(\tilde a -b)\right) \;.
\end{equation}
By using the symmetry of the interval of integration together with ${\rm atan}(-x) = - {\rm atan}(x)$, the derivative $F_{\chi}'(b)$ can be simply re-written as
\begin{equation} \label{Fprime2}
F_{\chi}'(b) = 4 \int_{-\infty}^\infty {\rm d} u \frac{{\rm atan}(u)}{1+(u-2b)^2} \;.
\end{equation}
To proceed, it is convenient to use the integral representation
\begin{equation} \label{int_rep}
\frac{1}{1+x^2} = \frac{1}{2}\,\int_{-\infty}^\infty {\rm d} k \, \e^{ {\rm i}  k x - |k|} \;.
\end{equation}
We then insert this integral representation into Eq.~(\ref{Fprime2})
and use the identity 
\begin{equation}\label{id_1}
\int_{-\infty}^\infty {\rm d}u\, {\rm atan}(u)\, \e^{{\rm i}k u} = \frac{{\rm i} \pi}{k}\,\e^{-|k|} 
\end{equation}
(which can be shown using integration by parts), to compute the integral over $u$ in Eq.~(\ref{Fprime2}). This leads to
\begin{equation}\label{Fprime3}
F_{\chi}'(b) = 2 \pi \int_{-\infty}^\infty \frac{\rm d k}{k} \sin(2 k \, b)\e^{-2 |k|} = 4 \pi {\rm atan}(b) \;,
\end{equation}
which can be easily 
proven by taking a derivative with respect to $b$. Finally, $F_\chi(b)$ can be obtained by integrating (\ref{Fprime3}) and using $F_{\chi}(0) = 0$, yielding
\begin{equation} \label{Ffinal}
F_{\chi}(b) = \int_0^b {\rm d}x \, {\rm atan}(x) = b\, {\rm atan}(b) - \frac{1}{2} \ln(1+b^2) \;.
\end{equation}
Using $\chi(x) \approx  \frac{c \, p_a(0)}{2 \pi x} F_{\chi}(x/(c \pi p_a(0)))$, this finally leads to Eq. (\ref{eq:comp_universal}). 

\section{Details of the replica calculation}
\label{app:replicas}

Here we provide details of the derivations presented in Sec.~\ref{sec:replica}.

\subsection{Density of states}

In this Appendix, we 
provide details on the derivation
of the density of states using the replica formalism.
We start by calculating the average of the partition function replicated $n$ times, according to \cref{eq:edwardsjonesZ,eq:replicatrick}:
\begin{align}
\langle \mathcal{Z}^n(\lambda) \rangle & = \Biggl \langle \int
 \prod_{\alpha = 1}^{n} \mathrm{d}\bm{r}_{\alpha} \; 
 \mathrm e^{-\frac{\rm i}{2} \sum\limits_{\alpha=1}^{n} 
 \bm{r}_{\alpha}^{\; T}(\lambda \mathbb{1} - \textbf{H}) \bm{r}_{\alpha} } \Biggr \rangle_{\textbf{H}}  
 \nonumber  \\
&=  \int
 \prod_{\alpha = 1}^{n} \mathrm{d}\bm{r}_{\alpha} \; 
 \mathrm e^{-\frac{\rm i}{2} \lambda \sum\limits_{i=1}^{N} \sum\limits_{\alpha=1}^{n}  r_{i \alpha}^2  } \; 
 \Biggl \langle	
 \mathrm e^{\frac{\rm i}{2} \sum\limits_{i,j=1}^{N} \sum\limits_{\alpha=1}^{n} r_{i \alpha} H_{ij} r_{j \alpha} } \Biggr \rangle_{\textbf{H}} 
 \, .
 \label{appeq:a1}
 \end{align}
Then we insert the definition of the model $H_{ij} = a_i \delta_{ij} + \nu M^{-\gamma} \sum_{k=1}^M W_{ik}W_{jk} $ 
(see Sec.~\ref{sec:WRP-model}),
where $W_{ij} \sim \mathcal{N}(0,1) $, and $a_i$ are i.i.d. random variables sampled from $p_a(a)$. 
Using that the matrices $\textbf{A}$ and $\textbf{W}$ are independent, we now
rewrite the term in 
brackets as
\begin{align}
     \Biggl \langle	\mathrm e^{\frac{\rm i}{2} \sum\limits_{i,j=1}^{N} 
    \sum\limits_{\alpha=1}^{n} r_{i \alpha} H_{ij} r_{j \alpha} } \Biggr \rangle_{\textbf{H}}  
    \! & =  \Biggl \langle \mathrm e^{\frac{\rm i}{2} \sum\limits_{i = 1}^N  \sum\limits_{\alpha=1}^{n} r_{i\alpha}^2 a_i } \Biggr \rangle_{\textbf{A}} \!
     \Biggl \langle \mathrm e^{\frac{\rm i \nu }{2M^{\gamma}} \sum\limits_{i,j=1}^{N} \sum\limits_{\alpha=1}^{n} r_{i \alpha} \sum\limits_{k=1}^{M} W_{ik}W_{jk} r_{j \alpha} } \Biggr \rangle_{\textbf{W}} \nonumber \\
    &  = \Biggl \langle \mathrm e^{\frac{\rm i }{2} \sum\limits_{i = 1}^N  \sum\limits_{\alpha=1}^{n} r_{i\alpha}^2 a_i } \Biggr \rangle_{\textbf{A}} \;
     \Biggl \langle \mathrm e^{\frac{\rm i \nu }{2M^{\gamma}} \sum\limits_{k=1}^{M}  \sum\limits_{\alpha=1}^{n} \left( \sum\limits_{i=1}^{N} r_{i \alpha} W_{ik} \right)^2} \Biggr \rangle_{\textbf{W}}.
\end{align}
Given that the measure is Gaussian, the average over $\textbf{W}$ could in principle be computed immediately (see e.g.~Ref.~\cite{Zavatone2023}, where this strategy is applied to the case of Wishart product matrices).
However, this gives rise to determinants, which are less convenient in our context in view of the following calculation.
An alternative strategy, applied for instance in Ref.~\cite{budnick2025eigenvaluestatisticsdilutedwishar} 
to
the case of diluted Wishart matrices, is to introduce
$M\times n $ Hubbard--Stratonovich transformations, with auxiliary variables $u_{k\alpha}$, to decouple the squared term before taking the average.
This gives (omitting the $(1/2\pi)^{nM/2} $ prefactor)
\begin{align}
    & 
    \Biggl \langle \mathrm e^{\frac{\rm i \nu }{2M^{\gamma}} \sum\limits_{k=1}^{M}  \sum\limits_{\alpha=1}^{n} \left( \sum\limits_{i=1}^{N} r_{i \alpha} W_{ik} \right)^2} \Biggr \rangle_{\textbf{W}} 
    \nonumber\\& \qquad\quad 
    \propto \int \prod_{k=1}^{M} \prod_{\alpha=1}^{n} \mathrm{d}u_{k \alpha} \, \mathrm e^{-\frac{u_{k \alpha}^2}{2} } \,
   \Biggl \langle \mathrm e^{ \frac{\sqrt{\nu }}{\sqrt{\mathrm i M^{\gamma}}}\sum\limits_{k=1}^{M} \sum\limits_{\alpha=1}^{n} u_{k \alpha} \sum\limits_{i=1}^{N} r_{i \alpha} W_{ik} } \Biggr \rangle_{\textbf{W}} 
    , 
\end{align}
where
the average over the Gaussian variables $W_{ik}$ 
can be easily computed as
\begin{align}
    \Biggl \langle \mathrm e^{ \frac{\sqrt{\nu }}{\sqrt{\mathrm i M^{\gamma}}}\sum\limits_{k=1}^{M} \sum\limits_{\alpha=1}^{n} u_{k \alpha} \sum\limits_{i=1}^{N} r_{i \alpha} W_{ik} } \Biggr \rangle_{\textbf{W}} 
    & = \mathrm e^{\frac{\nu }{2\mathrm i M^{\gamma}} \sum\limits_{i=1}^N \sum\limits_{k = 1}^M \left( \sum\limits_{\alpha = 1}^{n} u_{k \alpha} r_{i \alpha}  \right)^2 }.	
\end{align}
We now introduce the normalized densities $\phi(\vec{u}) $ and $\psi(\vec{r})$
defined as
\begin{align}
    \phi(\vec{u}) & = \frac{1}{M} \sum_{k = 1}^{M} \prod_{\alpha = 1}^{n} \delta(u_{\alpha} - u_{k \alpha}) = \frac{1}{M} \sum_{k = 1}^{M}  \delta(\vec{u} -\vec{u}_k) 
    \, , \\
    \psi(\vec{r}) & = \frac{1}{N} \sum_{i = 1}^{N} \prod_{\alpha = 1}^{n} \delta(r_{\alpha} - r_{i \alpha}) = \frac{1}{N} \sum_{i = 1}^{N}  \delta(\vec{r} -\vec{r}_i)
    \, ,
\end{align}
where $\vec{u}_k $ and $\vec{r}_i$ are the $n$-dimensional  vectors extracted from the lines of $u_{k \alpha}$ and $r_{i \alpha}$, i.e.~$\vec{u}_k = \left( u_{k \alpha}\right)_{\alpha = 1}^n$ and $\vec{r}_i = \left( r_{i \alpha}\right)_{\alpha = 1}^n$. 
(By contrast, in \cref{appeq:a1} we have used boldface to denote a vector $\bm r \in \mathbb{R}^N$.)
Inserting these expressions, and 
performing simplifications similar to the ones applied in Ref.~\cite{Venturelli_2023} to the case of the 
GRP model, we finally recover the path-integral representation reported in \cref{eq:Zn-lambda-RPW,eq:action_RP}.

We can now evaluate the average of the replicated partition function through the saddle-point method, in the limit $N, M \rightarrow \infty$ but with the ratio $c = N/M$ kept fixed. 
The condition of minimisation of the action, 
namely
\begin{equation}
    \frac{\delta \mathcal{S}_n}{\delta \psi}=\frac{\delta \mathcal{S}_n}{\delta \phi}=\frac{\delta \mathcal{S}_n}{\delta \hat{\psi}}=\frac{\delta \mathcal{S}_n}{\delta \hat{\phi}}=0 ,
    \label{eq:spc}
\end{equation}
yields the following four saddle-point equations:
\begin{align}
    \label{eq:saddlepoint_RP1}
    \hat{\psi}(\vec{r}) & = -\frac{1}{2} \eta  \int
    \mathrm{d}\vec{u} \; \phi(\vec{u}) \left( \vec{u} \cdot \vec{r} \right)^2 
    \, , 
    \\
    \label{eq:saddlepoint_RP2}
    \psi(\vec{r}) & =   
 \frac{ \varphi_a(-\vec{r}^{\, 2}/2) 
 \exp \left[-\frac{{\rm i} }{2} \lambda \vec{r}^{\, 2} + {{\rm i}}  \hat{\psi}(\vec{r}) \right]}{\int
    \mathrm{d}\vec{r}{\, '} \varphi_a(-\vec{r}{\, '}^2/2) \exp \left[-\frac{{\rm i} }{2} \lambda \vec{r}{\, '}^2 + {\rm i}  \hat{\psi}(\vec{r}{\, '}) \right]} 
    \, , 
    \\
    \label{eq:saddlepoint_RP3}
    \hat{\phi}(\vec{u}) & = -\frac{1}{2} c \eta  \int
    \mathrm{d}\vec{r} \; \psi(\vec{r}) \left( \vec{u} \cdot \vec{r} \right)^2 \, , 
    \\
    \label{eq:saddlepoint_RP4}
    \phi(\vec{u}) & = \frac{\exp \left[-\frac{1}{2} \vec{u}^{\, 2} + {\rm i}  \hat{\phi}(\vec{u}) \right]}{\int
    \mathrm{d}\vec{u}{\, '} \; \exp \left[-\frac{1}{2} \vec{u}{\, '}^2 + {\rm i}  \hat{\phi}(\vec{u}{\, '}) \right]}
    \, ,
\end{align}
where we recall the definition of $\eta = \nu M^{1-\gamma}$. Now we note that, according to the Edwards--Jones formula~\eqref{eq:edwardsJones}, the spectral density can be recovered as
\begin{equation}
    \label{eq:edwardsJones2}
    \rho(\lambda) = -\frac{2}{N\pi} \lim_{\epsilon \to 0^{+}} \operatorname{Im} \frac{\partial}{\partial \lambda} 
    \lim_{n \to 0} \frac{1}{n} \ln \langle \mathcal{Z}^n(\lambda) \rangle.
\end{equation} 
Using the saddle-point construction and \cref{eq:Zn-lambda-RPW}, 
we can express $\langle \mathcal{Z}^n(\lambda) \rangle$ 
as
\begin{align}
    \langle \mathcal{Z}^n(\lambda) \rangle  \simeq \exp \left\{ - \sqrt{NM} \mathcal{S}_n [ \phi^*, \hat{\phi}^*, \psi^*, \hat{\psi}^* ; \lambda]  \right\} , 
    \label{eq:Zn-lambda-RPW-2}
\end{align}
where the 
superscript ``star'' (introduced here for clarity, but omitted hereafter) indicates that these fields are solutions of the saddle-point equations~(\ref{eq:saddlepoint_RP1})--(\ref{eq:saddlepoint_RP4}).
In particular, we note that there is only one term in the action~\eqref{eq:action_RP} in which $\lambda$ appears explicitly, and it only involves the field $\hat{\psi}$. This implies that, in order to derive the spectral density, we only need to find the solution for $\hat{\psi}$ --- indeed, contributions coming from the implicit derivatives of the other fields with respect to $\lambda$ (calculated according the standard chain rule) vanish by construction at the saddle point due to \cref{eq:spc} (see Ref.~\cite{Venturelli_2023}). 
Thus, our
strategy will be to find a self-consistent equation for $\hat{\psi}$,
which can be achieved as follows:

\begin{enumerate}[(i)]
    \item First, we insert Eq.~(\ref{eq:saddlepoint_RP3}) into Eq.~(\ref{eq:saddlepoint_RP4}) to eliminate $\hat\phi$ 
and get an expression for $\phi$ as a functional of $\psi$:
\begin{equation}
    \label{eq:saddlepoint_RP5}
    \phi(\vec{u}) = \frac{1}{z_{\phi}} \exp \left [ -\frac{1}{2} \vec{u}^{\, 2} -{\rm i}  \frac{c\eta }{2} \int
   \mathrm{d}\vec{r} \; \psi(\vec{r}) \left( \vec{u} \cdot \vec{r} \right)^2 \right ]
    \, ,
\end{equation}
where $z_\phi = \int
    \mathrm{d}\vec{u} \; \exp \left [-\frac{1}{2} \vec{u}^2 -{\rm i}  \frac{c\eta }{2} \int
    \mathrm{d}\vec{r} \; \psi(\vec{r}) \left( \vec{u} \cdot \vec{r} \right)^2 \right ] $.

\item Second, we insert Eq.~(\ref{eq:saddlepoint_RP2}) into Eq.~(\ref{eq:saddlepoint_RP5}), to get an expression for 
$\phi$ as a functional of $\hat{\psi}$:
\begin{equation}
    \label{eq:saddlepoint_RP6}
    \phi(\vec{u}) = 
    \frac{1}{z_{\phi}}
    \exp \left[
    -\frac{1}{2} \vec{u}^{\, 2} -{\rm i}  \frac{c\eta }{2z_{\psi} } \int
   \mathrm{d}\vec{r} \; 
    \varphi_a\left(- \frac{ \vec{r}^{\, 2}}{2}\right) 
    \mathrm e^{
    -\frac{{\rm i} }{2} \lambda \vec{r}^{\, 2} + {\rm i}  \hat{\psi}(\vec{r}) 
    }
    \left( \vec u' \cdot \vec{r} \right)^2 
   \right] ,
\end{equation} 
where $z_{\psi} = \int \mathrm{d}\vec{r} \; 
    \varphi_a(-\vec{r}^{\, 2}/2) \exp \left[-\frac{{\rm i} }{2} \lambda \vec{r}^{\, 2} + {\rm i}  \hat{\psi}(\vec{r}) \right]$.
    
\item Finally, we insert Eq.~(\ref{eq:saddlepoint_RP6}) into Eq.~(\ref{eq:saddlepoint_RP1}), and deduce the self-consistency equation for 
$\hat{\psi}$:
\begin{align}
    \label{eq:saddlepoint_RP7}
     \hat{\psi}(\vec{r}) 
    =-\frac{\eta }{2z_{\psi}} \int
    \mathrm{d}\vec{u} \; 
    \mathrm e^{
    -\frac{ \vec{u}^{\, 2}}{2} -   \frac{{\rm i} c \eta }{2z_{\psi} } \int
    \mathrm{d}\vec{r} \; 
    \varphi_a\left(-\frac{ \vec{r}^{\, 2}}{2}\right) \exp \left [ -\frac{{\rm i} \lambda \vec{r}^{\, 2}}{2}  + {\rm i}  \hat{\psi}(\vec{r}) \right ]
    \left( \vec u' \cdot \vec{r} \right)^2 
    }
    \left( \vec{u} \cdot \vec{r} \right)^2
    \, .
\end{align}
\end{enumerate}

In the following,
we shall try to find a solution of 
\cref{eq:saddlepoint_RP7}
in the form of a rotationally invariant Ansatz, i.e.~$\hat{\psi}(\vec{r}) = \hat{\psi}(r)$, which only depends on the norm of $\vec{r}$ 
in replica space. 
Using the identity $\int \mathrm{d} \Omega_n \; (\vec{u} \cdot \vec{r})^2 = \frac{(ur)^2}{n} \int \mathrm{d} \Omega_n $,
where $\dd{\Omega_n}$ is the differential of the $n$-dimensional solid angle in spherical coordinates,
we obtain
\begin{align}
    & \hat{\psi}(r) 
    =  -r^2\frac{\eta \int \mathrm{d} \Omega_n}{2 n z_{\psi}}
    \int \mathrm{d}u \; u^{n+1} 
    \mathrm e^{-\frac{u^2}{2}  - \frac{{\rm i} c\eta\int \mathrm{d} \Omega_n}{2nz_{\phi}} u^2 \int \mathrm{d}r' r'^{\, n+1} 
    \varphi_a\left(-\frac{ r'^{\, 2}}{2}\right)  
    \exp \left [
    -\frac{{\rm i} \lambda r'^{\, 2}}{2}  + {\rm i}  \hat{\psi}(r')   
    \right ]  
    }.
     \label{eq:saddlepoint_RP8}
\end{align}
This expression can then be simplified by defining
\begin{equation}
    J(r;\lambda)  = \varphi_a(-r^2/2) \; \exp \left[ -\frac{{\rm i} }{2} \lambda r^2 + {\rm i}  \hat{\psi}(r) \right] 
    \, , 
    \label{eq:J_psihat}
\end{equation}
with $J'(r;\lambda)  = \partial_r J(r;\lambda)$. Rewriting $z_{\psi} = \int \mathrm{d} \Omega_n \int \mathrm{d}r \; r^{n-1} \; J(r;\lambda)$ and integrating by parts $ z_{\psi}= \frac{\int \mathrm{d} \Omega_n }{n} \int \mathrm{d}r \; r^{n} \; J'(r;\lambda)$, we obtain
\begin{equation}
    \hat{\psi}(r) = -r^2 \frac{\eta \int \mathrm{d} \Omega_n }{2 n z_{\phi}}   \int \mathrm{d}u \; u^{n+1} \exp \left [-\frac{1}{2} u^2 +{\rm i}  \frac{c\eta }{2}u^2 \frac{ \int \mathrm{d}r' \; r'^{n+1} J(r';\lambda) }{ \int \mathrm{d}r' r'^{n} J'(r';\lambda)} \right ] 
    \, .
\end{equation}
This expression can be further simplified by defining 
\begin{equation}
    F_n(u;\lambda) = \exp \left[ -\frac{1}{2} u^2 + {\rm i}  \frac{c\eta }{2} u^2 \frac{ \int \mathrm{d}r \; r^{n+1} J(r;\lambda) }{ \int \mathrm{d}r \; r^{n} J'(r;\lambda)} \right]
    \, , 
    \label{eq:Fn}
\end{equation}
and $F_n'(u;\lambda) = \partial_u F_n(u;\lambda)$. Again, rewriting $z_\phi =\int \mathrm{d} \Omega_n \int \mathrm{d}u \; u^{n-1} \; F_n(u;\lambda)$ and integrating by parts $z_\phi = \frac{\int \mathrm{d} \Omega_n}{n} \int \mathrm{d}u \; u^{n} \; F_n'(u;\lambda)$ we have 
\begin{equation}
    \hat{\psi}(r) = \frac{\eta}{2} r^2 \;  \frac{ \int \mathrm{d}u \; u^{n+1} F_n(u;\lambda)}{\int \mathrm{d}u \; u^{n} F_n'(u;\lambda)}
    \, .
\end{equation}
We can now take the limit $n \rightarrow 0$, and arrive at
\begin{equation}
    \label{eq:psihat}
    \hat{\psi}(r) = \frac{\eta}{2} \; r^2 \;  \frac{ \int \mathrm{d}u \; u \, F_0(u;\lambda)}{\int \mathrm{d}u  \; F_0'(u;\lambda)}.
\end{equation}

Let us now go back to the Edwards--Jones formula and the replica trick in 
\cref{eq:edwardsJones,eq:replicatrick},
which allow us to compute the spectral density as
\begin{align}
    \label{eq:EdwardsJones}
    \rho(\lambda) 
    & = - \frac{2}{\pi N} \lim_{\epsilon \rightarrow 0^+} \Im  \frac{\partial}{\partial \lambda}  \lim_{n \to 0}\frac{ \ln \langle \mathcal{Z}^n(\lambda_{\epsilon}) \rangle }{n} 
     =  \frac{2}{\pi N} \lim_{\epsilon \rightarrow 0^+} \Im \lim_{n \to 0}   \frac{1}{n} \sqrt{NM}  \frac{\partial}{\partial \lambda} \mathcal{S}_n^{\rm sp} 
     \nonumber \\
    & =  \frac{2 }{\pi\sqrt{c}} \lim_{\epsilon \rightarrow 0^+} \Im \lim_{n \to 0} \frac{-{\rm i}  \sqrt{c} }{2} \frac{1}{n} \frac{\int \mathrm{d}r \; r^{n+1} J(r;\lambda_\epsilon)}{\int \mathrm{d}r \; r^{n-1} J(r;\lambda_\epsilon)} \nonumber \\
    & =  \frac{1}{\pi} \lim_{\epsilon \rightarrow 0^+} \Im \mathrm{i}  \frac{\int \dd{r} r J(r;\lambda_\epsilon)}{\int \mathrm{d}r \; J'(r;\lambda_\epsilon)} \equiv \frac{1}{\pi} \lim_{\epsilon \rightarrow 0^+} \Im  G(\lambda_\epsilon)
    \, ,
\end{align}
where we have identified the resolvent
\begin{equation}
    \label{eq:Glambda}
    G(\lambda) \equiv \; {\rm i}  \; \frac{\int \mathrm{d}r \; r J(r;\lambda)}{\int \mathrm{d}r \; J'(r;\lambda)}.
\end{equation}
To 
obtain a closed equation for
the resolvent, we first rewrite 
\cref{eq:Fn} as
\begin{equation}
F_0(u;\lambda) =  \exp \left[ -\frac{1}{2} u^2  + \frac{c\,\eta}{2} u^2 G(\lambda) \right]
\, .
\end{equation} 
From this expression, if we assume that $G(\lambda)$ has a negative real part on the support of $\rho$ (to be checked a posteriori), then  we have that $F_0(u \to \infty ;\lambda) = 0$ and $F_0(u \to 0 ;\lambda) = 1 $. This allows us to simplify the denominator of Eq.~(\ref{eq:psihat}), and then by applying the change of variables $y = u^2/2$ we find
\begin{align}
    \hat{\psi}(r) & = \frac{\eta}{2} r^2  \int_0^{\infty} \mathrm{d}u \; u \exp \left[ -\frac{1}{2} u^2 + \frac{c\,\eta}{2} u^2 G(\lambda) \right] \nonumber \\
    & = \frac{\eta}{2} r^2  \int_0^{\infty} \mathrm{d}y \; \exp \left[ -y(1 - c \eta G(\lambda)) \right] 
    = \frac{\eta}{2} r^2  \frac{1}{1 -  c \,\eta G(\lambda)}.
    \label{appeq:hat-psi-G}
\end{align}
Next, since $G(\lambda)$ has a positive imaginary part, 
from \cref{appeq:hat-psi-G,eq:J_psihat} we deduce that
$J(r \to \infty ;\lambda) = 0$ and $J(r \to 0 ;\lambda) = 1 $. These limits let us simplify the denominator of Eq.~(\ref{eq:Glambda}),
so that by applying the change of variables $ z =r^2/2 $ 
and inserting \cref{eq:J_psihat} we get
\begin{align}
    G(\lambda) & =  {\rm i}  \int_0^{\infty} \mathrm{d}r \; r \; \varphi_a(-r^2/2) \exp \left[ -\frac{{\rm i} }{2} \lambda r^2 + {\rm i}  \hat{\psi}(r) \right] \nonumber \\
    & = {\rm i}  \int_0^{\infty} \mathrm{d}r \; r \; \varphi_a(-r^2/2) \exp \left[ -\frac{{\rm i} }{2} \lambda r^2 + \frac{{\rm i} }{2} r^2 \frac{\eta}{1 - c \, \eta G(\lambda)} \right] .
\end{align}
Finally, upon changing variables as $z=r^2/2$ we recover Eq.~\eqref{eq:resolvantWRP}.

\subsection{Full counting statistics and level compressibility}

To obtain the action~\eqref{eq:action_RP_levelcompress},
we follow very similar steps to the ones presented in the previous Section for the density of states:
we start from \cref{eq:cgf_replicas}, we perform the Hubbard--Stratonovich transformation, 
we average over the Gaussian measure and we insert the fields as in \cref{eq:saddlepoint_RP1,eq:saddlepoint_RP2,eq:saddlepoint_RP3,eq:saddlepoint_RP4} to obtain
\cref{eq:Qalphabeta}.

Now, to evaluate the action~(\ref{eq:action_RP_levelcompress}) at the saddle point, we first write the saddle-point equations for the fields, which read
\begin{align}
    \hat{\psi}(\vec{r}) & = -\frac{1}{2} \eta  \int
    \mathrm{d}\vec{u} \; \phi(\vec{u}) M( \vec{u},\vec{r} )
    \, , 
    \\
    \hat{\phi}(\vec{u}) & = -\frac{1}{2} c \eta  \int
    \mathrm{d}\vec{r} \; \psi(\vec{r}) M( \vec{u}, \vec{r} ) \, , 
    \\
    \psi(\vec{r}) & = \frac{1}{Z_{\psi}}  
\varphi_a \left( -\frac{1}{2} \vec{r}\hat{L} \vec{r} \right)  \exp \left[-\frac{{\rm i} }{2}  \vec{r} \hat{\Lambda} \vec{r}+ {\rm i}  \hat{\psi}(\vec{r}) \right]
    \, , 
    \\
    \phi(\vec{u}) & =  \frac{1}{Z_{\phi}} \exp \left[-\frac{1}{2} \vec{u}^{\, 2} + {\rm i}  \hat{\phi}(\vec{u}) \right]
    \, ,
\end{align} where we introduced $ M( \vec{u}, \vec{r} ) \equiv ( \vec{u} \hat{L} \vec{r} )^2 $.
Inserting the first two equations back into Eq.~(\ref{eq:action_RP_levelcompress}), the action at the saddle point simplifies to
\begin{align}
    \label{eq:simplified action}
    \mathcal{S}_{n_{\pm}} [ \phi, \psi ; \hat{\Lambda}] & =- \sqrt{c} \ln Z_{\psi} - \frac{1}{\sqrt{c}} \ln Z_{\phi} - \frac{{\rm i} }{2} \sqrt{c} \eta \int \mathrm{d}\vec{r} \mathrm{d}\vec{u} \text{ }\phi(\vec{u}) \psi(\vec{r}) M( \vec{u}, \vec{r} ) ,
\end{align}
where 
\begin{equation}
    \begin{aligned}
    Z_{\psi} &= \int \mathrm{d}\vec{r} \; \varphi_a \left( -\frac{1}{2} \vec{r}\hat{L} \vec{r} \right)  \exp \left[-\frac{{\rm i} }{2}  \vec{r} \hat{\Lambda} \vec{r}+ {\rm i}  \hat{\psi}(\vec{r}) \right] , 
    \\ 
    Z_{\phi}&= \int \mathrm{d}\vec{u} \; \exp \left[-\frac{1}{2} \vec{u}^{\, 2} + {\rm i}  \hat{\phi}(\vec{u}) \right]
    \, .
    \end{aligned}
    \label{eq:Zphipsi}
\end{equation}
On the other hand, by inserting the first two saddle-point equations into the last two, one can eliminate 
the dependence on the conjugated fields $\hat \psi$ and $\hat \phi$, and 
obtain
a set of two equations for $\psi$ and $\phi$ only: 
\begin{align}
    \psi(\vec{r}) & = \frac{1}{Z_{\psi}}  
\varphi_a \left( -\frac{1}{2} \vec{r}\hat{L} \vec{r} \right)  \exp \left[-\frac{{\rm i} }{2}  \vec{r} \hat{\Lambda} \vec{r}-\frac{{\rm i} }{2} \eta  \int  \mathrm{d}\vec{u} \; \phi(\vec{u}) M( \vec{u},\vec{r} ) \right]
    \, , 
    \label{eq:SPpsi0}
    \\
    \phi(\vec{u}) & =  \frac{1}{Z_{\phi}} \exp \left[-\frac{1}{2} \vec{u}^{\, 2} -\frac{{\rm i} }{2} c \eta  \int    \mathrm{d}\vec{r} \; \psi(\vec{r}) M( \vec{u}, \vec{r} )  \right]
    \, .
    \label{eq:SPphi0}
\end{align}
To make progress, we now introduce 
an $n \times n$ matrix $\hat{K}$ such that 
\begin{equation}
    \vec{u} \hat{K} \vec{u} \equiv\int \mathrm{d}\vec{r} \; \psi(\vec{r}) M( \vec{u}, \vec{r} ) ,
    \label{eq:K}
\end{equation}
which allows us to rewrite
\cref{eq:SPpsi0,eq:SPphi0} as
\begin{align}
    \phi(\vec{u}) & = \frac{1}{Z_{\phi}} \exp \left[-\frac{1}{2}    \vec{u}\left( \hat{\mathbb{1}} +iq \eta   \hat{K}  \right) \vec{u} \right] ,  \label{eq:SPphi} \\
\psi(\vec{r}) & = \frac{1}{Z_{\psi}}  
\varphi_a \left( -\frac{1}{2} \vec{r}\hat{L} \vec{r} \right)  
\exp \left\lbrace
-\frac{{\rm i} }{2}  \vec{r} \left[ \hat{\Lambda}+\eta \hat{L} \left( \hat{\mathbb{1}} +iq \eta   \hat{K}  \right) ^{-1} \hat{L}\right] \vec{r} 
\right\rbrace
    \, .
    \label{eq:SPpsi}
\end{align} 
Inserting \cref{eq:SPpsi} into \cref{eq:K} then 
gives the following self-consistent equation for $\hat{K}$:
\begin{align}
    \label{eq:selfconsistK}
    \vec{u}\hat{K}\vec{u} = \frac{1}{Z_{\psi}} \int \mathrm{d}\vec{r} \; \;
\varphi_a \left( -\frac{1}{2} \vec{r}\hat{L} \vec{r} \right)  
\mathrm e^{
-\frac{{\rm i} }{2}  \vec{r} \left[ \hat{\Lambda}+ \eta \hat{L} \left( \hat{\mathbb{1}} + iq \eta   \hat{K}  \right) ^{-1} \hat{L}\right] \vec{r} 
}
\;M( \vec{u}, \vec{r} ),
\end{align} 
which is actually equivalent to  Eq.~(\ref{eq:matrixSPeq}) after introducing the auxiliary matrix $\hat C$.

Using 
\cref{eq:SPphi,eq:SPpsi},
we can now express $Z_{\phi}$ and $Z_{\psi}$ 
in Eq.~\eqref{eq:Zphipsi} 
as 
\begin{align}
    Z_{\phi} & = (2\pi)^{n/2} \times \left[\det \left( \hat{\mathbb{1}} +iq \eta   \hat{K}  \right) \right]^{-1/2}, \\
    Z_{\psi} & = \int_{-\infty}^{\infty} 
     \mathrm{d}a \; p_a(a) \; \left[  \det \left(   \hat{C}^{-1}- ia \hat{L} \right)   \right]^{-1/2}.
\end{align}
Furthermore, the interaction term in the action~\eqref{eq:simplified action} can be rewritten as
\begin{align}
    &\int \mathrm{d}\vec{r} \mathrm{d}\vec{u} \text{ }\phi(\vec{u}) \psi(\vec{r}) M( \vec{u}, \vec{r} ) = 
     \int \mathrm{d}\vec{u} \text{ }\phi(\vec{u}) \int \mathrm{d}\vec{r} \text{ } \psi(\vec{r}) M( \vec{u}, \vec{r} ) = \int \mathrm{d}\vec{u} \text{ }\phi(\vec{u}) \; \vec{u} \hat{K} \vec{u} \nonumber \\
    & = \frac{1}{Z_{\phi}} \int \mathrm{d}\vec{u} \; \exp \left[-\frac{1}{2}    \vec{u}\left( \hat{\mathbb{1}} +iq \eta   \hat{K}  \right) \vec{u} \right] \; \vec{u} \hat{K} \vec{u}  
    = \Tr \left[ \hat{K} \left(  \hat{\mathbb{1}} +iq \eta   \hat{K}\right)^{-1} \right].
\end{align}
Combining these results, the expression in Eq.~(\ref{eq:simplified action}) finally simplifies 
to the one in Eq.~\eqref{eq:actionKC}.


\section{Details of the Dyson Brownian motion calculation}
\label{app:DBM}
Here we provide details of the derivations presented in Sec.~\ref{sec:Dyson}.

\subsection{First-order perturbation theory for the eigenvalues}
We begin by detailing the steps leading from \cref{eq:H_dbm} to the evolution 
equation~\eqref{eq:DBM_lambda} for the eigenvalues.
To this end, we resort to
perturbation theory. At first order, the latter
tells us that the perturbation $\delta \textbf{H}(t)$ shifts the eigenvalues of $\textbf{H}(t)$ by $\delta^{(1)} \lambda_i = \bra{\psi_i (t)} \delta \textbf{H}(t)\ket{\psi_i (t)}$, where $\ket{\psi_i (t)}$ is the $i$-th eigenvector of $\textbf{H}(t)$ at time $t$. 

First, let us study the effect of the term $\textbf{g}(t)\textbf{X}(t)^T$ in $\delta \textbf{H}(t)$ on the eigenvalues:
\begin{equation}
    \label{eq:DBMfirstorder1}
    \bra{\psi_i (t)}\textbf{g}(t)\textbf{X}(t)^T \ket{\psi_i (t)} = \sum_{n,m} \sum_l \psi_i (n; t) g_{nl}(t)X_{ml}(t) \psi_i (m; t) \equiv \xi_i(t),
\end{equation} 
where we introduced the random variable $\xi_i (t) $. As this random variable is itself a sum of random variables of finite variance, by virtue of the central limit theorem we expect it to be Gaussian distributed in the large-$N$ limit. Adopting the Itô convention, its average is given by
\begin{align}
    \langle \xi_i(t) \rangle & = \Bigl \langle \sum_{n,m} \sum_l \psi_i (n; t) g_{nl}(t)X_{ml}(t) \psi_i (m; t) \Bigr \rangle 
    \nonumber \\
    & = \sum_{n,m} \sum_l \psi_i (n; t) \langle g_{nl}(t) \rangle X_{ml}(t) \psi_i (m; t) = 0
    ,
\end{align}
while its variance reads
\begin{align}
    &\langle \xi_i(t) \xi_j (t')\rangle  = \Bigl \langle \sum_{
    n,m,n',m'
    } \sum_{l, l'} \psi_i (n; t) g_{nl}(t)X_{ml}(t) \psi_i (m; t)     \psi_i (n'; t') g_{n'l'}(t')X_{m'l'}(t') \psi_i (m'; t') \Bigr \rangle \nonumber \\
    & = \sum_{
    n,m,n',m'
    } \sum_{l, l'} \psi_i (n; t) \psi_i (m; t)  \psi_i (n'; t') \psi_i (m'; t') \langle g_{nl}(t)  g_{n'l'}(t') \rangle X_{ml}(t)  X_{m'l'}(t') \nonumber \\
    & = \sigma^2 \mathrm{d}t \; \delta (t-t') \sum_{
    n,m,n',m'
    } \sum_{l, l'} \psi_i (n; t) \psi_i (m; t)  \psi_i (n'; t') \psi_i (m'; t')  X_{ml}(t)  X_{m'l'}(t') \delta_{nl}  \delta_{n'l'} \nonumber \\
    & = \sigma^2 \mathrm{d}t \; \delta (t-t') \sum_{
    n,m,m'
    } \sum_{l} \psi_i (n; t)  \psi_i (m)  (t) \psi_i (n; t)  \psi_i (m'; t)  X_{ml} (t)  X_{m'l} (t).  
\end{align} 
We now note that, because of the normalization of the eigenvectors $\sum_n  \psi_j (n; t) \psi_i (n; t) = \braket{i(t)|j(t)} = \delta_{ij} $, the variance can be further simplified as
\begin{align}
    &\langle \xi_i(t) \xi_j (t')\rangle   = \sigma^2 \mathrm{d}t \; \delta_{ij} \; \delta (t-t') \sum_{m, m'} \sum_{l}  \psi_i (m; t)  \psi_i (m'; t)  X_{ml} (t)  X_{m'l} (t)  \nonumber \\
    & = \sigma^2 \mathrm{d}t \; \delta_{ij} \; \delta (t-t') \bra{\psi_i (t)} \textbf{X}(t)\textbf{X}(t)^T \ket{\psi_i (t)} 
     = \sigma^2 \mathrm{d}t \; \delta_{ij} \; \delta (t-t') \bra{\psi_i (t)} \textbf{H}(t) - \textbf{A}\ket{\psi_i (t)} \nonumber \\
     & = \sigma^2 \mathrm{d}t \; \delta_{ij} \; \delta (t-t') \left( \lambda_i(t) - \braket{a}_i(t) \right),
\end{align} 
where $\braket{a}_i(t)$ was given in \cref{eq:ai-def}.

Next, following similar steps as in Eq.~(\ref{eq:DBMfirstorder1}), 
one can show that the contribution of $\textbf{X}(t) \textbf{g}(t)^T $ is the same as the one of $\textbf{g}(t) \textbf{X}(t)^T $. Finally, the contribution of the term $ \textbf{g}(t)\textbf{g}(t)^T $ yields
\begin{align}
    \bra{\psi_i (t)}\textbf{g}(t)\textbf{g}(t)^T \ket{\psi_i (t)} 
    &= \sum_{n,m} \sum_l \psi_i (n; t) g_{nl}(t)g_{ml}(t) \psi_i (m; t) \nonumber \\
    &= M \sigma^2 \mathrm{d}t + \mathcal{O}( \mathrm{d}t ^{3/2} ).
    \label{eq:DBMfirstorder2}
\end{align}

\subsection{Second-order perturbation theory for the eigenvalues}

The second-order correction to the eigenvalues is given by
\begin{align}
\label{eq:2nd-eigen}
    \delta^{(2)} \lambda_i & = 
    \sum_{j \neq i} \frac{|\bra{\psi_j (t)}\delta \textbf{H}(t) \ket{\psi_i (t)} |^2}{\lambda_i - \lambda_j}\\
    &= \sum_{j \neq i} \frac{|\bra{\psi_j (t)}\textbf{g}(t)\textbf{X}(t)^T + \textbf{X}(t)\textbf{g}(t)^T  \ket{\psi_i (t)} |^2}{\lambda_i - \lambda_j} + \mathcal{O}(\mathrm{d}t^2) \nonumber \\
    & = \sum_{j \neq i} \biggl [ |\bra{\psi_j (t)}\textbf{g}(t)\textbf{X}(t)^T\ket{\psi_i (t)} |^2 + |\bra{\psi_j (t)} \textbf{X}(t)\textbf{g}(t)^T  \ket{\psi_i (t)} |^2 \nonumber \\
    & \quad\qquad + 2 \bra{\psi_j (t)}\textbf{g}(t)\textbf{X}(t)^T\ket{\psi_i (t)} \bra{\psi_j (t)} \textbf{X}(t)\textbf{g}(t)^T  \ket{\psi_i (t)}  \biggr ]  \frac{1}{\lambda_i - \lambda_j}  + \mathcal{O}(\mathrm{d}t^2). \nonumber
\end{align} 
In the second line, the second term is the same as the first one upon switching $i \leftrightarrow j$ in the numerator, while the term in the last line will give a $\delta_{ij}$ upon averaging, thus giving no contribution to the sum over $j \neq i$. 
Retaining only the terms up to order $\mathrm{d}t$ and henceforth, unless otherwise stated, omitting the time-dependence at $t$, we obtain
\begin{align}
    &\delta^{(2)} \lambda_i  = \sum_{j \neq i}  \Bigg[ 
    \sum_{n,m,n',m'}
    \psi_j (n) g_{nl}X_{ml} \psi_i (m)     \psi_j (n')  g_{n'l'}X_{m'l'} \psi_i (m') + \bigl( i \leftrightarrow j \bigr)   \Bigg]    \frac{1}{\lambda_i - \lambda_j}  \nonumber \\
    & =  \sigma^2 \mathrm{d}t \sum_{j \neq i} \left[ \sum_{
    n,m,m'
    } \sum_{l} \psi_j (n) X_{ml} \psi_i (m)     \psi_j (n)  X_{m'l} \psi_i (m')   + \bigl( i \leftrightarrow j \bigr) \right ]   \frac{1}{\lambda_i - \lambda_j} \nonumber \\
    & = \sigma^2 \mathrm{d}t \sum_{j \neq i} \Bigg[  \sum_{\substack{m,m'}} \sum_{l}  X_{ml} \psi_i (m)     X_{m'l} \psi_i (m')  + \bigl( i \leftrightarrow j \bigr) \Bigg]   \frac{1}{\lambda_i - \lambda_j} \nonumber \\
    & = \sigma^2 \mathrm{d}t \sum_{j \neq i}   \frac{\bra{\psi_i} \textbf{XX}^T \ket{\psi_i} + \bra{\psi_j} \textbf{XX}^T \ket{\psi_j}}{\lambda_i - \lambda_j} 
    = \sigma^2 \mathrm{d}t \sum_{j \neq i}   \frac{\bra{\psi_i} \textbf{H} - \textbf{A}\ket{\psi_i} + \bra{\psi_j} \textbf{H} - \textbf{A}\ket{\psi_j}}{\lambda_i - \lambda_j} \nonumber \\
    & = \sigma^2 \mathrm{d}t \sum_{j \neq i}   \frac{\lambda_i + \lambda_j  - \braket{a}_i - \braket{a}_j}{\lambda_i - \lambda_j}.
    \label{eq:DBMsecondorder}
\end{align}  
Gathering all the contributions in Eqs.~\eqref{eq:DBMfirstorder1}, \eqref{eq:DBMfirstorder2} and~\eqref{eq:DBMsecondorder}, we finally get Eq.~(\ref{eq:DBM_lambda}).

\subsection{First-order perturbation theory for the eigenvectors}

We now detail the steps leading from \cref{eq:H_dbm} to the evolution 
equation~\eqref{eq:DBM_psi} for the eigenvectors.
The correction to the $n$-th component of the $i$-th eigenvector at first order reads
\begin{align}
     \delta^{(1)}\psi_i (n) = \sum_{j \neq i} \psi_j (n) \frac{\bra{\psi_j}\textbf{g}\textbf{X}^T + \textbf{X}\textbf{g}^T + \textbf{gg}^T \ket{\psi_i}}{\lambda_i - \lambda_j}.
\end{align} 
Let us then focus on the first two terms, 
\begin{align}
    \label{eq:DBMfirstorderpsi}
    \sum_{j \neq i} \psi_j (n) \frac{\bra{\psi_j}\textbf{g}\textbf{X}^T  
     + \textbf{Xg}^T\ket{\psi_i}}{\lambda_i - \lambda_j} \equiv \sum_{j \neq i} \psi_j (n) \frac{\xi_{ij}}{\lambda_i - \lambda_j},
\end{align} 
where we defined the random variable $\xi_{ij} \equiv \bra{\psi_j}\textbf{gX}^T+ \textbf{Xg}^T \ket{\psi_i}$. Again, 
since this
is a sum of random variables with finite variance,
we expect that in the large-$N$ limit it will converge to a Gaussian random variable of mean value
\begin{align}
    \langle \xi_{ij} (t)\rangle &  =  \sum_{n,m} \sum_l \psi_j (n; t) \langle g_{nl}(t) \rangle X_{ml}(t) \psi_i (m; t) + \bigl( i \leftrightarrow j \bigr) = 0,
\end{align} 
and variance
\begin{align}
    &\langle \xi_{ij}(t) \xi_{kl} (t')\rangle  =  \Bigl \langle \bra{\psi_j}\textbf{gX}^T\ket{\psi_i} \bra{\psi_l}\textbf{gX}^T\ket{\psi_k} + \bra{\psi_j}\textbf{gX}^T\ket{\psi_i} \bra{\psi_l}\textbf{Xg}^T\ket{\psi_k} + \bigl( i \leftrightarrow j \bigr)\Bigr \rangle\nonumber \\
    & = \sigma^2 \mathrm{d}t \; \delta (t-t') \; \left [ \delta_{jl}(\lambda_i \delta_{ik}- \langle a \rangle_{ik}) + \delta_{jk}(\lambda_i \delta_{il}- \langle a \rangle_{il}) + \bigl( i \leftrightarrow j \bigr) \right ] 
    \label{eq:noise-eigenvectors}
    \\
    & = \sigma^2 \mathrm{d}t \; \delta (t-t') \; \Bigl [ (\lambda_i + \lambda_j)\bigl( \delta_{ik}\delta_{jl} + \delta_{il}\delta_{jk} \bigr) - \bigl( 
    \delta_{ik}\langle a\rangle_{jl} + 
    \delta_{il}\langle a\rangle_{jk} + 
    \delta_{jk}\langle a\rangle_{il} + 
    \delta_{jl}\langle a\rangle_{ik} 
    \bigr) \Bigr], 
    \nonumber
\end{align} 
where 
$\braket{a}_{ik}(t)$ was given in \cref{eq:aik-def}.

We finally consider the contribution of $\bra{\psi_j}\textbf{g}\textbf{g}^T \ket{\psi_i} $, finding
\begin{align}
    &\sum_{j \neq i} \psi_j (n) \frac{\bra{\psi_j}\textbf{gg}^T \ket{\psi_i}}{\lambda_i - \lambda_j}  = \sum_{j \neq i} \psi_j (n) \frac{\sum_{m,m'} \sum_l \psi_j (m)  \psi_i (m')  g_{ml} g_{m'l}}{\lambda_i - \lambda_j}  \\
    & = M \sigma^2 \mathrm{d}t \sum_{j \neq i} \psi_j (n) \frac{\sum_{m} \psi_j(m)  \psi_i (m) }{\lambda_i - \lambda_j} + \mathcal{O}(\mathrm{d}t^{3/2}) 
    = M \sigma^2 \mathrm{d}t \sum_{j \neq i} \psi_j (n) \frac{\braket{\psi_i|\psi_j} }{\lambda_i - \lambda_j}  = 0. \nonumber
\end{align}

\subsection{Second-order perturbation theory for the eigenvectors }

The second-order calculation for the eigenvectors has three contributions:
\begin{align}
\label{eq:2nd-vectors}
    \delta ^{(2)} \psi_i(n) & = \sum_{k \neq i} \sum_{l \neq i} \psi_k (n) \frac{\bra{\psi_k} \delta \textbf{H}(t)\ket{\psi_l}  \bra{\psi_l} \delta \textbf{H}(t) \ket{\psi_i} }{(\lambda_i - \lambda_k)(\lambda_i -\lambda_l)}  \\ 
    &- \sum_{k \neq i}  \psi_k (n) \frac{\bra{\psi_k} \delta \textbf{H}(t) \ket{\psi_i}  \bra{\psi_i} \delta \textbf{H}(t) \ket{\psi_i} }{(\lambda_i - \lambda_k)^2 } - \frac{1}{2} \psi_i (n) \sum_{k \neq i} \frac{ |\bra{\psi_k} \delta \textbf{H}(t) \ket{\psi_i} |^2 }{(\lambda_i - \lambda_k)^2 }. \nonumber
\end{align}
The first one is given by
\begin{align}
    \label{eq:DBM2ndorderpsi1}
   & \sum_{k \neq i} \sum_{l \neq i} \psi_k (n) \frac{\bra{\psi_k} \textbf{g}\textbf{X}^T + \textbf{X}\textbf{g}^T \ket{\psi_l}  \bra{\psi_l}  \textbf{g}\textbf{X}^T + \textbf{X}\textbf{g}^T  \ket{\psi_i} }{(\lambda_i - \lambda_k)(\lambda_i -\lambda_l)} + \mathcal{O}(\mathrm{d}t^{3/2}) \nonumber \\
   & = \sigma^2 \mathrm{d}t \sum_{ 
   \substack{k \neq i \\ l \neq i}}
    \frac{\delta_{kl} \bra{\psi_l}\textbf{XX}^T \ket{\psi_i} + \delta_{ik} \bra{\psi_l}\textbf{XX}^T \ket{\psi_l} + \delta_{il} \bra{\psi_k}\textbf{XX}^T \ket{\psi_l} + \delta_{ll} \bra{\psi_k}\textbf{XX}^T \ket{\psi_i}}{[\psi_k (n)]^{-1}(\lambda_i - \lambda_k)(\lambda_i -\lambda_l)} \nonumber \\
   & = \sigma^2 \mathrm{d}t \sum_{k \neq i } \psi_k (n) \frac{\lambda_i \delta_{ik} -\braket{a}_{ik} }{(\lambda_i - \lambda_k)^2 } +  \sigma^2 \mathrm{d}t \sum_{k \neq i } \sum_{l \neq i}\psi_k (n) \frac{\lambda_i \delta_{ik} -\braket{a}_{ik} }{(\lambda_i - \lambda_k)(\lambda_i -\lambda_l) } \nonumber \\
   & = - \sigma^2 \mathrm{d}t \sum_{k \neq i } \psi_k (n) \frac{\braket{a}_{ik} }{(\lambda_i - \lambda_k)^2 } - \sigma^2 \mathrm{d}t  \sum_{k \neq i } \psi_k (n) \frac{ \braket{a}_{ik} }{(\lambda_i - \lambda_k) }  \sum_{l \neq i} \frac{1}{(\lambda_i -\lambda_l) } \, . 
\end{align}
The second contribution
reads
\begin{align}
    \label{eq:DBM2ndorderpsi2}
    & - \sum_{k \neq i}  \psi_k (n) \frac{\bra{\psi_k} \textbf{gX}^T + \textbf{Xg}^T \ket{\psi_i}  \bra{\psi_i}  \textbf{gX}^T + \textbf{Xg}^T  \ket{\psi_i} }{(\lambda_i - \lambda_k)^2 } + \mathcal{O}(\mathrm{d}t^{3/2}) \nonumber \\
    & = - \sigma^2 \mathrm{d}t\sum_{k \neq i}  \psi_k (n) \frac{2\delta_{ik} \bra{\psi_i}\textbf{XX}^T \ket{\psi_i} + 2\delta_{ii} \bra{\psi_k}\textbf{XX}^T \ket{\psi_i}  }{(\lambda_i - \lambda_k)^2 } \nonumber \\
    & = - 2\sigma^2 \mathrm{d}t \sum_{k \neq i}  \psi_k (n) \frac{ \lambda_i \delta_{ik} - \braket{a}_{ik}  }{(\lambda_i - \lambda_k)^2 } 
    = 2\sigma^2 \mathrm{d}t \sum_{k \neq i}  \psi_k (n) \frac{  \braket{a}_{ik}  }{(\lambda_i - \lambda_k)^2 }.
\end{align}
Finally, the last term in \cref{eq:2nd-vectors} can be simplified using
\begin{align}
    \label{eq:DBM2ndorderpsi3}
    \sum_{k \neq i} \frac{ |\bra{\psi_k} \textbf{gX}^T + \textbf{Xg}^T \ket{\psi_i} |^2 }{(\lambda_i - \lambda_k)^2 } 
    =  
    \sum_{k \neq i} \frac{ \lambda_i + \lambda_k -\braket{a}_i - \braket{a}_k }{(\lambda_i - \lambda_k)^2 } .
\end{align}
%
Gathering all the contributions in Eqs.~\eqref{eq:DBMfirstorderpsi}, \eqref{eq:DBM2ndorderpsi1}, \eqref{eq:DBM2ndorderpsi2} and \eqref{eq:DBM2ndorderpsi3},
we finally 
obtain Eq.~(\ref{eq:DBM_psi}).
The noise variance in Eq.~\eqref{eq:DBM_noise_psi} follows instead from \cref{eq:noise-eigenvectors}.

\section*{References}

\bibliographystyle{iopart-num}
\bibliography{references} 

\end{document}